\documentclass[fleqn,usenatbib]{mnras}

\usepackage[T1]{fontenc}
\usepackage{ae,aecompl}
\usepackage[titletoc]{appendix}
\usepackage{lipsum}
\usepackage{dblfloatfix}
\usepackage{float}

\usepackage{adjustbox}
\usepackage{amssymb}
\usepackage{amsmath} 
\usepackage{amsfonts}

\usepackage{array}
\usepackage{caption}
\usepackage{color}
\usepackage{graphicx}	
\usepackage{float}
\usepackage[switch]{lineno}
\usepackage{multirow}
\usepackage{subcaption}
\usepackage{mwe}
\usepackage[normalem]{ulem}
\usepackage[dvipsnames]{xcolor}

\usepackage{newtxtext,newtxmath}

\newcommand{\YUS}[1]{{\color{Black}#1$^{\text{}}$}}

\usepackage{newtxtext,newtxmath}

\usepackage[T1]{fontenc}
\usepackage{ae,aecompl}

\usepackage{adjustbox}
\usepackage{amsmath} 
\usepackage{amssymb}
\usepackage{amsfonts}
\usepackage{array}
\usepackage{color}
\usepackage{graphicx}	
\usepackage[switch]{lineno}
\usepackage{multirow}
\usepackage[normalem]{ulem}
\usepackage[dvipsnames]{xcolor}
\usepackage[colorinlistoftodos]{todonotes}

\title[A porous dust layer of big aggregates]{Cometary surface dust layers built out of millimetre scale aggregates: Dependence of modelled cometary gas production on the layer transport properties. }

\author[Yu. Skorov et al.]{
Yu. Skorov$^{1,2}$,\thanks{E-mail: skorov@mps.mpg.de}
J. Markkanen$^{2,1},$
V. Reshetnyk$^{3,4},$
S. Mottola$^{5},$
M. K\"uppers$^{6},$
S. Besse$^{6},$ \and
M. R. El-Maarry$^{7},$ 
P. Hartogh$^{1}$
\\
\\
$^{1}$Max-Planck-Institut f\"ur Sonnensystemforschung, Justus-von-Liebig-Weg 3, D-37077 G\"ottingen, Germany\\
$^{2}$Institut f{\"u}r Geophysik und extraterrestrische Physik, Technische Universit{\"a}t Braunschweig, \\
\enspace Mendelssohnstr. 3, D-38106 Braunschweig, Germany\\
$^{3}$Taras Shevchenko National University of Kyiv, Glushkova ave. 2, Kyiv, Ukraine\\
$^{4}$Main Astronomical Observatory of National Academy of Science of Ukraine, Akademika Zabolotnoho Str. 27, 03680 Kyiv, Ukraine\\
$^{5}$DLR Institute of Planetary Research, Rutherfordstrasse, 2, 12489 Berlin, Germany\\
$^{6}$ European Space Agency (ESA), ESAC, Camino Bajo del Castillo s/n, 28692 Villanueva de la Cañada, Madrid, Spain\\
$^{7}$Space and Planetary Science Center, and Department of Earth Sciences, Khalifa University, PO Box 127788 Abu Dhabi, UAE\\
\\
}

\date{Accepted ?. Received YYY; in original form ZZZ}

\pubyear{2022}

\hypersetup{draft} 
\DeclareUnicodeCharacter{2212}{-}
\begin{document}

\setlength{\abovedisplayskip}{0pt}
\setlength{\belowdisplayskip}{12pt}

\label{firstpage}
\pagerange{\pageref{firstpage}--\pageref{lastpage}}
\maketitle

Monthly Notices of the Royal Astronomical Society, Volume 522, Issue 3, July 2023, Pages 4781–4800
\begin{abstract}
{The standard approach to obtaining knowledge about the properties of the surface layer of a comet from observations of gas production consists of two stages. First, various thermophysical models are used to calculate gas production for a few sets of parameters. Second, a comparison of observations and theoretical predictions is performed. This approach is complicated because the values of many model characteristics are known only approximately. Therefore, it is necessary to investigate the sensitivity of the simulated outgassing to variations in the properties of the surface layer. This problem was recently considered by us for aggregates up to tens of microns in size. For millimetre-size aggregates, a qualitative extension of the method used to model the structural characteristics of the layer is required. It is also necessary to study the role of radiative thermal conductivity, which may play an important role for such large particles. We investigated layers constructed from large aggregates and having various thicknesses and porosity and evaluated the effective sublimation of water ice at different heliocentric distances. For radiative conductivity, approximate commonly used models and the complicated model based on the Dense Medium Radiative Transfer theory were compared. It was shown that for millimetre-size aggregates careful consideration of the radiative thermal conductivity is required since this mechanism of energy transfer may change the resulting gas productivity by several times. We demonstrate that our model is more realistic for an evolved comet than simple models parameterising the properties of the cometary surface layer, yet maintains comparable computational complexity.}
\end{abstract}





\section{Introduction}
The theoretical study of gas production in comets is based on the use of some thermal model that describes the process of absorption of solar energy (which is essentially the only source of activity), leading to the sublimation of the volatile components of the nucleus. All modern models of energy distribution in the near-surface active region of the cometary nucleus somehow consider the presence of an insulating dust layer on the surface.  The assumption of an insulating layer is supported by space missions (e.g., \citealp{Keller87};  \citealp{Sunshine:2006};  \citealp{A'Hearn:2011}; \citealp{Thomas2021cometary}), in situ observations (\citealp{Spohn:2015}) and the realisation that comets composed of sublimating ice at the surface would result in gas production rates incompatible with the observations. This layer plays a crucial role in the entire process of energy redistribution. It is its structural (e.g., porosity, thickness, characteristic particle size) and thermophysical (e.g., material density, heat capacity, thermal conductivity) characteristics that determine what fraction of the absorbed solar energy can be used for sublimation and what fraction of sublimation products can escape creating the observed gaseous activity. 

Our knowledge of the properties of the near-surface region of comets is far from complete, and without exception, all the characteristics included in the models are known to us with a greater or lesser degree of accuracy. This unavoidable uncertainty or incompleteness of our knowledge raises the question of how sensitive the results of theoretical modelling are to this. \citet{Skorov2023a} for the first time attempted a comprehensive analysis of the sensitivity of the simulated gas formation to the uncertainties in the values of the model parameters, both structural and thermophysical. Layers with different structures (created from solid monomers and porous aggregates), porosity, and thickness were considered. Cases of sublimation of the most characteristic cometary volatiles ($\mathrm{H_2O, CO, CO_2}$) were studied. Uncertainties in the effective thermal conductivity due to the composition of dust particles were examined. 

We have shown that within the physically justified limits of parameter changes, gas production can vary by tens of percent. Such a high sensitivity makes the approach of selecting one or a few model parameters for the best fit with observations dubious and of little use. It should be emphasized that it is this approach to the analysis of observations that is the most commonly used so far. Instead of choosing the "best" set, we proposed to consider the totality of solutions, taking into account the existing uncertainties.

One of the important restrictions of our previous work was the fact that considering hierarchical porous layers consisting of aggregates, we had to limit our analysis to small particle sizes. Considering that spherical monomers are about submicron-micron size, we considered aggregates containing up to a thousand monomers, i.e. having a typical size of the order of tens of microns. The typical size of voids in a porous medium with a filling factor of a few tens of per cent is comparable to the particle size, i.e. this is much less than a millimetre for such small particles. In the general case, the effective thermal conductivity of the porous irradiated nucleus surface containing sublimation products includes solid, radiative, and gas components. Understanding this, we have touched on all these mechanisms of energy transfer in past work. At the same time, the radiative  thermal conductivity in the zeroth approximation is proportional to the size of voids in the medium. Therefore, their role was not great in the study of small grains, and we did not focus our attention on them. Thus, an important part of the analysis has not been considered so far. This work fills this gap. For the size of the particles considered in this study (about mm-size) the uncertainties present in the estimates of radiative thermal conductivity and the corresponding sensitivity to gas production become important. 

Not only thermophysical but also structural parameters (void size and layer permeability) directly affect the modelling of gas production. Therefore, for the first time, we present an approach to assess these characteristics in hierarchical porous layers consisting of very large aggregates, containing billions of monomers. This approach provides an estimate of the size of voids in the layer, which is required for thermal conductivity analysis.  We then consider various models for estimating thermal conductivity and analyse the sensitivity of gas production to uncertainties in the model parameters. {\color{black} Finally, we compare our model to a frequently used simplified model of cometary activity, and show that it is physically more meaningful without a significant increase in computational cost. }

\vspace{-10pt}

\section{Structural characteristics of the layer.}

Earlier we considered free molecular diffusion of gas in porous layers having a hierarchical structure, carrying out the simulation of each elementary scattering of test particle within the particulate medium (\citealp{Skorov:2022}, \citealp{Reshetnyk:2022SoSyR}). Such layers were created from relatively small clusters of the BAM2 type (i.e. ballistic agglomerates with two  migrations)  containing up to several thousand monomers (\citealp{Shen2008ApJ}). The characteristic internal porosity of such aggregates is about 60\%. In the cited papers, one can find a detail description of the structures and characteristics of the previously studied layers. Usually, the total number of monomers (elementary spheres - scatters) in the modelled layer was several million. This total number was determined from the requirements for the size of the simulated region, which should be large enough so that the boundary effects do not play any significant role. To obtain statistically reliable estimates of the transport characteristics of the layers (permeability, mean free path, and others), we used millions of test particles. One calculation required several hours of CPU time on a regular  desktop PC.

Obviously, the direct application of the deterministic modelling used before is not feasible for layers containing very large aggregates (mm-scale), and it is precisely such layers that are the focus of this study. Indeed, if we assume that the size of the monomer is submicron, then the typical size of the adequate aggregates considered in our earlier studies does not exceed tens of microns. The hypothesis about  monomer size of submicrons to microns is based both on the results of the MIDAS instrument onboard Rosetta and on extensive ground-based observations of comet dust where polarization observations play a particularly important role (\citealp{Tazaki2022}, \citealp{Mannel2019}, \citealp{Guettler2019}, \citealp{Levasseur-Regourd2018}). Dense porous aggregates of millimetre size and larger should already contain billions of monomers. The total number of monomers in a layer of such aggregates is several orders of magnitude higher, that is, $10^{11-12}$ spheres. Therefore, it is necessary to describe scattering and diffusion with a computationally more efficient algorithm.

A detailed description of the model upgrade and test results is given in Appendix A. Here we only briefly summarize the two basic ideas used. Previous simulations have shown that the diffusion of test particles in hierarchical layers is naturally divided into internal (when the particle moves inside the aggregate) and external (when the particle moves between the aggregates) parts. This division reflects two "pore scales": voids inside aggregates and voids between aggregates. The second general observation is associated with a rapid decrease in the permeability of dense porous layers: already for a porous layer with a thickness of about one hundred monomer radii, the permeability is several percent only. It means that the vast majority of particles are scattered back, i.e. molecules outbound of the bed fly out from the same side, but, of course, in a different place. Based on these observations, we propose to replace the deterministic description of the elementary scattering act with an approximate one, based on two scales of porosity and the "non-locality of scattering", i.e. difference between entering and exit. As a result of such model modernization, we obtain estimates of the average transport characteristics required primarily for use in thermophysical models. We are talking about the permeability $\Pi$ determining the energy loss due to the evacuation of sublimation products, and the mean free path $\mathit{MFP}$ of the test particle underlying the estimates of the radiative thermal conductivity of the porous layer.  

Hereafter, we present calculations of $\Pi$ and $\mathit{MFP}$ of a test particle for homogeneous random porous layers constructed from pseudo-monomers. Each pseudo-monomer is a very large mm-scale aggregate containing billions of elementary spheres. For the layers under consideration, we use the packings created earlier for homogeneous monodisperse layers of different porosity (\citealp{Skorov:2021}, \citealp{Reshetnyk:2021SoSyR}).  From the analysis of previous model layers of aggregates (\citealp{Skorov:2022}, \citealp{Reshetnyk:2022SoSyR}), we know that the description of hierarchical layers using layers from pseudo-monomers is very successful and completely adequate for our goals. We have shown that the transport characteristics of hierarchical layers built with careful control of layer connectivity (such layers we call "layers with control over contacts") differ only slightly from analogues calculations for layers from pseudo-monomers (for which there is no "control over contacts" between aggregates). This conclusion is especially strong in relation to the averaged characteristics, like permeability $\Pi$ and a particle mean free path $\mathit{MFP}$.

\begin{table}
\caption{Basic characteristics of some hierarchical layers built from pseudo-monomers without contact control and corresponding layers of monomers.}
\label{table:1}
\begin{tabular}{lllllllll}
Layer         & N collisions & MFP    & $\Delta z$      & $|\Delta x|$    & $|\Delta y|$    & $|\Delta z|$ \\
\hline
0.65spe       & 39878514     & 1.34  & 0.09 & 0.68 & 0.68 & 0.65      \\
0.65dif       & 68681394     & 1.34 & 0.07 & 0.68 & 0.68 & 0.65      \\
0.65por-R     & 64757292     & 1.36 & 0.07 & 0.69 & 0.69 & 0.66      \\
0.65por-0.5R  & 67499610     & 1.34 & 0.07 & 0.68 & 0.68 & 0.65     \\
0.65por-0.25R & 68028640     & 1.34 & 0.07 & 0.68 & 0.68 & 0.65      \\
0.75dif       & 26785823     & 2.17 & 0.17 & 1.10  & 1.10  & 1.05       \\
0.75por-R     & 21954243      & 2.20 & 0.17 & 1.11  & 1.11 & 1.06      \\
0.85dif       & 7848249      & 4.18 & 0.54 & 2.13  & 2.13  & 2.00     \\
0.85por-R     & 7476526      & 4.22 & 0.55 & 2.15  & 2.15  & 2.02      
\end{tabular}
\end{table}

A summary of the basic simulation results is given in Table \ref{table:1}. The layers are characterised by "conditional"\footnote{ For the notation of the new hierarchical layers from mm-scale pseudo-monomers we use the values of the corresponding porosity of the parental layers of monomers because it is the packing features of these layers that determine the voids between the aggregates and the resulting transport layer properties.} porosity (for the hierarchical layers) and real porosity (for the layers of monomers) and type of scattering (specular or diffuse). Shown are the total number of collisions in the layer, the mean free path, the absolute values of the average displacements along the three axes, and the average vertical displacement between collisions. All these values are normalised to the corresponding size of the monomer or pseudo-monomer. The influence of the size of the region from which particles can scatter back (i.e. the spherical segment on the pseudo-monomer) was tested when the properties of porous layers of big aggregates were estimated (see details in Appendix A). The calculations were carried out for cases when the "conditional" layer porosity was 65\% and the size of the scattering  region is equal to the  radius ($0.65-R$), half-radius ($0.65-0.5R$) and a quarter of the radius ($0.65-0.25R$) of pseudo-monomer. It can be seen that the segment size of a quarter of the radius is sufficient: the results obtained for this case differ very little from the case when the area of the "probable departure" region is increased 16 times. This allows one to speed up the simulation procedure significantly. In the table, the results obtained for homogeneous layers of monomers with porosity of 75 and 85 per cent, and the corresponding hierarchical layers built on their basis from pseudo-monomers are also shown. 
 
A direct comparison shows that the results obtained for the parental homogeneous layers built from monomers and the corresponding "child" hierarchical layers built from pseudo-monomers (mm-size aggregates) differ insignificantly. One should remember that the presented estimates are dimensionless, i.e. length is given in units of grain size. The inclusion of "non-locality" in the scattering by pseudo-monomers only slightly increases the mean free path (~2\%). The same can be said about other properties. 
 
\begin{figure} 
\includegraphics[width=.45\textwidth]{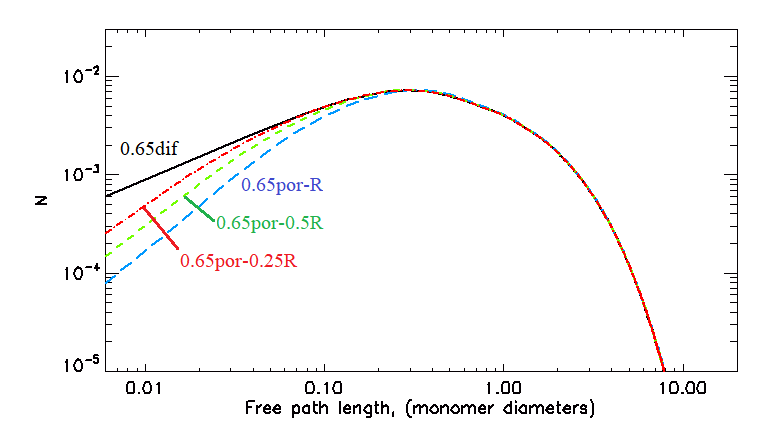}
\caption{\YUS{Free path distribution functions of hierarchical layers built from aggregates as a function of the free path length measured in pseudo-monomer size. Results were obtained for the random homogeneous layers with the different model regions of scattering non-locality (see details in the text). The porosity of the "parent" layer of the solid spheres is 65\%. Results for this layer of monomers with the same real porosity are shown for comparison. Scattering is diffuse.}} 
\vspace{-6pt}
\label{fig:Fig_1}
\end{figure}

 To better understand the obtained quantitative agreement, we show illustrative examples of the distribution of path lengths for different layers in Fig.~\ref{fig:Fig_1}. It is clear that variations in the distribution of chords are observed only for very short spans (up to about one-tenth of the particle size). The total contribution of such spans to the calculated average is negligible, which is why these differences become so small for all averages presented in Table 1. For layers with higher porosity (75\% and 85\%), there are no noticeable new features. The picture remains the same. As might be expected, changing the scattering model from diffuse to specular does not change the behavior of functions either.

\begin{figure} 
\includegraphics[width=.45\textwidth]{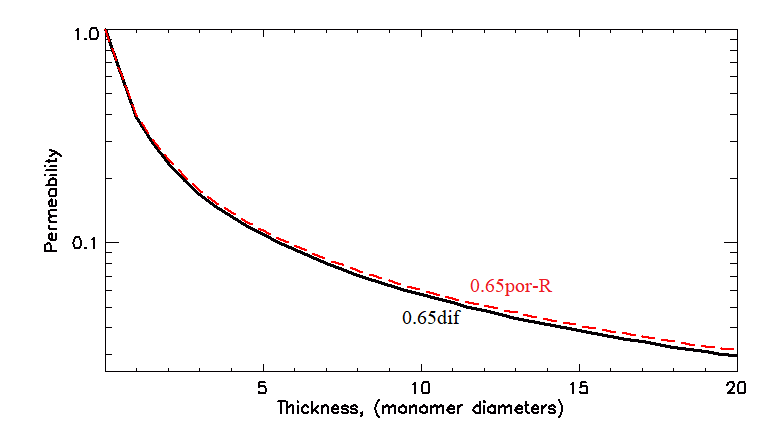}
\caption{\YUS{Permeability as a function of layer thickness in particle size. The results are shown for the parent layer of monomers and child hierarchical layers. The porosity of the monomer layer is 65\%. Scattering is diffuse. The size of the non-locality region for layers of pseudo-monomers is equal to its radius.}} 
\vspace{-6pt}
\label{fig:Fig_2}
\end{figure}

Since the permeability is a linear fractional function of the mean free path with good accuracy (\citealp{Skorov:2022}), the changes for it are also insignificant as follows from the results in Table \ref{table:1}. For permeability, the results for the "parent" layers of monomers and "child" hierarchical layers of pseudo-monomers are very close. A comparison of a layer of monomers with a real porosity of 65\% and a layer of pseudo monomers based on this case with the corresponding "conditional" porosity is shown in Fig.~\ref{fig:Fig_2}. As expected from the results shown for the free path distribution, the permeability is also insensitive to the choice of the scattering model and the model region of scattering non-locality (cutoff region size, see Appendix A). The difference is a few per cent only when the area of this region varies sixteen times. 

Detailed calculations have shown that the relative deviation of the permeability of the hierarchical layer from the reference permeability does not exceed 6\% for a layer with a thickness of twenty pseudo-monomer sizes (which means a thickness of  2 cm for millimetre aggregates). As before, reducing of the area of non-local scattering from the pseudo-monomer radius to a quarter does not increase this deviation. This allows a smaller value to be used in the calculations, speeding up the simulation.  

The size of  large aggregates is much larger than the penetration depth of the test particles. Our simulations show that their internal porosity does not significantly affect the average transport characteristics of the passage of free gas molecules through the hierarchical layer. Therefore we can define an outer boundary layer, where the vast majority of scattering events occurs. For relatively small porous aggregates, where some of the molecules can pass through the quasi-monomers, it is no longer possible to isolate such a layer. Therefore, the proposed separation of collisions inside the boundary layer and between pseudo-monomers (aggregates) becomes invalid. 

Linking these results with those obtained earlier (\citealp{Skorov:2022}, \citealp{Reshetnyk:2022SoSyR}), we can conclude that the proposed models satisfactorily describe Knudsen diffusion for the limiting cases of small (up to tens of microns) and very large (on the order of a millimetre) aggregates consisting of micron- and sub-micron sized monomers. The range of average sizes remains unexplored since both proposed modelling methods do not work well in this case evaluation of structural characteristics is difficult. The good agreement between the transport characteristics allows  us to directly use the results obtained in previous studies for layers of monomers to the corresponding cases of layers of very large aggregates. Thus, we can conclude that the limiting cases of hierarchical layers built from small porous particles (when the aggregate contains up to a thousand monomers and has a typical size of up to tenths microns) or layers built from very large porous particles (when the aggregate contains up to billions of monomers and has a typical millimetre size)  have studied. Such layers can now be analysed in thermophysical models, for example, to evaluate the possible uncertainty of modelled gas production due to the incompleteness (inaccuracy) of our knowledge of the micro-structural properties of the surface layers. 

Finally, note that the described  coincidence of transport characteristics does not mean at all that the results of thermophysical models for the layers of monomers can also be applied to mm-size grains. The heat transfer problem is not scaled like the one discussed in this section. 
The case of very large aggregates requires a revision not only of the estimates of the structural transport characteristics of the hierarchical layer (such as permeability and mean free path) but also of various types of thermal conductivity directly dependent on the layer properties. In the previous work, we focused on solid-state conductivity and only touched on the contribution of radiative thermal conductivity. 
For the gas conductivity, the simplest estimates were presented. The radiative thermal conductivity was calculated directly following the scheme suggested by \citep{Gundlach:2012}. Hereafter we consider the radiative transfer mechanism in more detail, focusing on the case when the layer consists of grains with mm-scale sizes. This is the subject of the next  section.

\vspace{-10pt}

\section{Radiative heat conductivity estimated from simple models}

The thermal conductivity of gas usually did not attract much attention in thermophysical models of cometary nuclei. At the same time, radiative thermal conduction, that is, the transfer of energy by photons is, unexpectedly, the focus of the pioneering work of \citet{Whipple:1950}. Whipple gives a simple analysis for a homogeneous grey medium under a number of simplifications, considering the energy balance for a set of plane-parallel layers, and derives a basic formula for thermal conductivity as a function of the cube of local temperature $T$, bond albedo $A_b$, Stefan-Boltzmann constant $\sigma $, and bed thickness $L_b$

\begin{equation}
K_{\mathit{rad}}= 4 \sigma (1-A_b) T^3 L_b.  
\label{eq:K_rad_Whipple}
\end{equation}

Later, \citet{MendisBrin1977} considered this kind of energy transfer, already calling it the radiative conductivity of a non-isothermal porous medium consisting of particles. In their work, the  conductivity is   proportional to the average distance between the particles, i.e. $L_b$ is the average void size. 

Without any significant changes, this formula was used in cometary publications until recently. Attention to the role of radiative thermal conduction in the energy balance of the cometary nucleus has grown markedly over the past ten years (see, e.g., \citealp{Gundlach:2012}, \citealp{Blum:2017}, \citealp{Krause2011}, \citealp{Arakawa2017thermal} \citealp{Arakawa2019geometrical}, \citealp{Sakatani2017thermal}). This interest is caused by the hypothesis of the surface layer of the cometary nucleus consisting of large particles and the observed high porosity.  \citet{Gundlach:2012} showed qualitatively that for comet-like conditions, the contribution of radiative heat conduction can be dominant in the surface layer of very big ($\gtrsim$ mm) aggregates.  At the same time, the presented formula (Eq. \ref{eq:K_rad_Whipple}) is usually used without any analysis and criticism.

Hereafter, we give a very brief overview of the approaches used to estimate radiative conductivity and discuss the existing uncertainties in the resulting estimates. Any extended overview of the state of the issue is beyond our goals, and we focus on cases applicable to cometary conditions. For a detailed treatment of the global problem see, for example, \citet{Kaviany:2014}, \citet{sparrow2018radiation}, \citet{balaji2014essentials}, \citet{delgado2011heat}.  

Models usually used in space physics to estimate the radiative heat conduction in a porous medium can be classified into two groups. The first type is based on the idealised geometry of the medium and the static radiation balance. For example, plane parallel layers of homogeneous grey medium perpendicular to the transport propagation were studied by Whipple. The second type is based on a  consideration of radiative  transfer as a random walk process. Using formally the same approach as we presented for a study of transport characteristics of random porous media the well-known Russell's type formula for the radiative conductivity can be obtained \citep{Russell:1935}:

\begin{equation}
K_{\mathit{rad}}= 4 \sigma A_{em} T^3 l_{ph},   
\end{equation}

where $A_{em}$ is a so-called exchange factor - a constant depending upon emissivity and the geometric factor, $l_{ph}$ is the mean path length of photons (or the $\mathit{MFP}$ of a test particle in our notation). \citet{Merrill:1969} 
provided a list of formulas derived for granular media (see, Table~\ref{table:Merrill}) and obtained an expression for $K_{\mathit{rad}}$ treating the radiation as a photon gas and based on the gas kinetic theory

\begin{table}
\caption{$A_{em}$ from different models reviewed in (\citealp{Merrill:1969}). All these models assume that the average photon path $l_{ph}$ equals particle size.}
\label{table:Merrill}
\addtolength{\tabcolsep}{12pt}    

\begin{tabular}{lll}
  
\hline
$\frac { \epsilon}{(2-\epsilon)\phi}$ & \citealp{wesselink1948heat} & \\

$\frac { \epsilon}{(2-\epsilon)\phi}[1-\phi^{2/3}+\phi^{4/3}]$ & \citealp{laubitz1959thermal} & \\

$\phi$    & \citealp{schotte1960thermal} & \\
\end{tabular}

\addtolength{\tabcolsep}{1pt}

\end{table}

\begin{equation}
K_{\mathit{rad}}= \frac{16 \sigma  T^3 l_{ph}} {3} .   
\end{equation}

Note that the formulas for the exchange factor given in the table are obtained for models based on the so-called unit cell approach. Strictly speaking, an understanding of the shape, size, location, and conductivity of each particle and their interaction is needed for a quantitative assessment of the radiative thermal conductivity of a granular layer. This is difficult to do even for regular packings of spheres of the same size. Therefore, the usual approach to the problem is to represent the layer by its geometrically simplified unit cell and calculate the conductivity of this representative cell. In such approximation, radiation is considered a local process that occurs between adjacent surfaces of particles in a unit cell and the long-range action of radiation is not taken into account. The main model constraints of this approach are the following (\citealp{vortmeyer1978radiation}):
1) particle size is much larger than wavelength; 2) radiating surfaces are grey and opaque; 3) relative temperature change across the particle layer is much smaller than one. More exhaustive summaries of such models can be found, for example, in (\citealp{vortmeyer1978radiation}, \citealp{Tien1988})

All these models are based on the same general approach and differ in the  estimate of the factor that somehow connects the emissivity $\epsilon$, layer porosity $\phi$, and photon mean free path $l_{ph}$. This simplicity of the approach makes it really very attractive. Estimating the $l_{ph}$ is a non-trivial problem. For example, \citet{Bosworth1952} proposed to consider it equal to particle size, and this assumption was used in all models from Table~\ref{table:Merrill}. \citet{Chan1974} estimated this parameter  for a regular dense packing of specular scattering spheres of the same size. \citet{Tien:1979} studied the limits of variation (upper and lower bounds) of this characteristic for random packings of identical spheres assuming the porous bed to be composed of four different regular structures randomly distributed throughout the bed. A Monte-Carlo computer model was used in (\citealp{Yang1983}) to evaluate the properties of random packings. 

The test particle method that we used above to calculate the $\mathit{MFP}$ can be applied directly to obtain a quantitative estimate of the photon path in various types of layers studied herein and in (\citealp{Skorov:2022}; \citealp{Skorov:2021}), and therefore obtain formulas for the corresponding radiative thermal conductivity.
Note that when a ray tracing method or diffusion approach is used, the limits of geometrical optics must be satisfied: a size parameter $S_{go}$ defined as a ratio of grain size to the wavelength of the radiation should be $\gg 1$. For the monodisperse layer composed of spheres the lower limit for geometric optics, where diffraction effects are negligible, is about $S_{go} = 115/\pi$ (\citealp{Tien1988}). Under typical cometary conditions (where the thermal wavelength is roughly 10 $\mu m$ for 300K) a particle radius should be at least above 100 $\mu m$. Thus, the application of this type of model for small spheres leads to unpredictable inaccuracy in the conductivity estimations. 

\vspace{-10pt}

\section{Radiative conductivity estimated from RT solution in particulate medium}

\begin{figure*} 
\centering
\includegraphics[width=\textwidth]{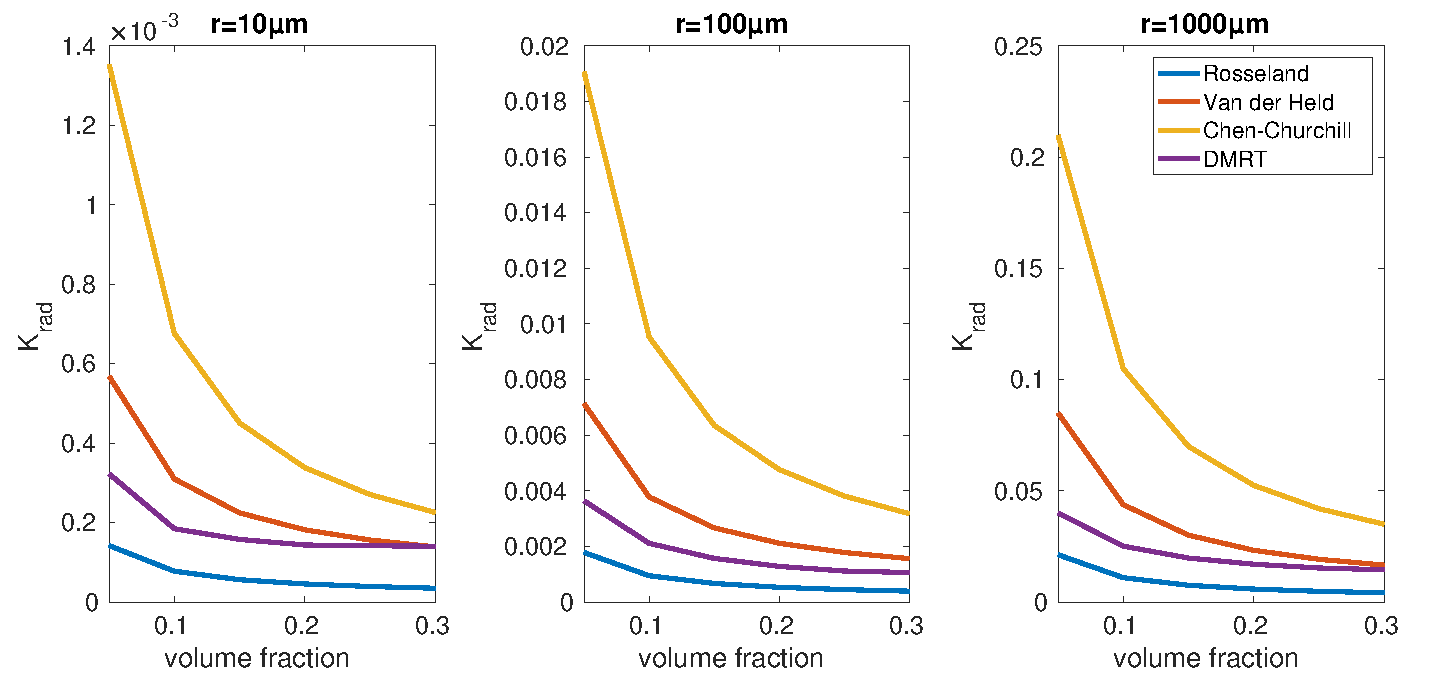} 
\caption{Radiative thermal conductivity as a function of the medium filling factor. The results are shown for three particle sizes (10, 100, and 1000 microns) under various models (see legend). The heat capacity and density for spheres are $750 \ \mathrm{Jkg^{-1} K^{-1}}$ and $1000 \ \mathrm{kg m^{3}}$. These numbers do not include macroporosity. The optical characteristics of the medium and the description of the models are presented in the text.}
\label{fig:K_rad_JM}
\end{figure*}

Above, when estimating the radiative conductivity, the particles were assumed to be opaque and the optical properties were not taken into account. An approach that considers these characteristics was proposed by \citet{Rosseland:1936} and has become widespread along with the two other approaches discussed above. Originally it was suggested for the case of the optically thick isotropically scattering and absorbing medium in a vacuum.
It consists of replacing the radiative transfer  integro-differential equations with a single heat diffusion equation including a nonlinear diffusion coefficient. In this case, the radiative thermal conductivity is expressed by the formula  \citep{vanderHeld1952}

\begin{equation}
K_{\mathit{rad}}= \frac{16 n_r^2 \sigma  T^3}{3 \beta}.   
\label{eq:K_rad_vanderHeld}
\end{equation}

where $n_r$ is the effective index of refraction, $\beta = \kappa_{sca} + \kappa_{abs}$  is the mean extinction coefficient calculated over the entire wavelength range ($\kappa_{abs}$ is the absorption coefficient, $\kappa_{sca}$ is the scattering coefficient). Below we will consider in more detail the approaches that take into account the optical properties of the medium and the radiation transfer equations. 

Later, this approach was developed further by \citet{ChenChurchill1963} by using a two-fluxes representation of RT, as   earlier proposed by \citep{Hamaker1947}. Description of the radiant intensity by forward and  backward fluxes reduces the general integro-differential equation for RT to the two differential equations, and the radiative conductivity for locations sufficiently far
from the boundaries of an optically thick bed is expressed by the formula 

\begin{equation}
K_{\mathit{rad}}= \frac{8 \sigma  T^3}{\kappa_{sca} + 2 \kappa_{back}},   
\end{equation}

where $\kappa_{back}$ is the backscattering cross-section per unit volume of packing.  
 
In all these studies an analytical expression for conductivity was obtained based on a steady-state energy balance for a differentical volume and applying certain simplifications. We use the numerical method of estimating the radiative conductivity from the RT solution for the particulate medium presented in Appendix B to  obtain quantitative estimates and compare them with the approximate analytical   solutions.  

With this model, we calculate radiative heat transfer coefficients for media consisting of randomly deposited equal-sized spherical particles by solving the heat  equation with the DMRT (dense media radiative transfer) model. We used four different sphere sizes r=1, 10, 100, and 1000 microns with varying volumetric filling factors from 0.05 to 0.3. We assumed a wavelength-independent refractive index of $m=1.6 + i0.1$ for the spheres. The one-dimensional simulation domain of length $l_d$ was discretized into 160 equal-sized elements. To minimise the boundary effects, the physical size of  $l_d$ and the total simulation time $t_{max}$ were selected such that the heat wave could not reach the boundaries. Yet, the simulation time was long enough to retrieve the heat conductivity for all volume fractions. As before the initial temperature distribution was a step function of width $w$ in the middle of the domain with the maximum temperature $T = 300$ K and minimum $T = 100$ K. The simulations were run until $t_{max}$ was reached and then the temperature distribution was fitted to the  simple model in which the radiative heat conduction coefficient was a free parameter. The parameters of the computational model are collected in Table \ref{tab:JM_label}.

\begin{table}
    \centering
     \caption{Simulation parameters}
    \begin{tabular}{llll} 
     $r [\mu m] $ & $l [m]$ & $t_{max}$ [s] & $w [m]$\\
        \hline
        1  & 0.01 & 5.0 & 0.00125\\
        10 & 0.02 & 50.0 & 0.0025\\
        100 & 0.08 & 50.0 & 0.01\\
        1000 & 2.0 & 500.0 & 0.25\\
    \end{tabular}
    \label{tab:JM_label}
\end{table}

To estimate the heat conduction coefficients with the standard models mentioned above (models of Rosseland, Chen-Churchill, and Van der Held), we calculated the scattering and absorption coefficients using the Mie theory and the static structure factor (SSF) correction. The SSF-corrected scattering, backscattering, and absorption cross-sections are calculated as follows 

\begin{equation}
C^{SSF}_{sca} = \frac{C_{sca}^{Mie}}{4\pi}\int_{4\pi} S_M(\theta) M^{Mie}_{11}(\theta) \, d\Omega,     
\end{equation}

\begin{equation}
C^{SSF}_{back} = C_{back}^{Mie}S_M(180^\circ), 
\end{equation}

\begin{equation}
C^{SSF}_{abs} = C^{Mie}_{abs}.
\end{equation}

Here, $M_{11}^{Mie}$ is the scattering phase function from the Mie solution. The explicit expression for $S_M(\theta)$ is given in \citep{Tsang2001} (Eqs. 8.4.19–8.4.22). Since the scattering and absorption cross sections are functions of wavelength, we weighted them by the Planck function $B(\lambda, T)$. Hence the scattering and absorption coefficients used in Van der Held's, and Chen-Churchill's expressions for the radiative heat conduction coefficients \citep{ChenChurchill1963, vanderHeld1952} are given by

\begin{equation}
    \kappa_{x} = \frac{\Phi}{4/3\pi r^3} \frac{\int^\infty_0 C^{SSF}_{x}(\lambda) B(\lambda, T)\, d\lambda}{\int^\infty_0 B(\lambda, T)\, d\lambda}. 
\end{equation}

Here the subscript $x$ denotes $sca$, $back$, or $abs$, and $\Phi$ is the volume fraction and $r$ is the radius of particles. In our calculations,  temperature $T$ was 300 K. The mean free path in Merrill's formulation is 

\begin{equation}
    l = \frac{1}{\kappa_{sca} + \kappa_{abs}}.
\end{equation}

This is equivalent to Van der Held model Eq. (\ref{eq:K_rad_vanderHeld}) in a vacuum where the medium refractive index is 1.

The results of calculating the radiative thermal conductivity for particles of different sizes (10, 100 and 1000 microns) using various approximate models and their comparison with the model are shown in Fig. \ref{fig:K_rad_JM}. Porosity varies from 70\% to 95\%. An average effective refractive index was selected that is related to the porous (60\% microporosity) mixture of amorphous carbon (~70 vol\%) and silicates (~30 vol\%). In reality, it should be wavelength dependent but the dependence is quite weak. However, the monomer size and macroporosity have a much stronger effect on the heat conduction coefficient than the small changes in the refractive index. The values obtained in the Chen-Churchill model are always higher than the results of the DMRT model, and the values obtained in the Rosseland model are always lower than the results of this model. The best choice among approximate models is the Van der Held model. The relative difference between the approximate models and the reference DMRT model does not change qualitatively with a hundredfold increase in particle size: from 10 to 1000 microns.

\section{Sensitivity of modelled gas production}


\begin{figure*}
\centering
 \begin{subfigure}{\columnwidth}
   \centering
  \includegraphics[width=\columnwidth]{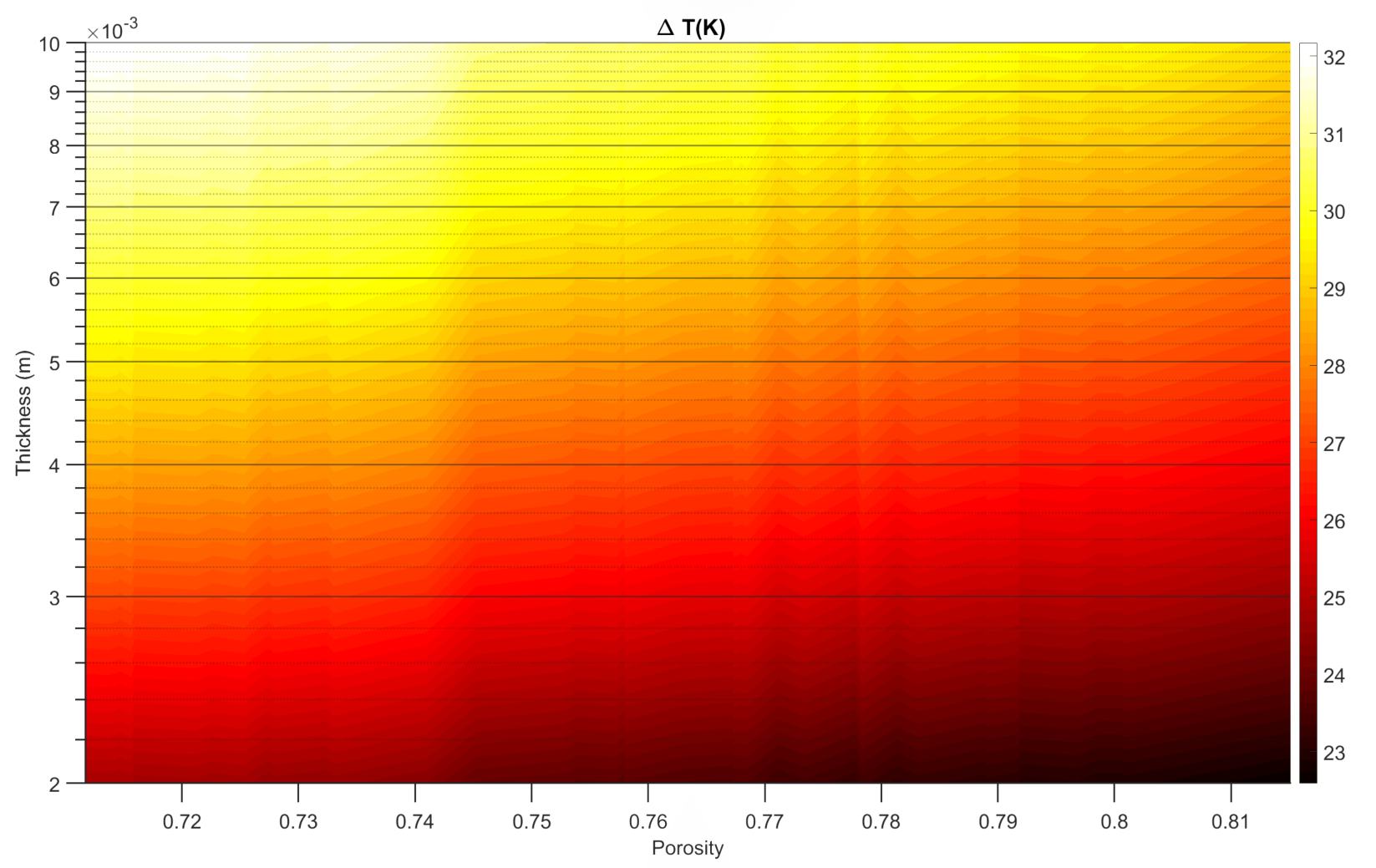}
  \end{subfigure}
  \hfill
    \begin{subfigure}{\columnwidth}
      \centering
  \includegraphics[width=\columnwidth]{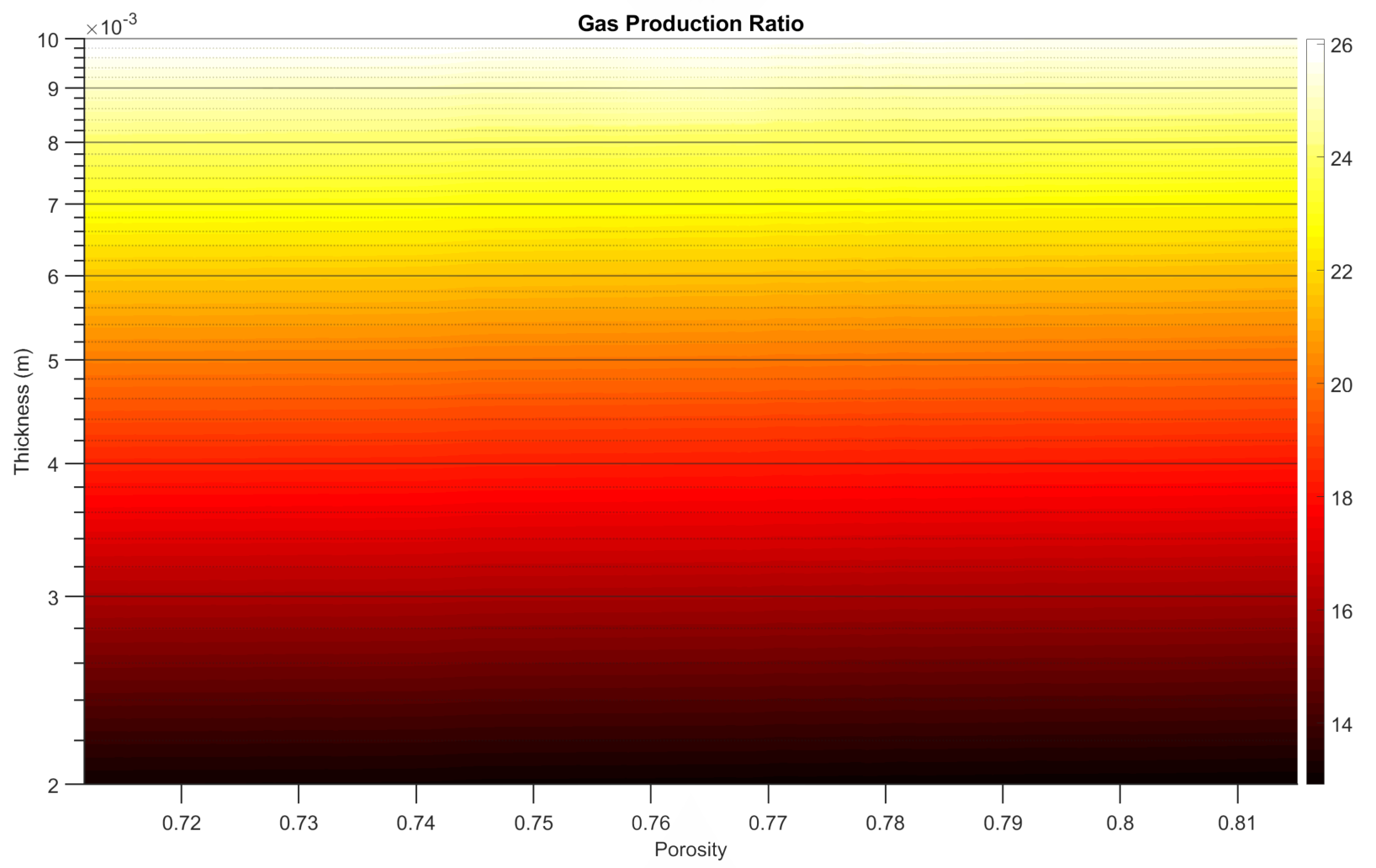}
  \end{subfigure}
   \begin{subfigure}{\columnwidth}
   \centering
  \includegraphics[width=\columnwidth]{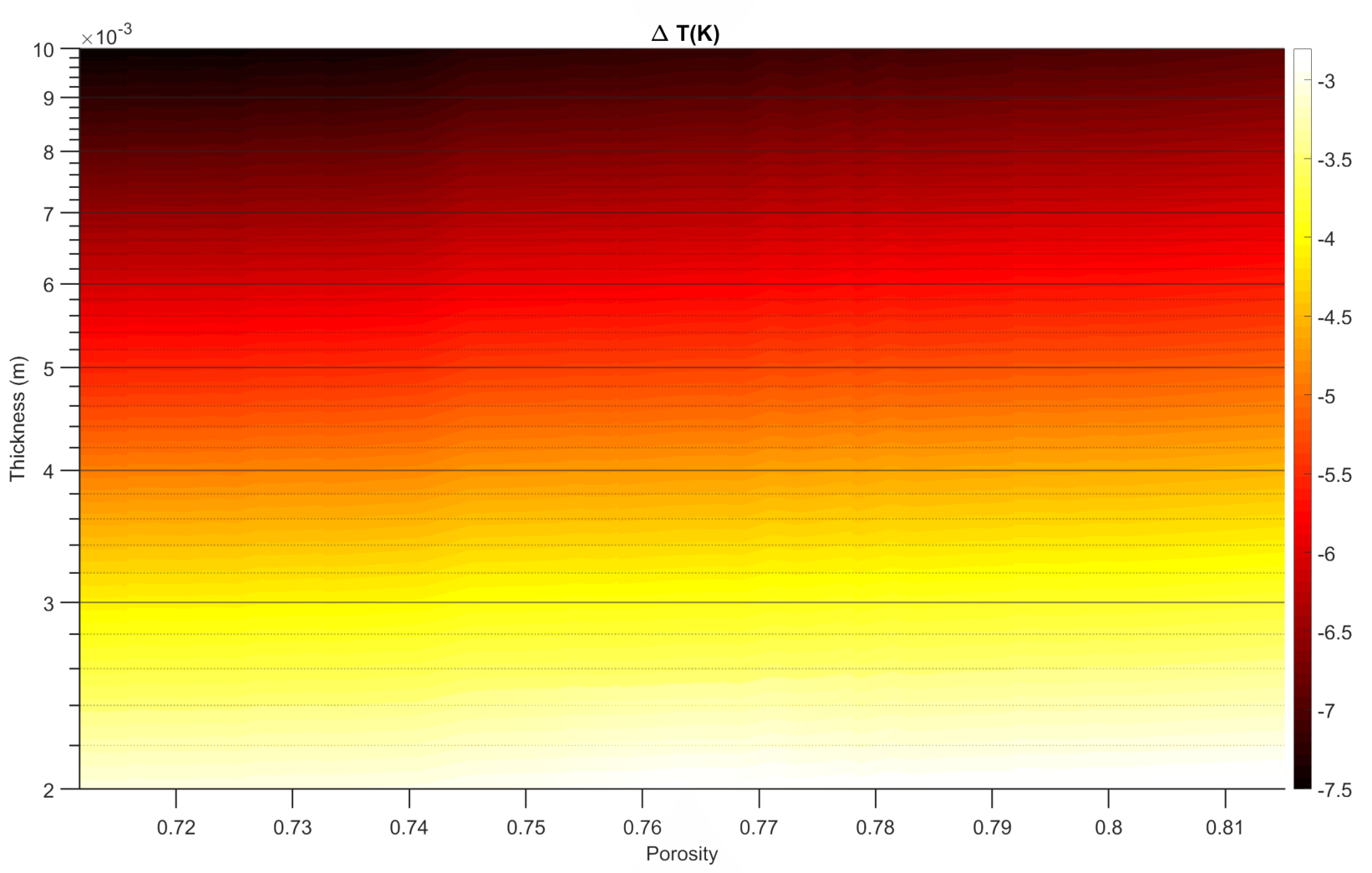}
  \end{subfigure}
  \hfill
    \begin{subfigure}{\columnwidth}
      \centering
  \includegraphics[width=\columnwidth]{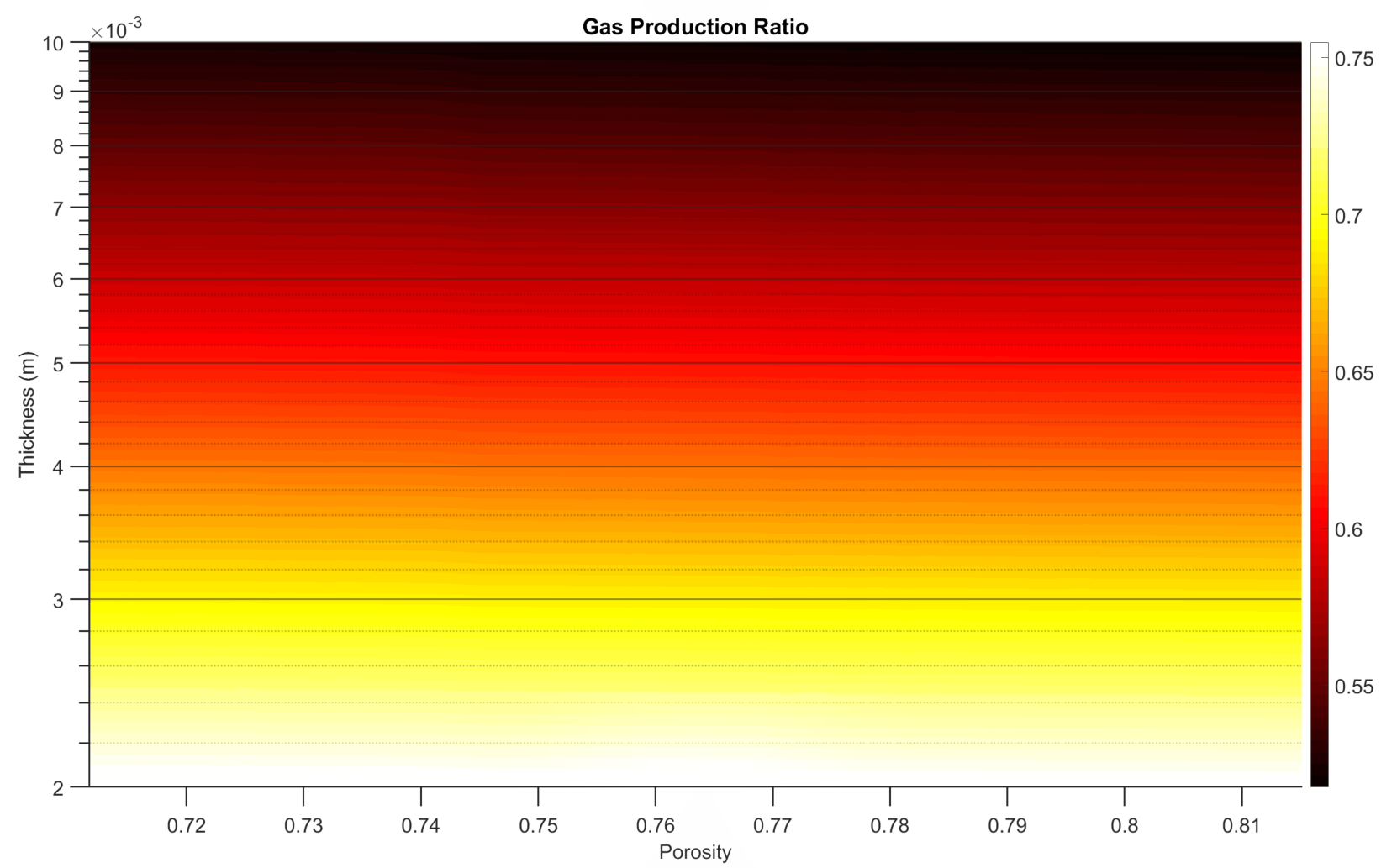}
  \end{subfigure}
     \begin{subfigure}{\columnwidth}
   \centering
  \includegraphics[width=\columnwidth]{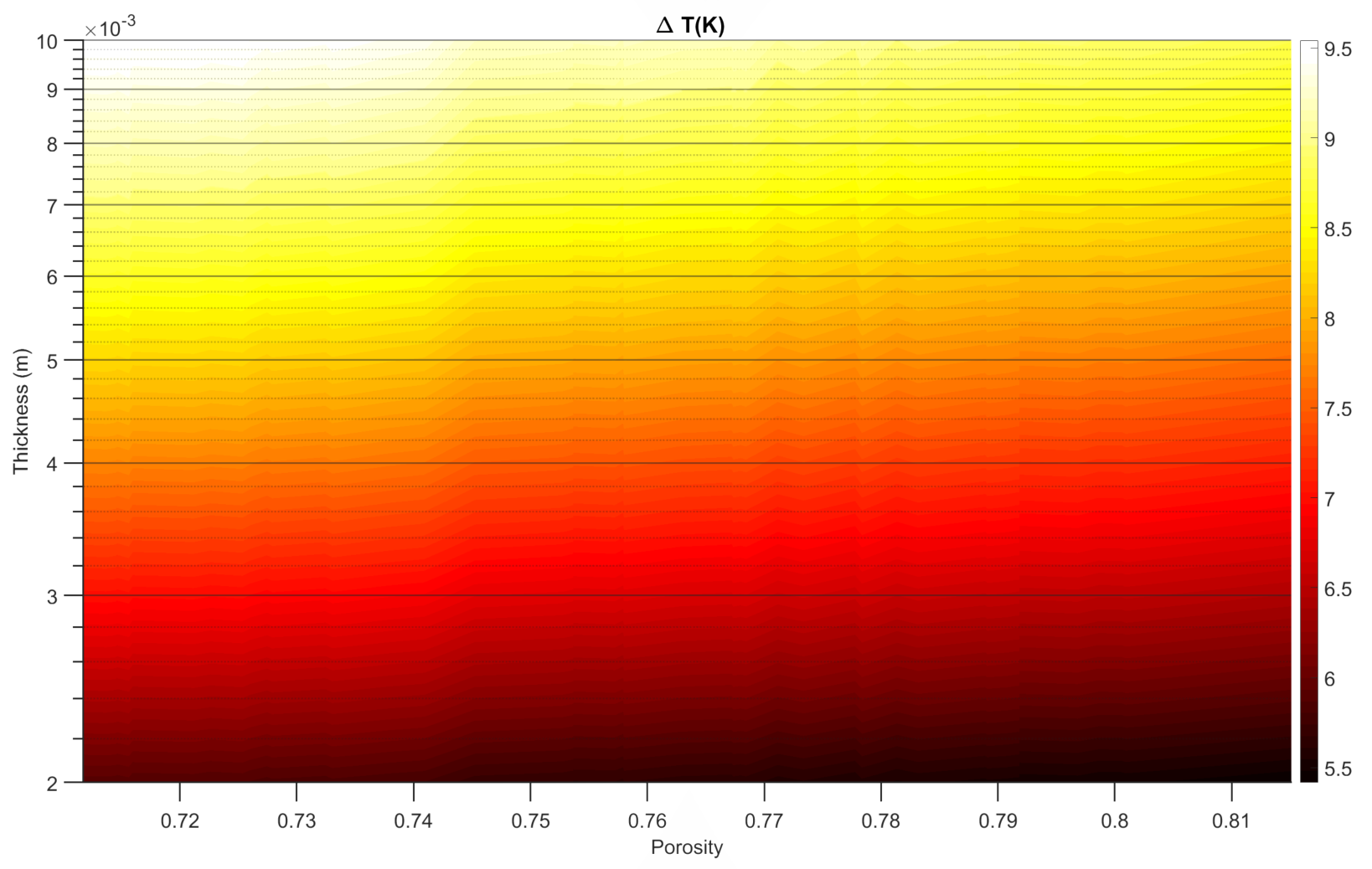}
  \end{subfigure}
  \hfill
    \begin{subfigure}{\columnwidth}
      \centering
  \includegraphics[width=\columnwidth]{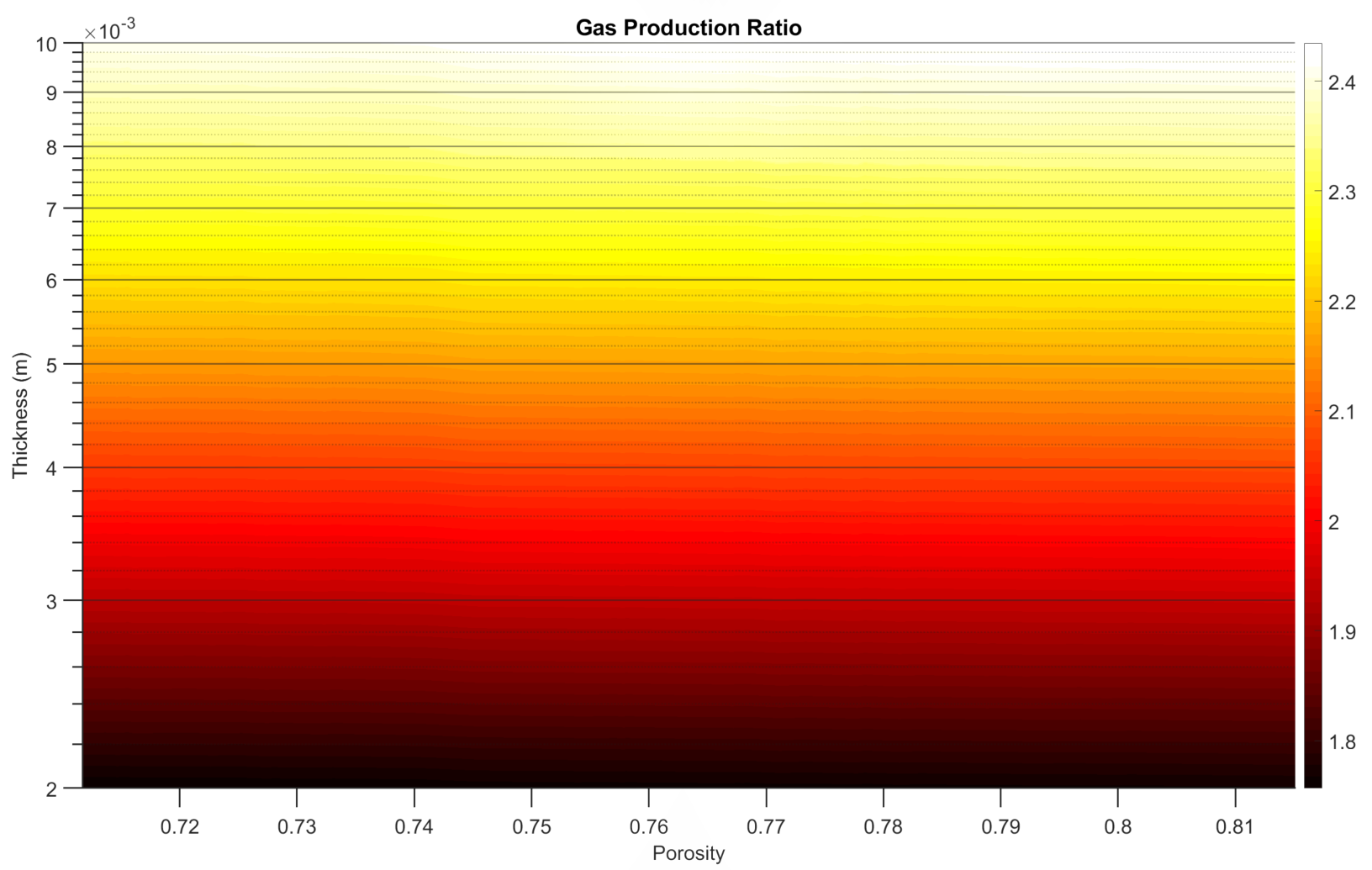}
  \end{subfigure}
\caption{\YUS{The temperature difference at the $\mathrm{H_2O}$ ice boundary (left column) and gas production ratio (right column) of the porous layer as functions of the porosity of the layer and its thickness. The particle size is 1mm.  The heliocentric distance is $R_H=1.243$ au. First row: DMRT model vs model without radiative conductivity. Second row: DMRT model vs Chen-Churchill model of radiative conductivity. Third row: DMRT model vs Rosseland model of radiative conductivity.}} 
\vspace{-6pt}
\label{fig:RA500_RH12}
\end{figure*}

\begin{figure*}
\centering
 \begin{subfigure}{\columnwidth}
   \centering
  \includegraphics[width=\columnwidth]{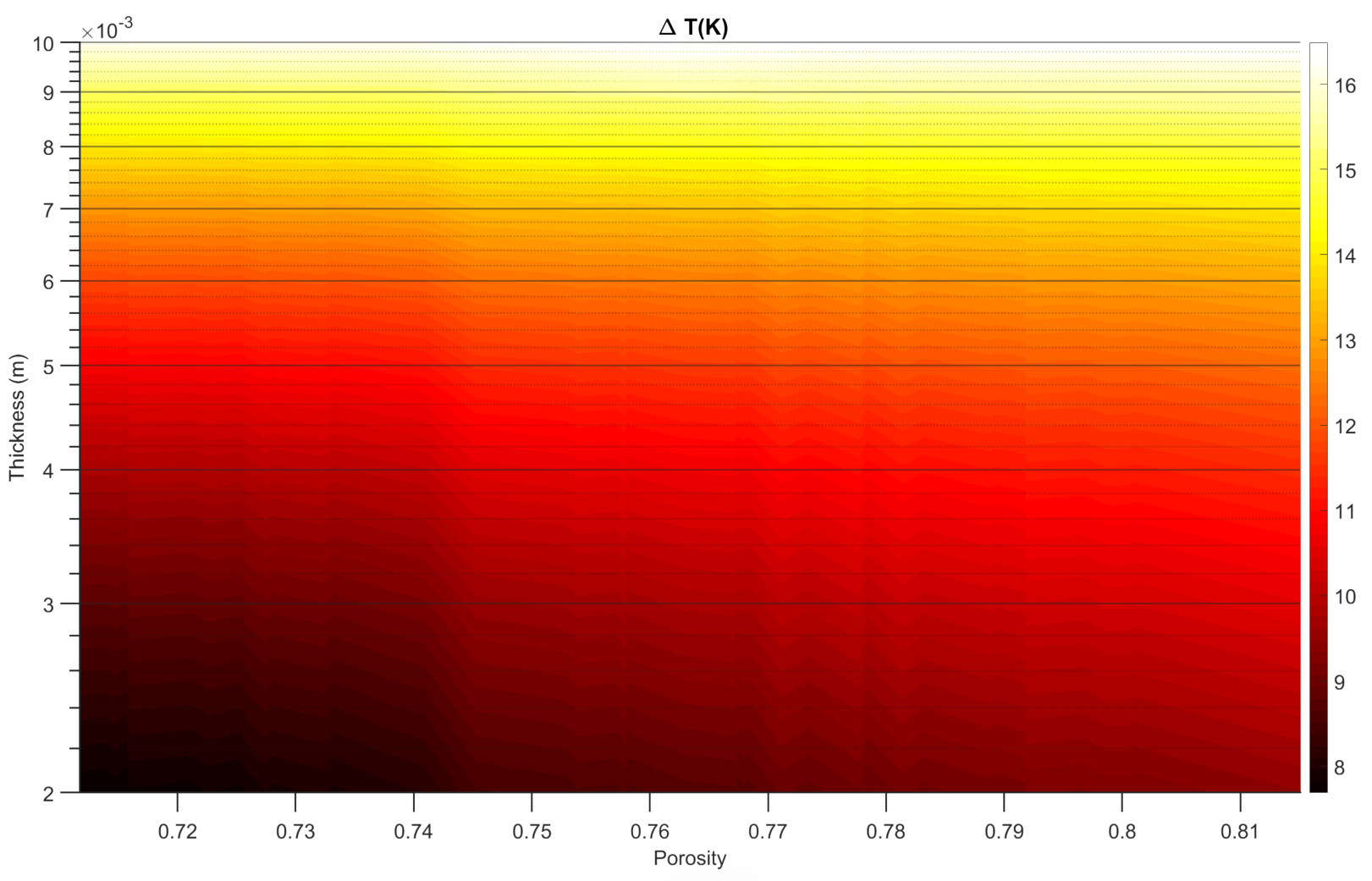}
  \end{subfigure}
  \hfill
    \begin{subfigure}{\columnwidth}
      \centering
  \includegraphics[width=\columnwidth]{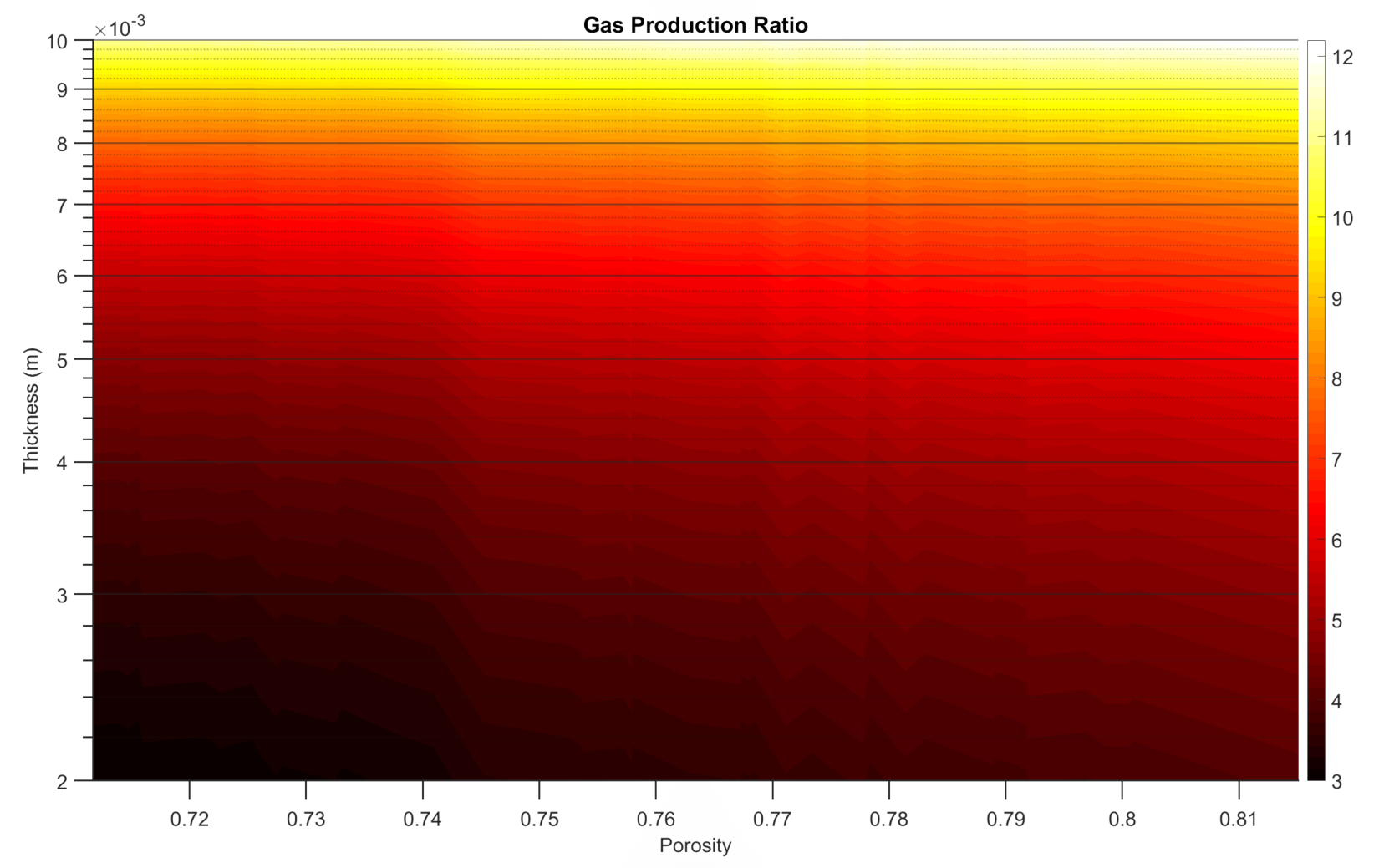}
  \end{subfigure}
   \begin{subfigure}{\columnwidth}
   \centering
  \includegraphics[width=\columnwidth]{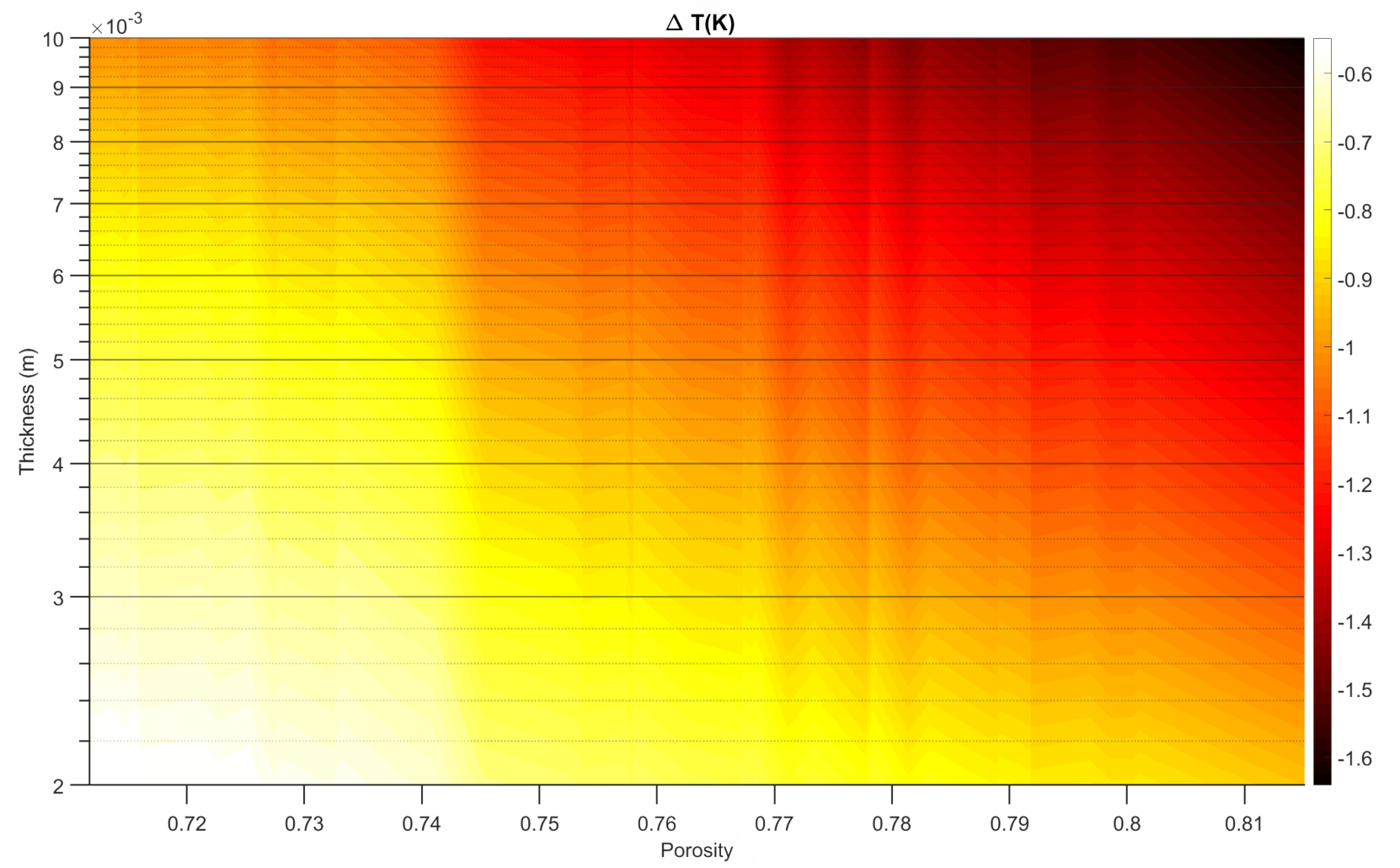}
  \end{subfigure}
  \hfill
    \begin{subfigure}{\columnwidth}
      \centering
  \includegraphics[width=\columnwidth]{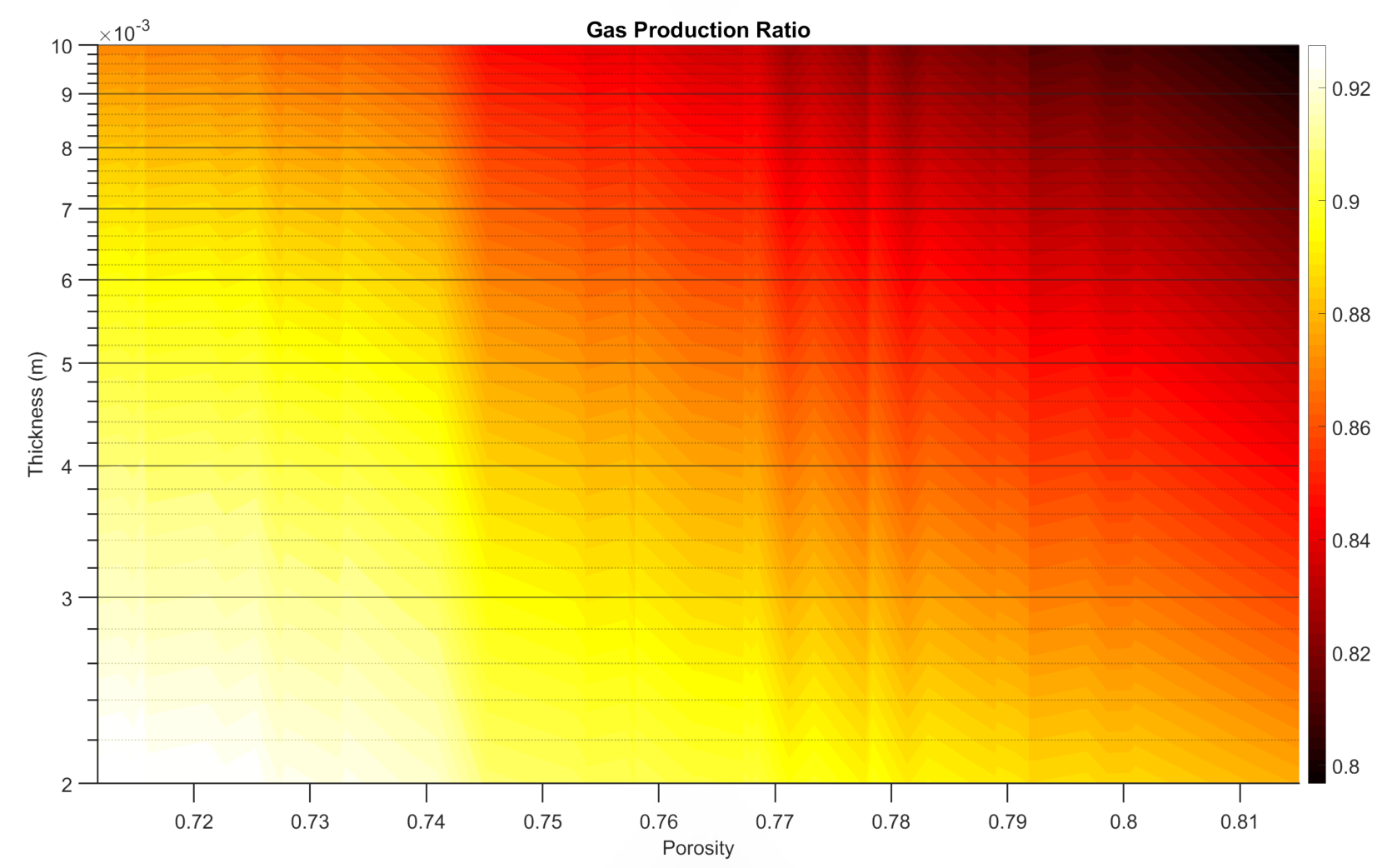}
  \end{subfigure}
     \begin{subfigure}{\columnwidth}
   \centering
  \includegraphics[width=\columnwidth]{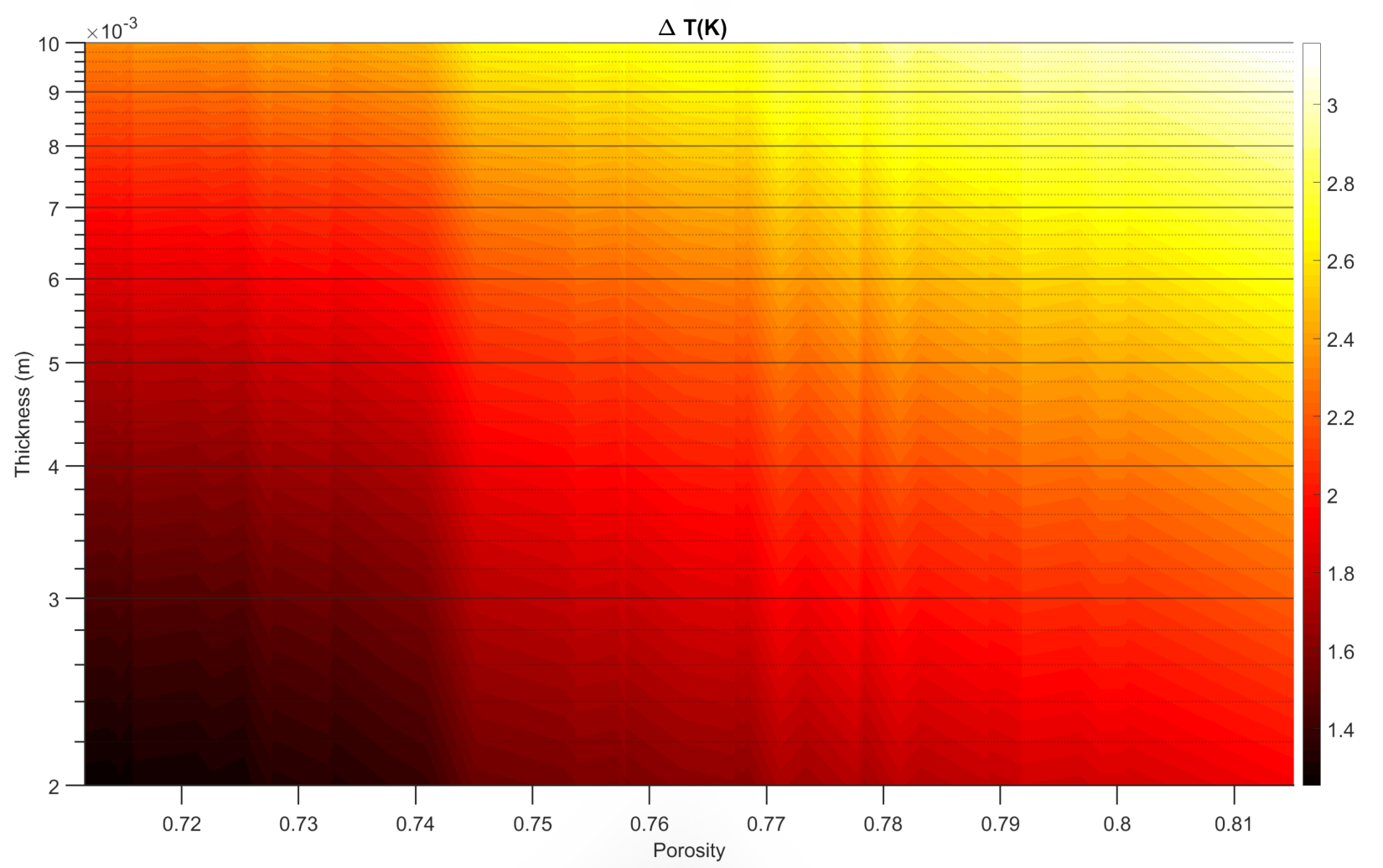}
  \end{subfigure}
  \hfill
    \begin{subfigure}{\columnwidth}
      \centering
  \includegraphics[width=\columnwidth]{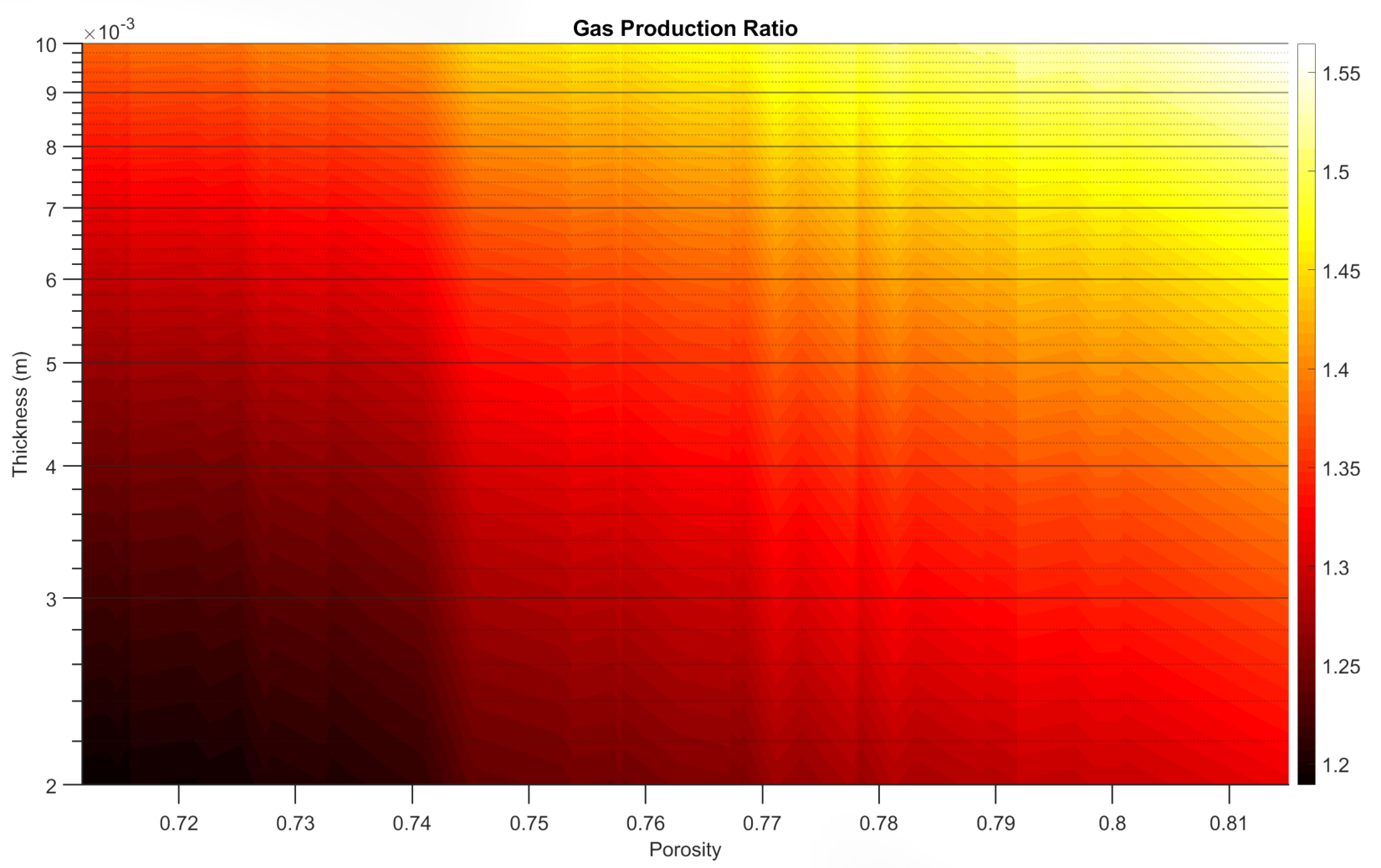}
  \end{subfigure}
\caption{\YUS{The temperature difference at the $\mathrm{H_2O}$ ice boundary (left column) and gas production ratio (right column) of the porous layer as functions of the porosity of the layer and its thickness. The particle size is 1mm.  The heliocentric distance is $R_H=3.45$ au. First row: DMRT model vs model without radiative conductivity. Second row: DMRT model vs Chen-Churchill model of radiative conductivity. Third row: DMRT model vs Rosseland model of radiative conductivity.}} 
\vspace{-6pt}
\label{fig:RA500_RH345}
\end{figure*}

\begin{figure*}
\centering
 \begin{subfigure}{\columnwidth}
   \centering
  \includegraphics[width=\columnwidth]
  {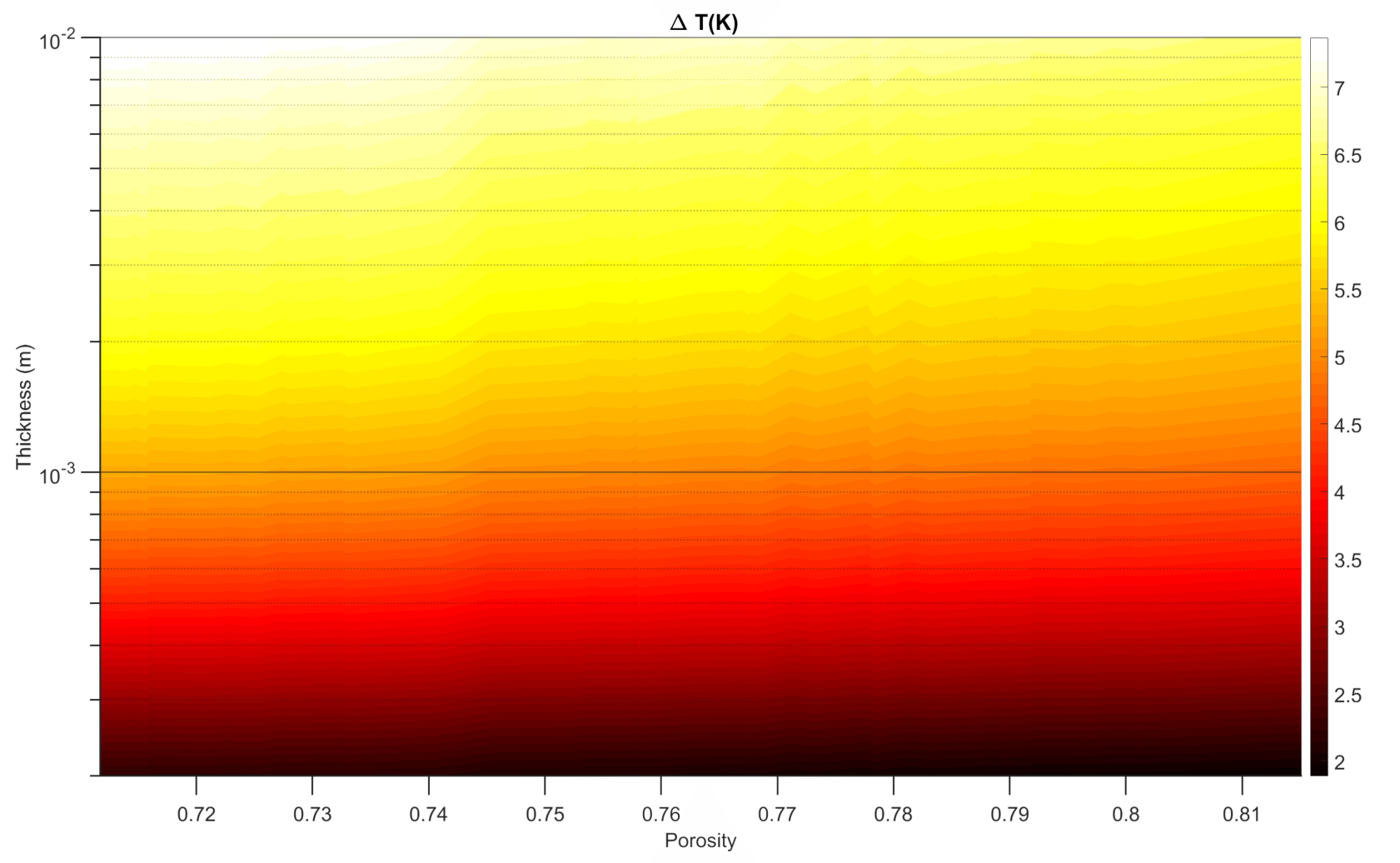}
  \end{subfigure}
  \hfill
    \begin{subfigure}{\columnwidth}
      \centering
  \includegraphics[width=\columnwidth]
  {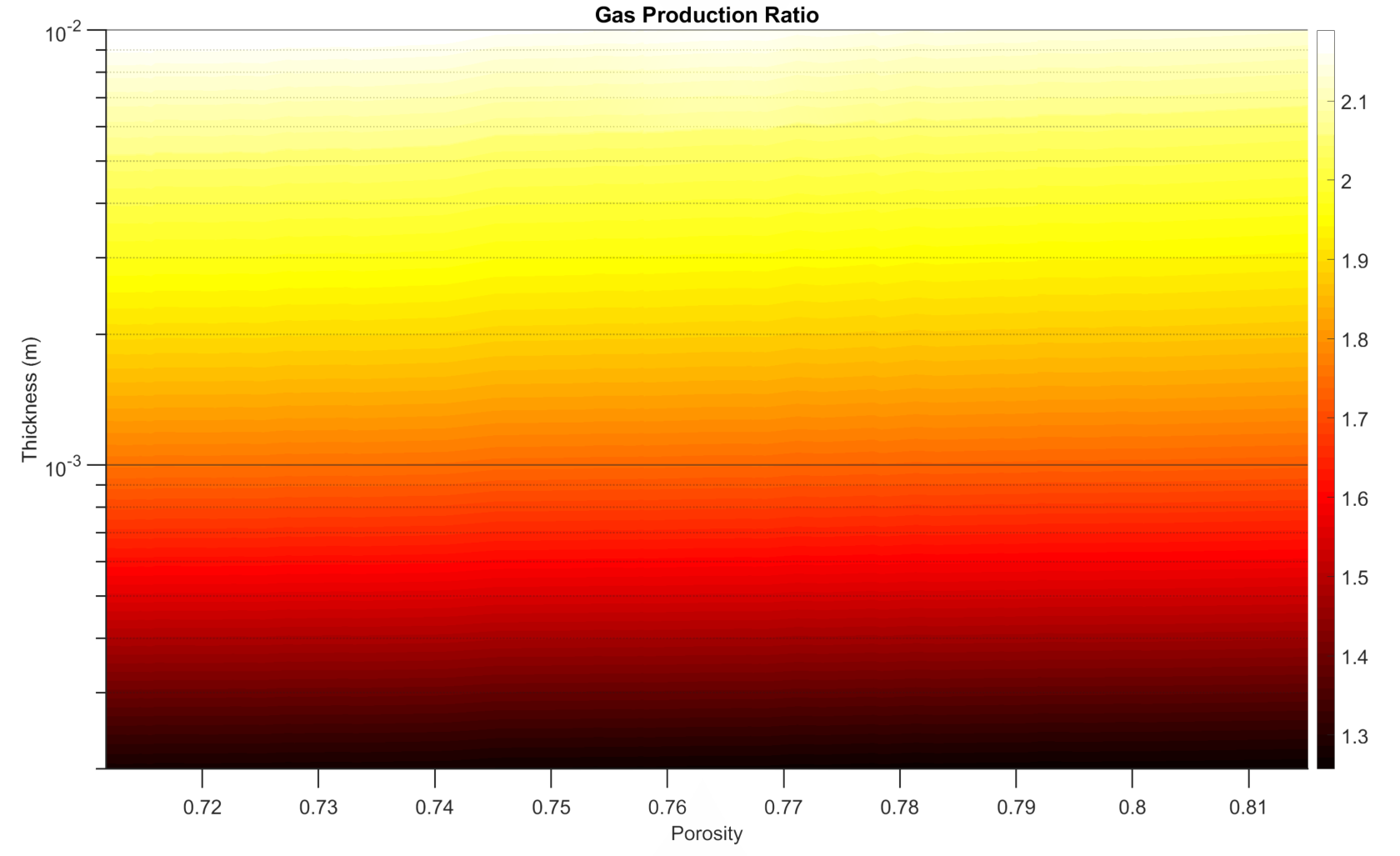}
  \end{subfigure}
   \begin{subfigure}{\columnwidth}
   \centering
  \includegraphics[width=\columnwidth]
  {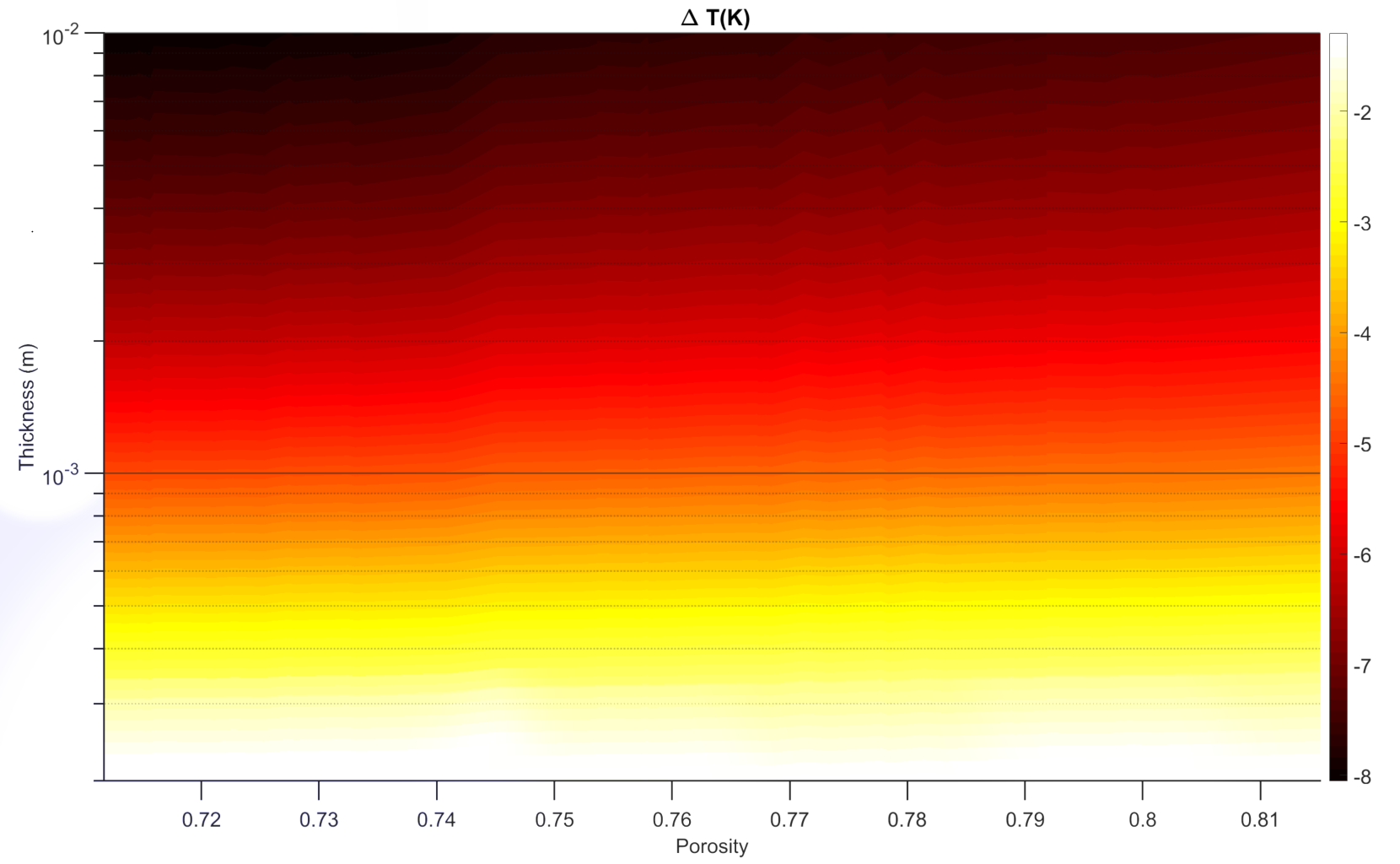}
  \end{subfigure}
  \hfill
    \begin{subfigure}{\columnwidth}
      \centering
  \includegraphics[width=\columnwidth]{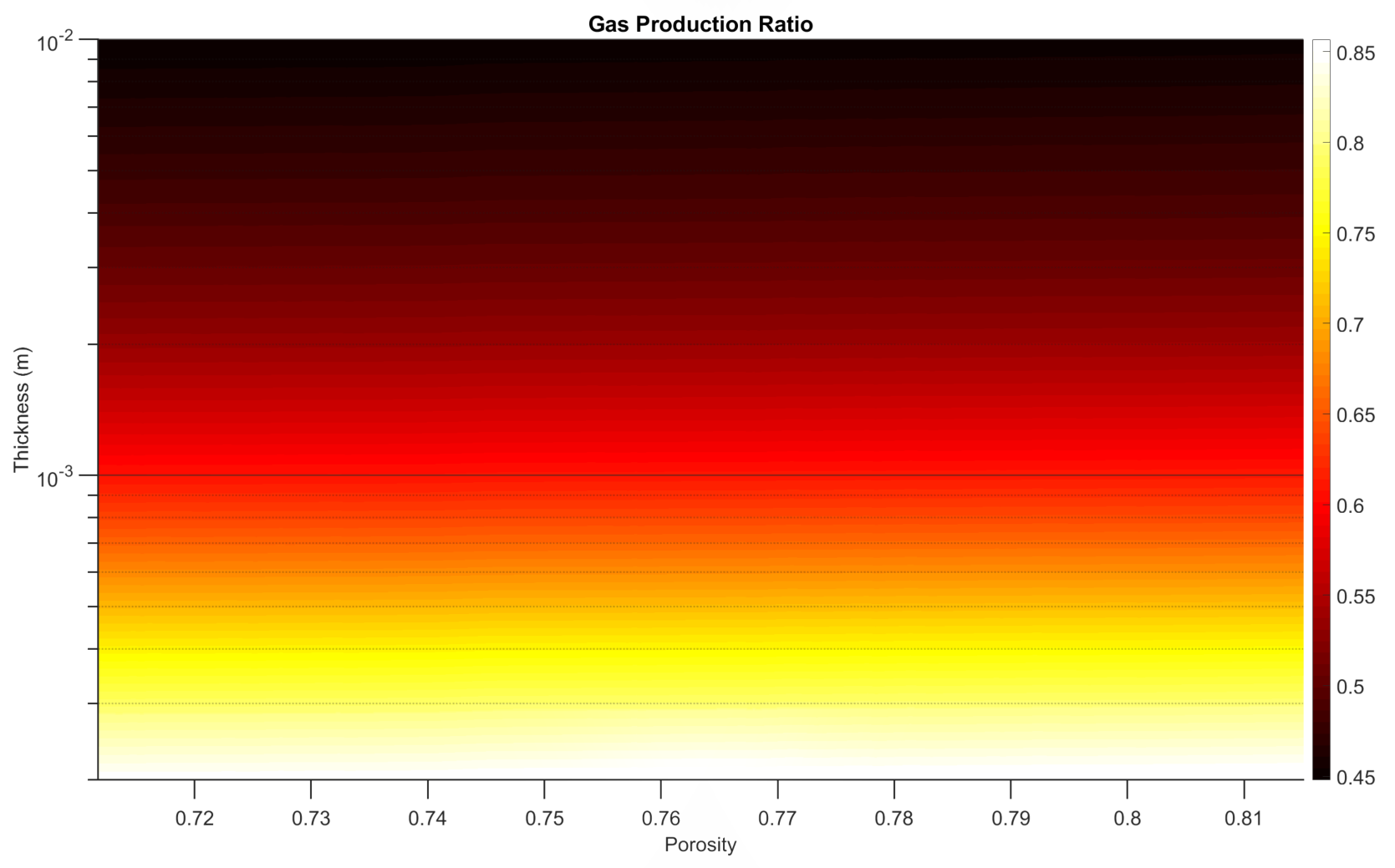}
  \end{subfigure}
     \begin{subfigure}{\columnwidth}
   \centering
  \includegraphics[width=\columnwidth]{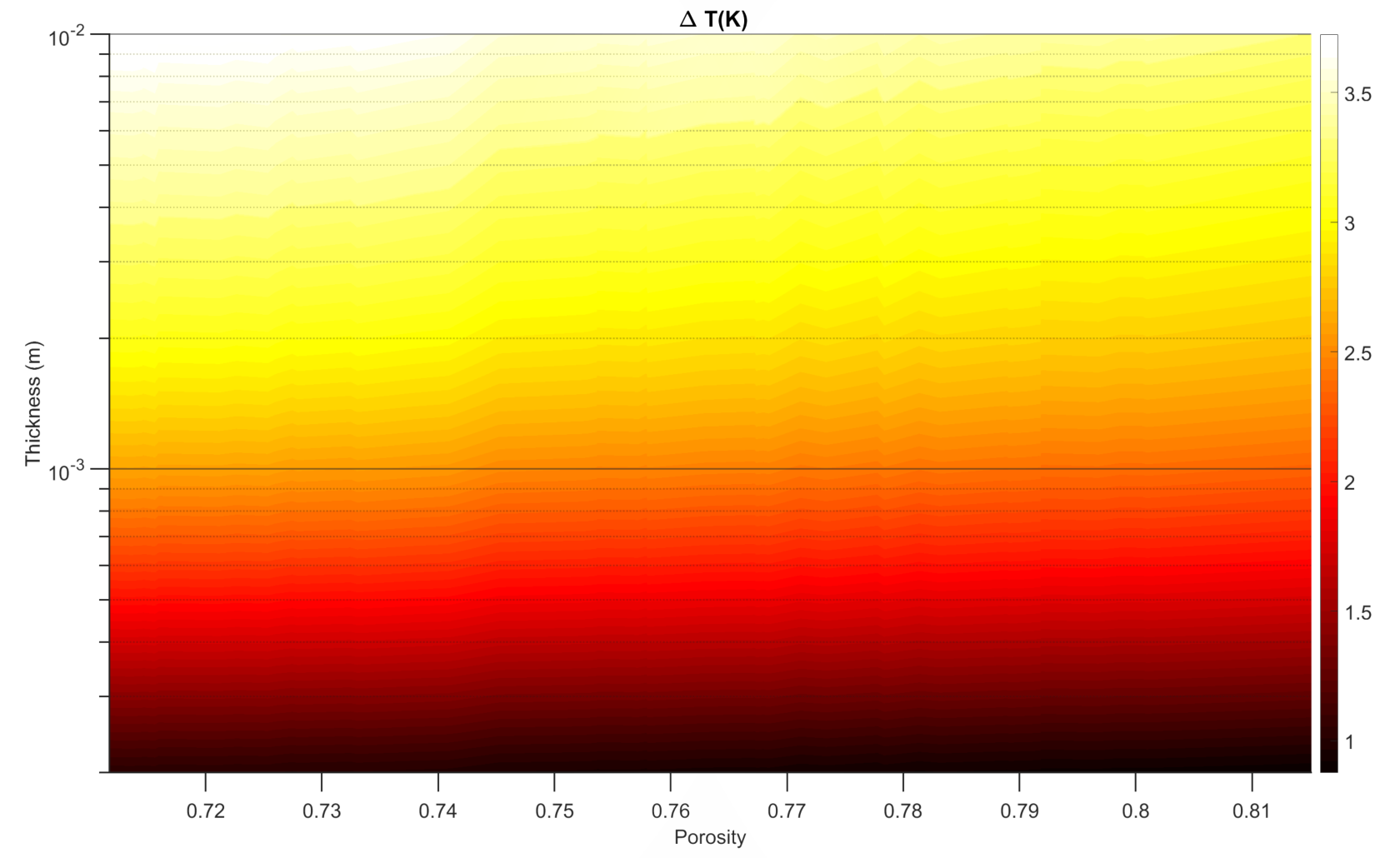}
  \end{subfigure}
  \hfill
    \begin{subfigure}{\columnwidth}
      \centering
  \includegraphics[width=\columnwidth]{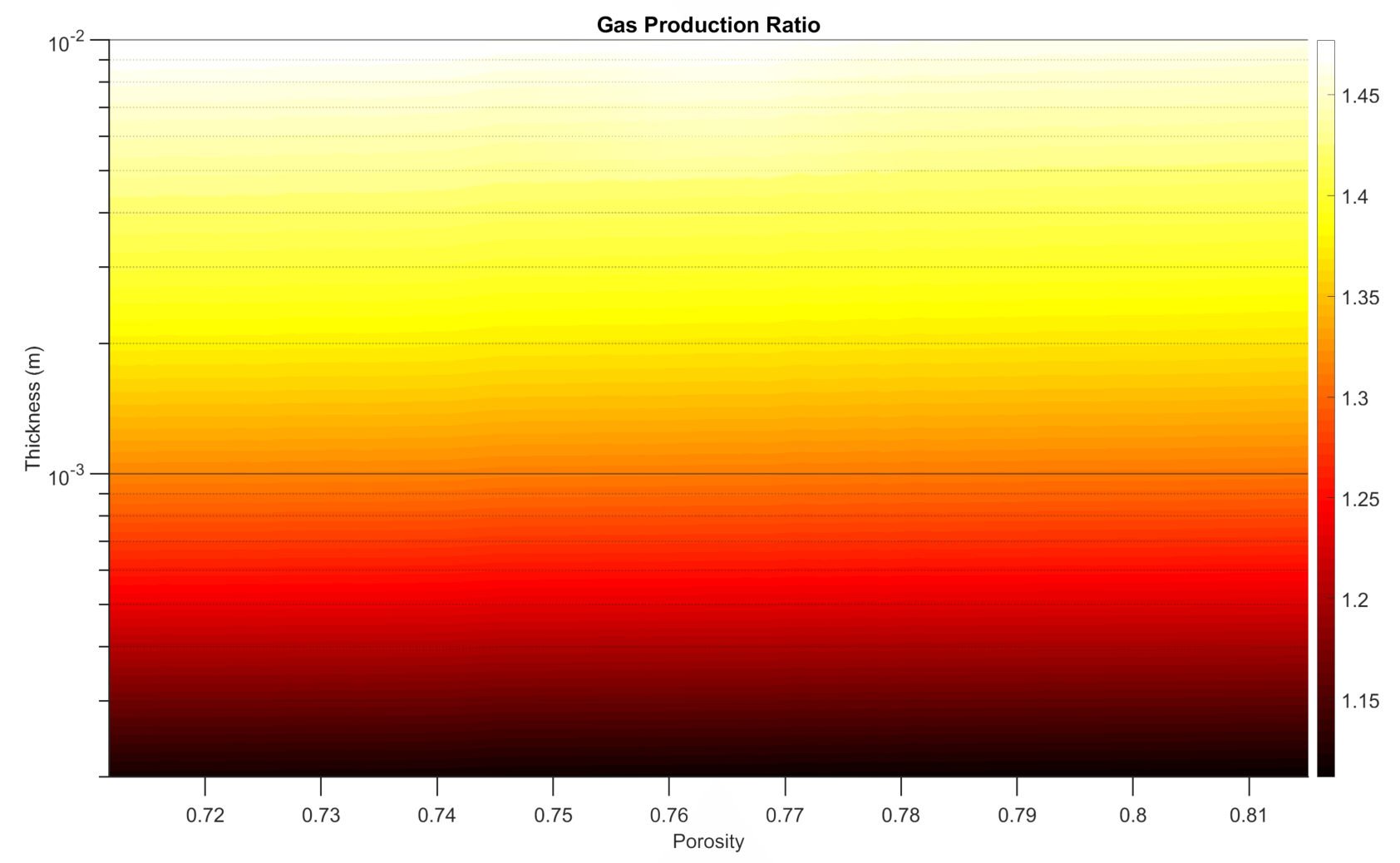}
  \end{subfigure}
\caption{\YUS{The temperature difference at the $\mathrm{H_2O}$ ice boundary (left column) and gas production ratio (right column) of the porous layer as functions of the porosity of the layer and its thickness. The particle size is 0.1mm.  The heliocentric distance is $R_H=1.243$ au. First row: DMRT model vs model without radiative conductivity. Second row: DMRT model vs Chen-Churchill model of radiative conductivity. Third row: DMRT model vs Rosseland model of radiative conductivity.}} 
\vspace{-6pt}
\label{fig:RA50_RH12}
\end{figure*}

\begin{figure*}
\centering
 \begin{subfigure}{\columnwidth}
   \centering
  \includegraphics[width=\columnwidth]{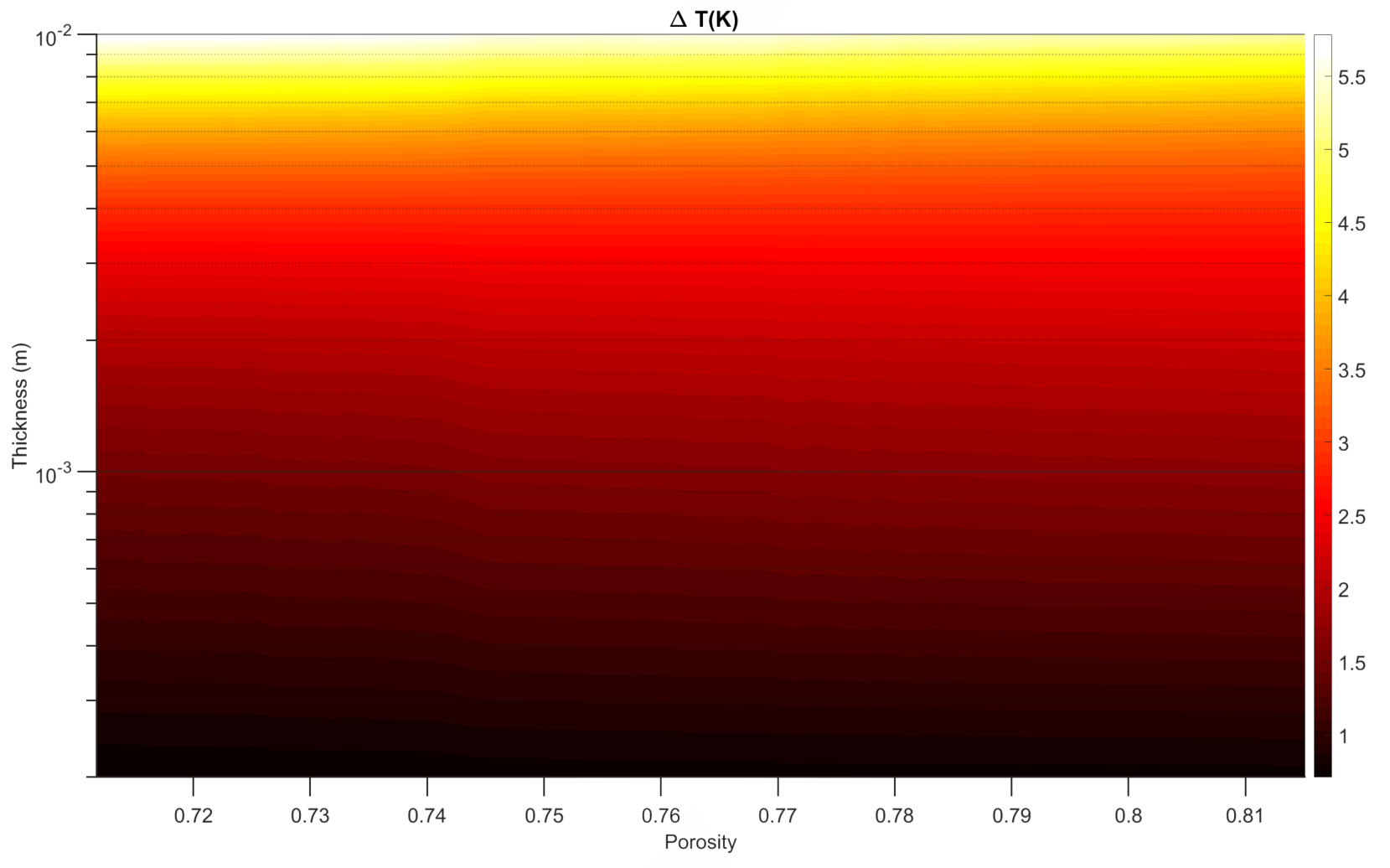}
  \end{subfigure}
  \hfill
    \begin{subfigure}{\columnwidth}
      \centering
  \includegraphics[width=\columnwidth]{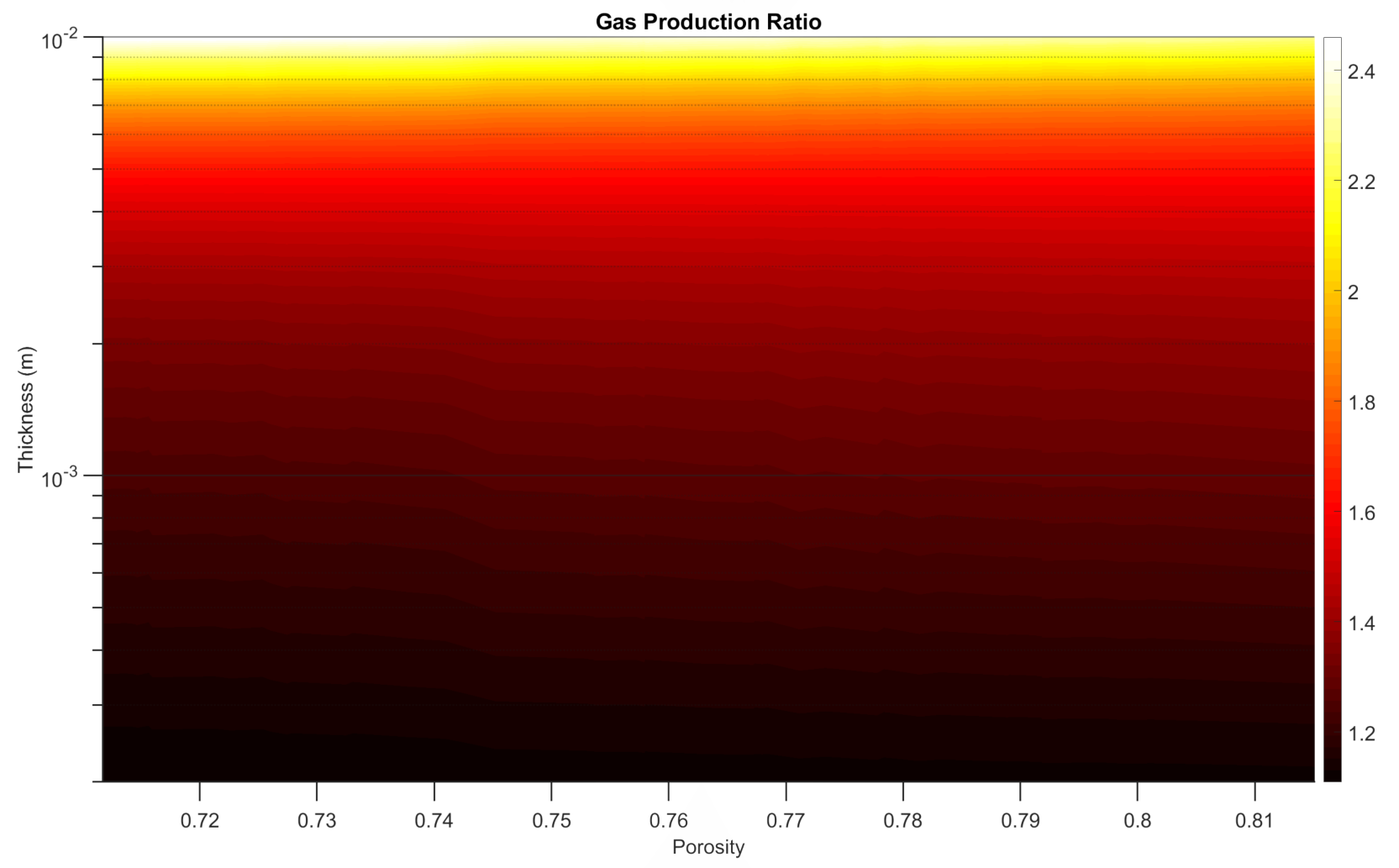}
  \end{subfigure}
   \begin{subfigure}{\columnwidth}
   \centering
  \includegraphics[width=\columnwidth]{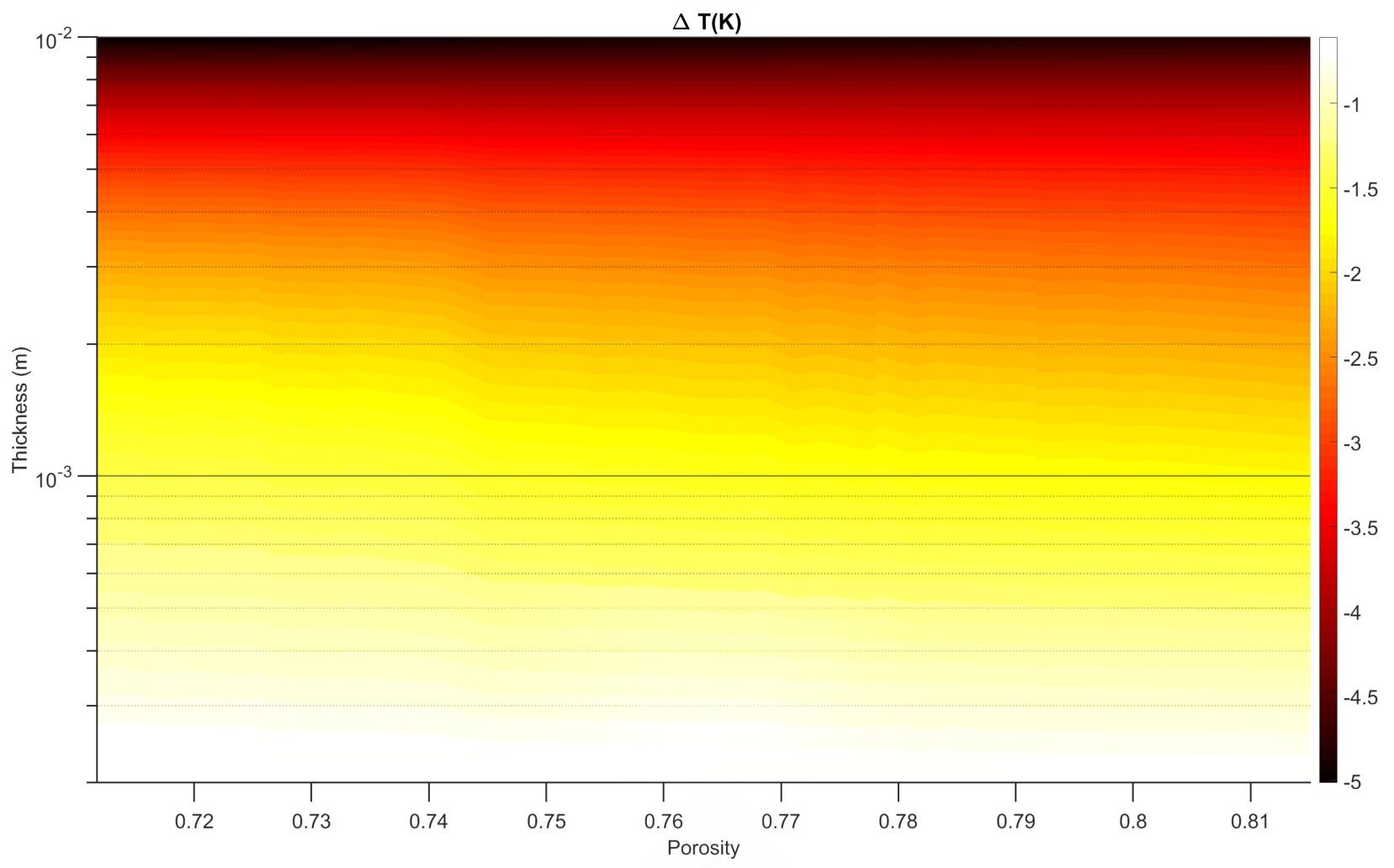}
  \end{subfigure}
  \hfill
    \begin{subfigure}{\columnwidth}
      \centering
  \includegraphics[width=\columnwidth]{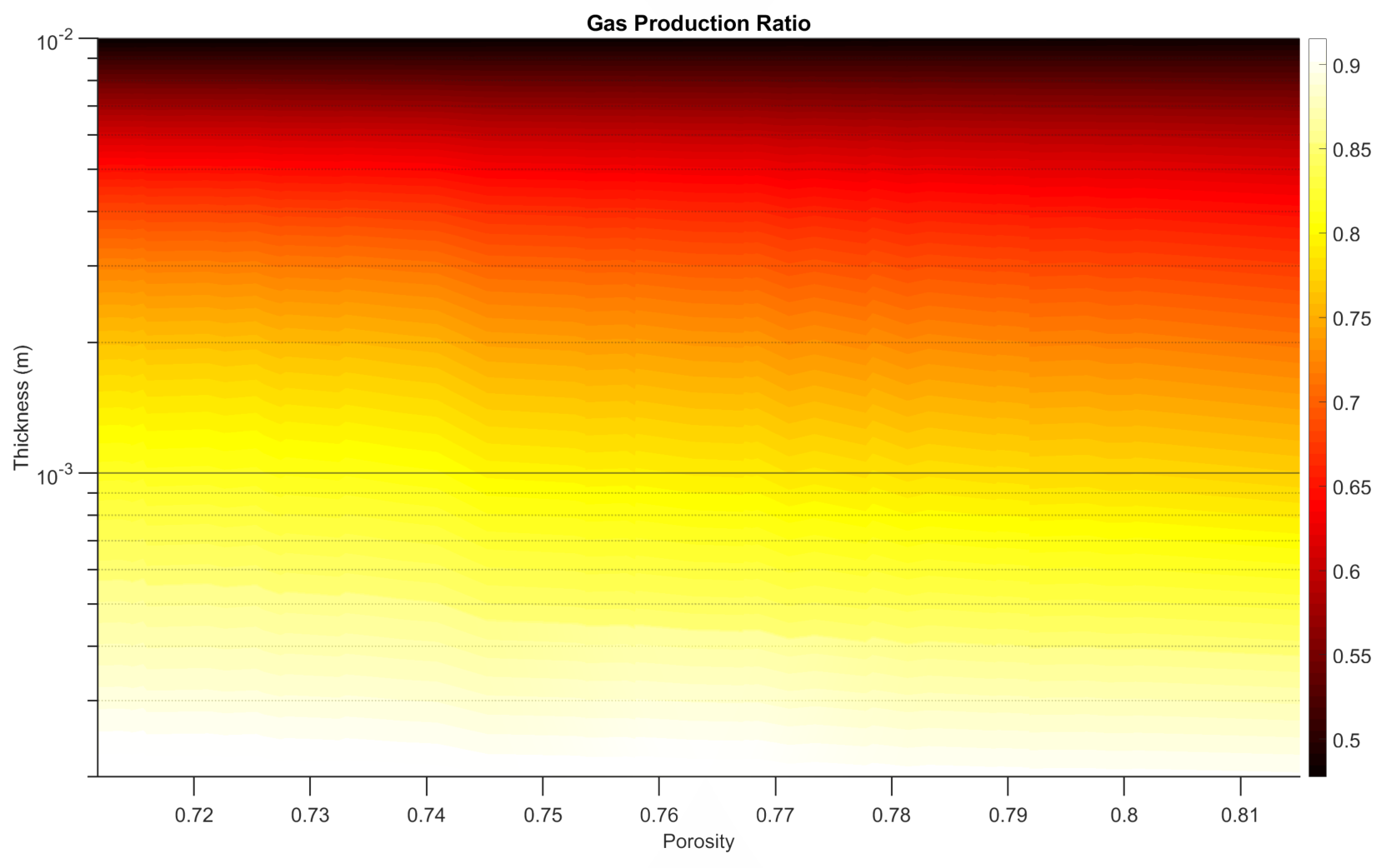}
  \end{subfigure}
     \begin{subfigure}{\columnwidth}
   \centering
  \includegraphics[width=\columnwidth]{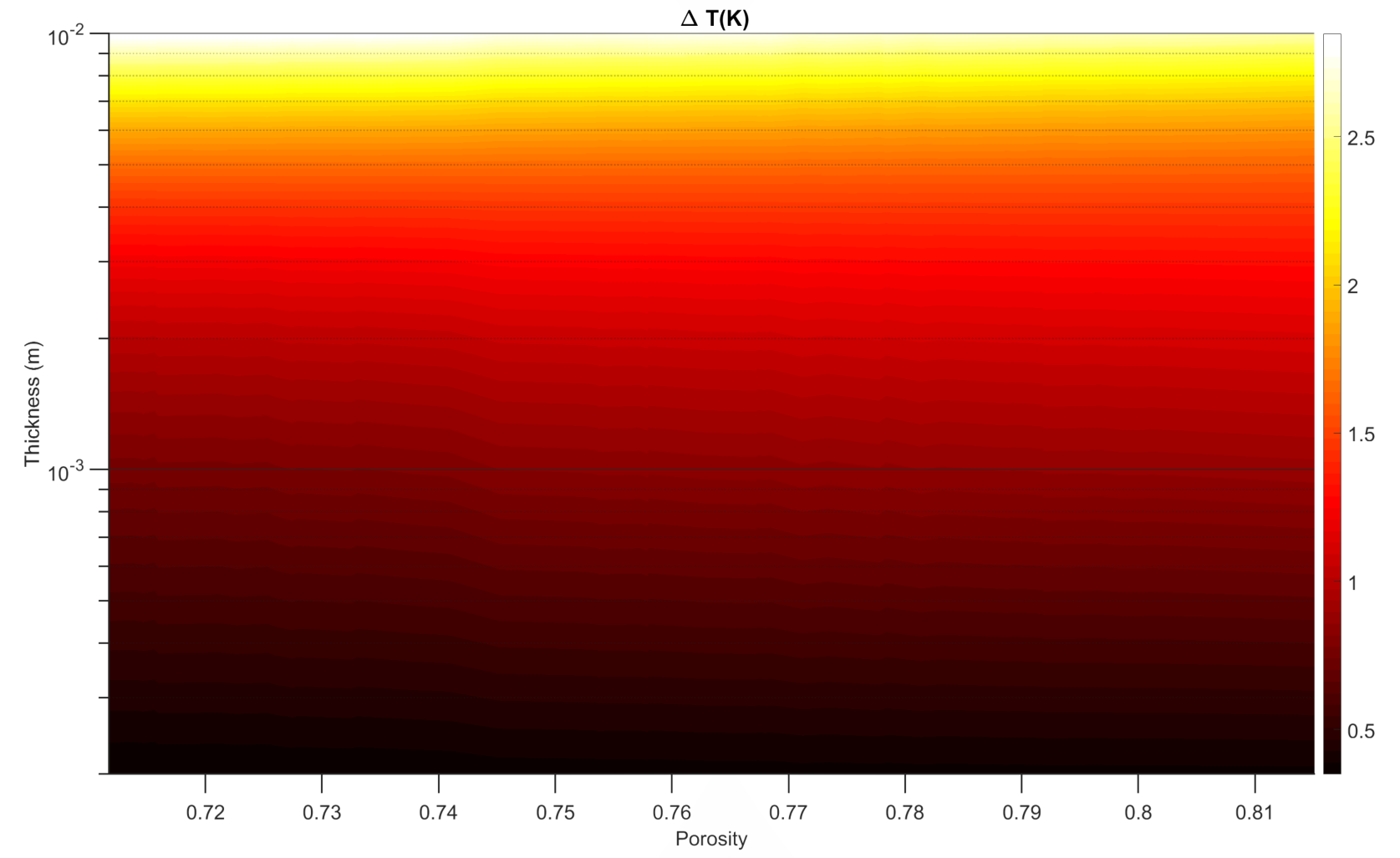}
  \end{subfigure}
  \hfill
    \begin{subfigure}{\columnwidth}
      \centering
  \includegraphics[width=\columnwidth]{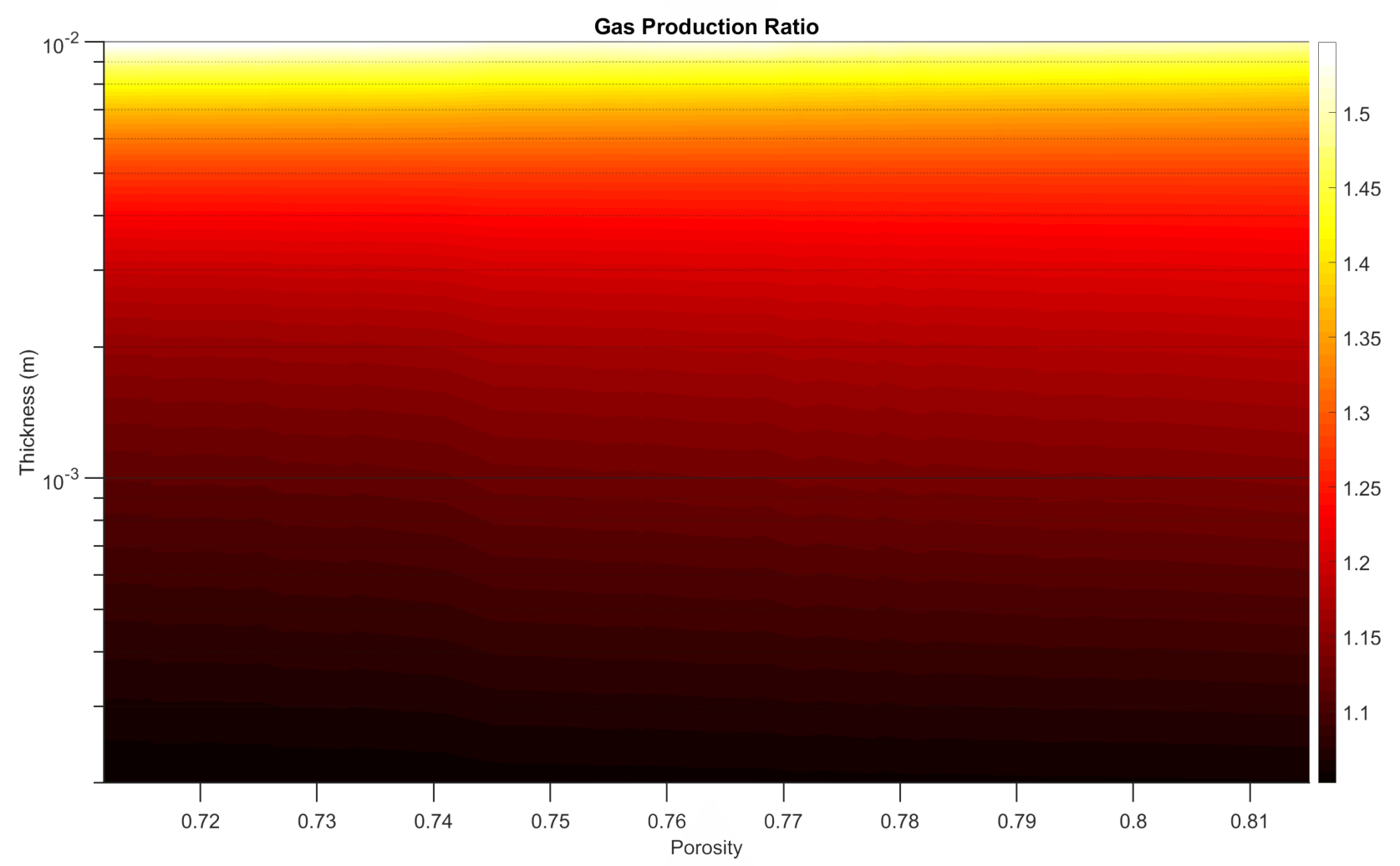}
  \end{subfigure}
\caption{\YUS{The temperature difference at the $\mathrm{H_2O}$ ice boundary (left column) and gas production ratio (right column) of the porous layer as functions of the porosity of the layer and its thickness. The particle size is 1mm.  The heliocentric distance is $R_H=3.45$ au. First row: DMRT model vs model without radiative conductivity. Second row: DMRT model vs Chen-Churchill model of radiative conductivity. Third row: DMRT model vs Rosseland model of radiative conductivity.}} 
\vspace{-6pt}
\label{fig:RA50_RH345}
\end{figure*}

The general scheme for analysing the role of radiative thermal conductivity 
reflects the general dependence of this energy transfer mechanism on temperature and the size of voids in the layer: we test aggregates of two sizes (0.1 and 1 mm) and two heliocentric distances (1.243 and 3.45 au). To show the difference in the simulation results, we use the temperatures at the sublimation front, their difference ($\Delta T$) and the ratio of the corresponding gas production.

Since various approaches and models are used to estimate the radiative thermal conductivity, we focus on the scatter of the results due to the choice of one or another method for calculating this characteristic. 

The results are shown in Figs.~\ref{fig:RA500_RH12} - \ref{fig:RA50_RH345}.
In the first step (upper row of figures), we always compared the model that takes into account the radiative thermal conductivity (the DMRT model is used as a reference model) and the model in which the radiative thermal conductivity is ignored. In the latter case, only the thermal conductivity of the aggregates was taken into account, and for its calculation, we strictly followed the method proposed in (\citealp{Gundlach:2012}). We then compare our reference DMRT model with two other popular models, namely the Chen-Churchill model (\citealp{ChenChurchill1963}) (middle row) and the classical Rosseland model (bottom row). Temperature differences are shown on the left and gas flow ratios on the right. 

Although the results shown in the first row of Fig.~\ref{fig:RA500_RH12} are very different, this is not a big surprise. Indeed, it was already (\citealp{Gundlach:2012}) emphasised that for millimetre particles, the radiative thermal conductivity may be several times greater than the solid thermal conductivity of porous aggregates made of micron monomers. Nevertheless, a numerical comparison is useful and illustrative: the temperatures of the sublimation front are about twenty degrees higher in the model with radiative heat conduction. Since the comet is at a small distance from the sun ($R_H=1.243$ au), this difference in temperatures finds an enhanced expression in the difference in gas production: the estimates differ by more than an order of magnitude over the entire range of tested parameters. 

Accounting for radiative thermal conductivity immediately narrows the scatter of results (middle and bottom rows). In these cases, the gas production varies from about \YUS{50\% to 250\%}. It is interesting to note that the two tested simple models differ from the reference one in different directions: the Chen-Churchill model gives a relatively higher gas production value, while the Rosseland model gives a lower gas production value, i.e. these models differ more from each other.

Comparison with the results obtained for $R_H=3.45$ au confirms our expectations. Both the temperature difference and the gas production ratio decrease (when comparing models with and without radiative thermal conduction (top row) but remain clearly visible. The smallest difference in gas production is observed for the thinnest layers, as anticipated. In these layers, the ice is closer to the surface and sublimation cools the dust layer more actively. But even in these cases, the difference is above hundreds of percent, reaching for a centimetre layer, regardless of porosity, already ten times difference. Comparing models with different descriptions of radiative thermal conductivity (rows two and three), we see that the nature of the deviations has changed not only quantitatively but also qualitatively. For both tested models (Chen-Churchill and Rosseland models), the maximum deviations in temperature are observed in the corners of the diagonal: from the thinnest and densest to the thickest and most porous variants. This picture is predictably transferred to the picture showing the ratio of gas production (right column). The decrease in absolute temperature differences leads to the fact that the  Chen-Churchill model differs from the reference one by only twenty percent, while for the Rosseland model the maximum difference is almost two times. 

Figures \ref{fig:RA50_RH12} and \ref{fig:RA50_RH345} show the results for 100 micron aggregates, i.e. a size ten times less. Since the radiative thermal conductivity is roughly proportional to the size of the voids in the layer, which in turn are approximately proportional to the size of the particles (at least for quasi-spherical ones), we should expect a noticeable weakening of all the differences and effects described above. Calculations fully confirm this. Even at a small heliocentric distance (Fig.~\ref{fig:RA50_RH12}), although the allowance for radiative thermal conductivity increases with increasing layer thickness, the ratio of gas production values is 210-220\% (top row). As before, the  Chen-Churchill model gives slightly higher gas production values compared to the reference model, and the Rosseland model gives smaller values. In the first case, the ratio differs by about a factor of two, and in the second - by one and a half times. The dependence on porosity is insignificant in the considered range of values. The characteristic temperature differences (left column) are close to those obtained for particles whose size was ten times larger at the heliocentric distance about ten times larger (left column, middle and bottom rows of figures ~\ref{fig:RA500_RH345} and ~\ref{fig:RA50_RH12}). Thus, in the variants under consideration, a decrease in the particle size (and hence the size of the void) is approximately compensated by an increase in temperature, the cube of which is included in the formula for radiative thermal conductivity. This comparison is only qualitative, but it gives a sense of the importance of the different model parameters.
Moving on to the results obtained at a distance of 3.45 au (Fig.~\ref{fig:RA50_RH345}), we see that the maximum deviations have not changed significantly. As before, taking into account radiative thermal conductivity increases the gas production to about two and a half times (top row). But now this difference is much more concentrated in the region of thick layers ($\sim 5$ mm); in the remaining region, the difference between the models does not exceed approximately 50\%. The same characteristic concentration of differences is also observed for comparing different models including radiative heat conduction (middle and bottom rows). For layers thinner than $\sim 5$ mm, the difference in gas production does not exceed about 30\%, and only for thick layers does it sharply increase. The influence of porosity on the results is difficult to distinguish.

Summing up, we can say that 1) for the considered values of the parameters (particle size, layer thickness and porosity, irradiation flux), taking into account the radiative thermal conductivity dramatically affects the assessment of gas production of water at all considered heliocentric distances
2) the choice of a specific model describing this type of energy transfer is also of great importance, and the results differ by tens and hundreds (for large particles and small distances) per cent.

\section{Application of the model}

\begin{figure*}
\centering
\includegraphics[width=\textwidth]{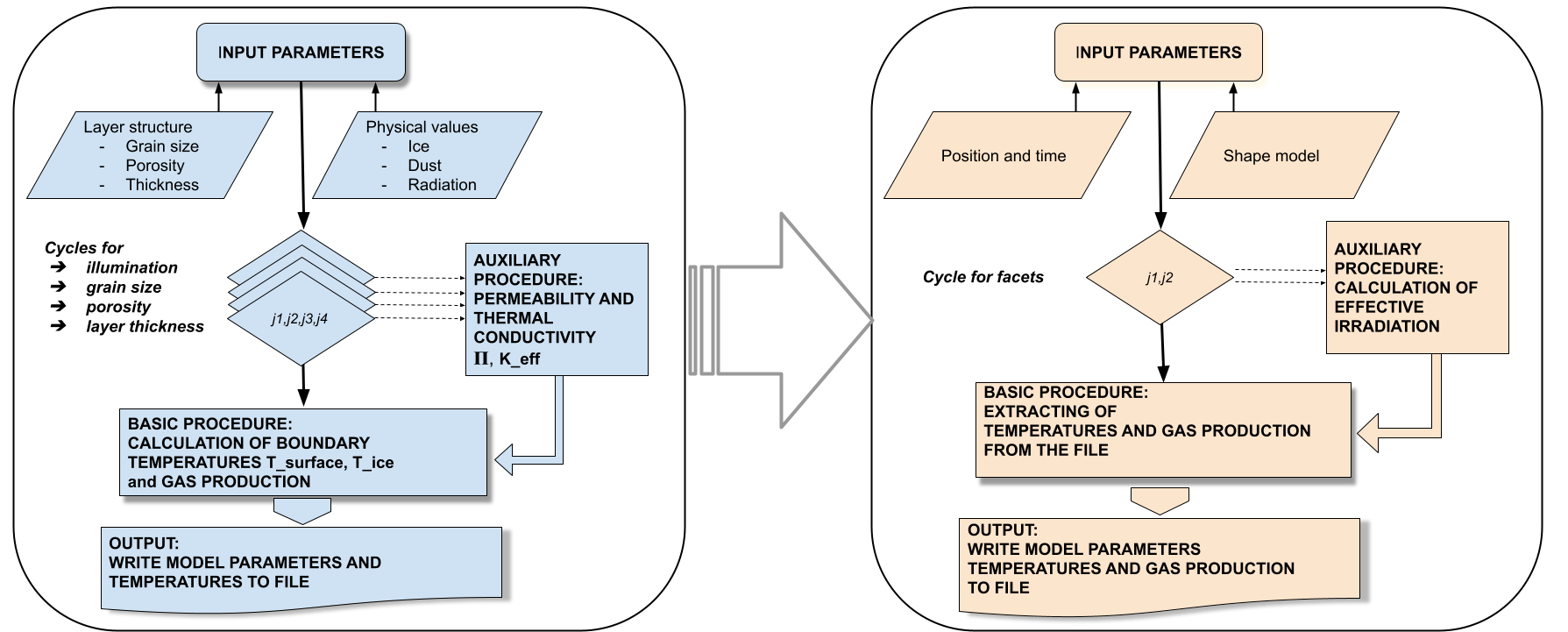}
\caption{Flowchart describing the estimation of the gas production using \emph{Model B} and a realistic shape model. The left panel shows the general structure of the first computational block where, based on the given input parameters, the temperatures at the layer boundaries, the corresponding pressure and the effective gas production estimated for the specific layer permeability are computed and recorded in a multidimensional matrix. Structural input parameters: layer thickness, porosity, size and type of constituent dust particles. Thermophysical input parameters: solar radiation, physical characteristics of ice (e.g. sublimation energy) and dust (e.g. bulk thermal conductivity of the solid phase). The right panel shows the general structure of the second computational block where the pre-computed corresponding gas production is extracted for the set of chosen structural parameters based on the local irradiation. The effective local irradiation is computed for the realistic shape model taking into account surface roughness.
}
\vspace{-6pt}
\label{fig:Fig_flowchart}
\end{figure*}

\begin{figure*} 
\centering
\includegraphics[width=\textwidth]{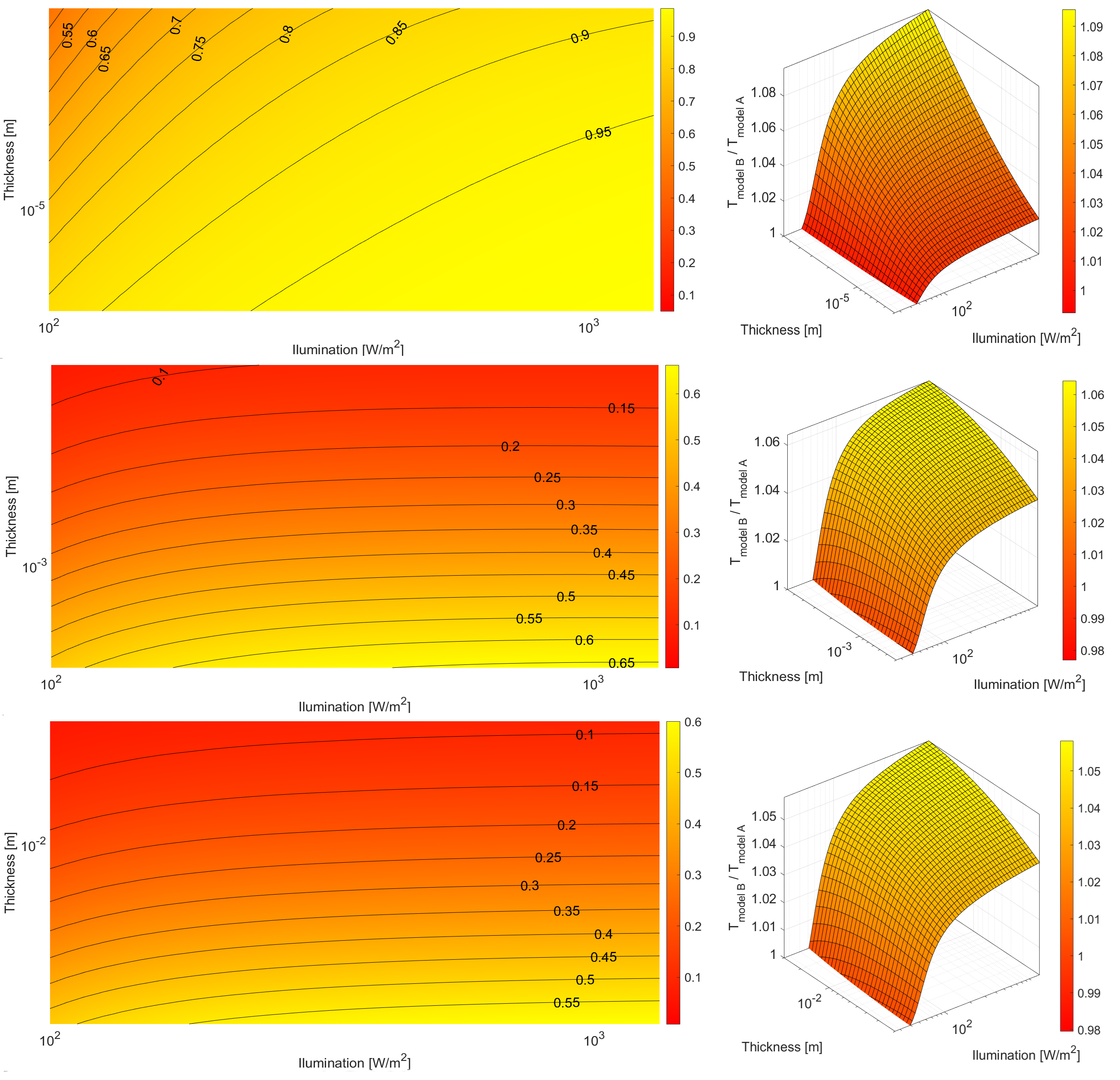} 
\caption{Comparison of  \emph{Models A} and \emph{B}. Left column: ratio of gas productions calculated in \emph{Models A} and \emph{B} as functions of illumination (x-axis) and layer thickness (y-axis). Grain size from top to bottom: 10, 100 and 500 microns. The right column shows the ratios of the respective temperatures for sublimating water ice.}
\vspace{-6pt}
\label{fig:models_A2B}
\end{figure*}

In the previous sections, we examined the structural characteristics of layers made of millimetre-scale particles. Much attention was paid to the estimates of radiative thermal conductivity and the sensitivity of the simulated gas production to the choice of the modelling method for this characteristic. The proposed approaches as well as approximations can and should be applied in thermophysical models to analyse the observed activity of comets. First of all, we should consider their usage in calculating the gas productivity of a comet.

The investigation of the sensitivity of the simulated gas production to the choice of the model parameters was already started in (\citealt{Skorov2023a}). 
All estimates were performed using a one-dimensional two-layer thermophysical model (hereafter called $\emph{Model B}$\footnote{There is a slight difference in the way we use the term $\emph{Model B}$ hereafter. Originally we used this term for indicating a 2-layer model with a specific set of parameters. For referring to a 2-layer model with another set of parameters the letter $\emph{C}$ was used. In this research, we use one term for a 2-layer model with any choice for the parameter set.}) first presented in (\citealp{Keller:2015a}) and later reused (\citealp{Keller:2015b}; \citealp{Keller:2017}; \citealp{Skorov:2017}). The basic assumption of this model is that the uppermost surface layer of a slowly rotating comet nucleus is in a quasi-stationary state, i.e. the heat transfer equation can be considered in the stationary approximation. The estimates given by (Keller et al. 2015a) showed the applicability of such a simplification. The detailed model description can be found in the cited works and the Appendixes below. Note that the gas production in $\emph{Model B}$ depends on the effective thermal conductivity and the permeability of the dust layer. 
We studied the layers built from small grains or aggregates of sizes up to about ten microns. It was shown that, despite the parameter range having been narrowed down by results from the Rosetta mission, the unavoidable uncertainty in the values of some model parameters (e.g. thermal conductivity) blurs the theoretical simulation estimates. We emphasised that analysing the entire range of possible solutions is desirable, instead of presenting a narrow set of specific solutions munching the observations.


When it is necessary to analyse dozens of combinations of model parameters, the calculation speed of the model becomes one of the key properties. {\color{black} This explains the popularity of simple models like \emph{Model A}  \citep{Skorov2023a}} (where the local gas production is determined from a single algebraic equation for the surface temperature) which is attractive because of its simplicity. This allows one to use it, for example, to calculate the global and local gas production, taking into account the realistic shape model of the nucleus with a lot (up to hundreds of thousands) of illuminated facets. However, this model is inconsistent with proven observations showing the absence of ice on the surface and the presence of a hot, dry nonvolatile crust. It contains too many physical oversimplifications which are the
price to pay for its high performance.

Let's ask ourselves, is \emph{Model B} 
{\color{black} computationally too expensive? }
Below we compare \emph{Models A} and \emph{B} 
{\color{black} in detail}
and 
{\color{black} discuss}
the computational complexity 
{\color{black} and the advantages of the latter. }

The absence of a dust layer as well as all mechanisms of heat conduction in \emph{Model A} makes it possible to calculate a surface temperature $T_{\mathit{Model \ A}}$ from the instantaneous energy balance. Neglecting the energy sinks due to thermal conductivity and thermal emission of a hot dust crust results in the gas production of \emph{Model A} 
being {\color{black} higher than in most}
other model estimates. Usually, the model gas production $GP_{\mathit{Model \ A}}$ is much larger than the observed one, 
so that an additional model parameter - the surface active fraction - is generally introduced to reconcile this discrepancy.
This is just a fitting parameter that has no physical meaning if the comet does not possess any exposed ice. We have paid attention to this circumstance in (\citealt{Keller:2015b})\footnote{For completeness, we note that one could conceive a comet with patches of exposed ice on the surface. In this case, the active fraction would represent exactly that: the fraction of the surface covered by fresh ice. This would be a physical interpretation.}. 

\emph{Model B} 
{\color{black} is} much more realistic for the case of an evolved comet. It includes two non-linear algebraic equations expressing the energy balance written at the upper and lower boundaries of the porous dust layer above 
{\color{black} a}
sublimating ice/dust mixture. In the case when there are no sinks or energy sources inside the layer, the heat flux is preserved and these expressions are exact solutions for the stationary heat equation. This model contains two unknowns: the surface temperature and the temperature of the sublimation ice front, which give a much more physical description of a nucleus covered with an inactive crust than \emph{Model A}. It also contains expressions for heat fluxes, that is, the solutions depend on thermal conductivity. In the appendix it is shown that the general approach can be extended to the important case when radiative thermal conductivity, depending on temperature, is included in the model. This generalisation appears important because the radiative energy transfer mechanism  may be dominant for large particles and high temperatures (see analysis in \citealp{Gundlach:2012}). Thus, \emph{Model B} is more closely related to the classical Fourier heat transfer model and consistent with observations. 
We emphasise that all the observations we have point towards comets that have an insulating layer on top of a sublimating icy interior.  Therefore \emph{Model B} is probably a more physical description at least for periodic comets. It is not excluded, however, that comets could exist (for example a new comet coming directly out of the Oort cloud) that are covered by fresh ice at the surface and that are therefore adequately described by \emph{Model A}.

How about the complexity 
and speed of this model? Does it lead to significant extra computational costs?

As for \emph{Model A}, there is no analytical solution for \emph{Model B}. The solution methods are the same in both cases. Usually, Newton Raphson's method is used to find the roots of a nonlinear equation. However, we recommend using the standard Levenberg-Marquardt algorithm modified by  \citet{Fletcher1971AMM}  for finding the optimal solution of a system of nonlinear equations in the least-squares sense. 
The advantage of this method includes less sensitivity to the choice of initial values for the required variables. In our case, obtaining such initial values is not difficult. The corresponding black body temperature value can be used for the estimation of dust surface temperature. And for the temperature of subliming ice, one can use the solution from 
\emph{Model A}. This approach allows us to compare the results of \emph{Models A} and \emph{B} without additional costs.  

\begin{figure*} 
\centering
\includegraphics[width=\textwidth]{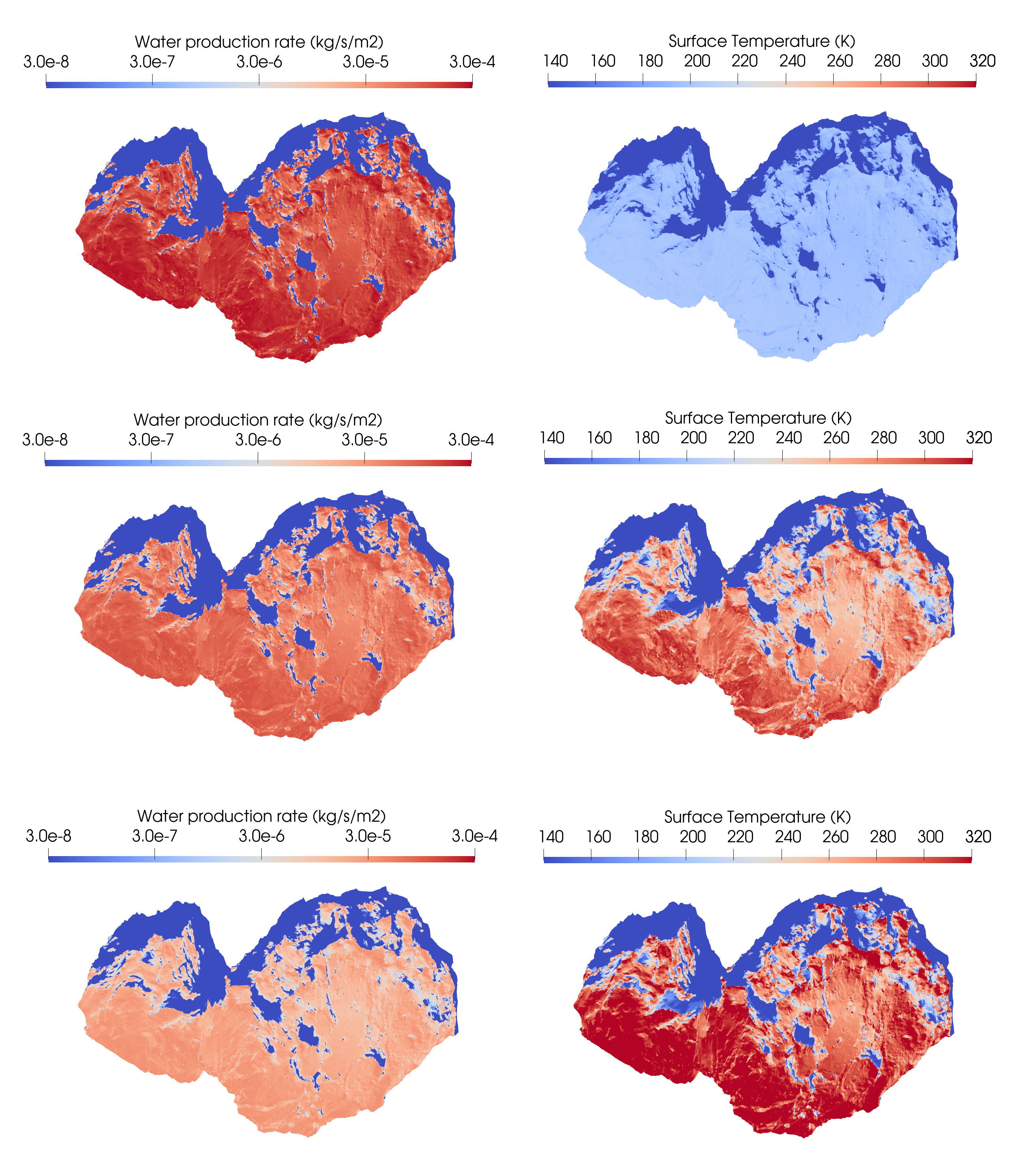} 
\caption{Comparison of the results of \emph{Models A} and \emph{B} at perihelion. The left column shows the gas production, and the right column shows the surface temperatures for a shape nucleus model containing 125,000 facets. The size of the aggregate in \emph{Model   B} equals 0,1 mm. Top row - \emph{Model  A}, middle row - \emph{Model  B} with a layer thickness of ten aggregate radii (1 mm), bottom row - \emph{Model  B} with a layer thickness of fifty aggregate radii (5 mm). }
\label{fig:models_A2B_peri}
\end{figure*}

\begin{figure*} 
\centering
\includegraphics[width=\textwidth]{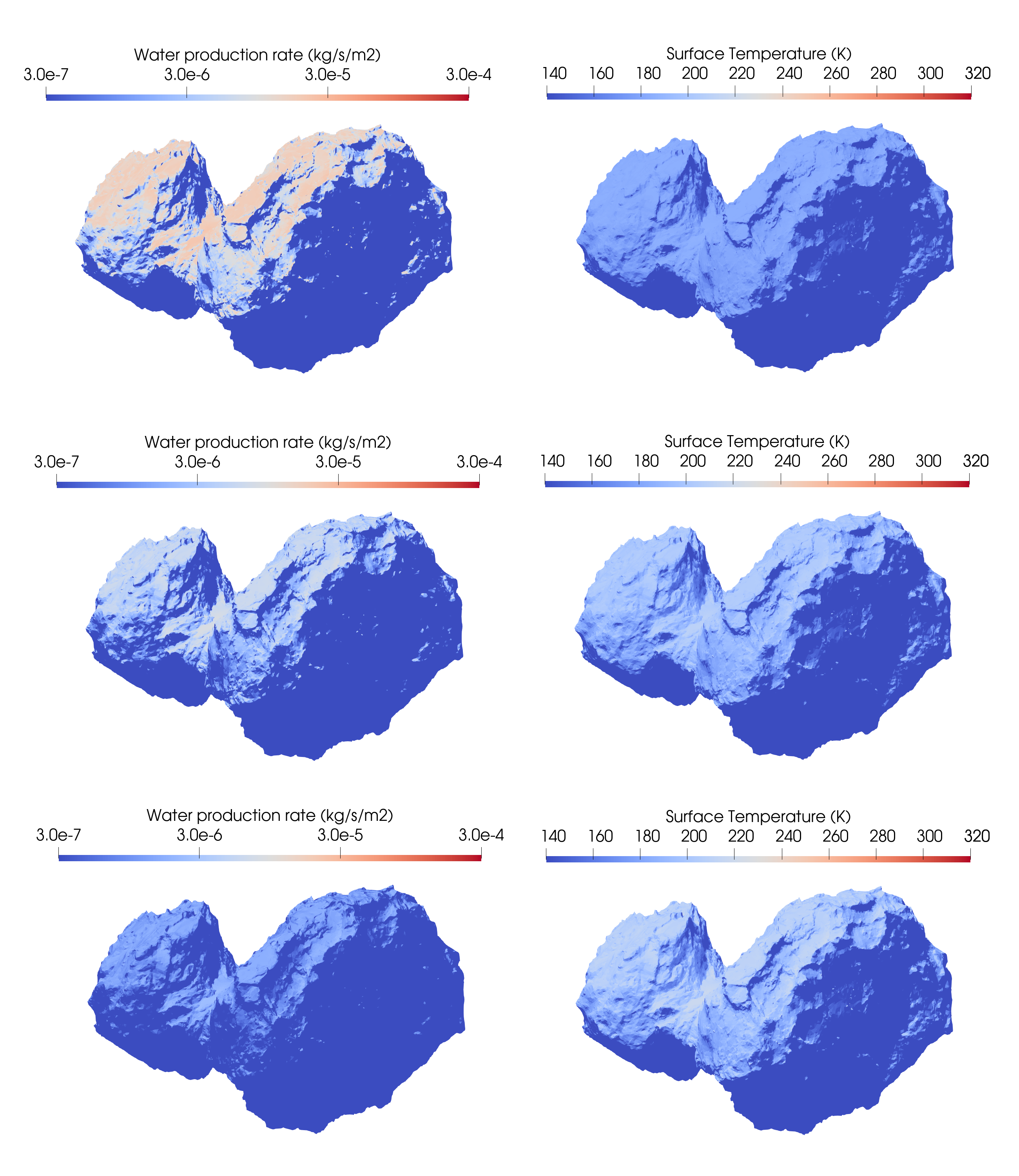} 
\caption{Same as in Fig. \ref{fig:models_A2B_peri} for the heliocentric distance $R_H$ = 3.45 au.}
\vspace{-6pt}
\label{fig:models_A2B_345au}
\end{figure*}

For the applied tasks (such as an estimation of the gas production or the lifting force) it is necessary to calculate the algebraic functions of known temperature. In \emph{Models A} and \emph{B} such estimation is performed in 
the same {\color{black} two} stages. {\color{black} Therefore the computational complexity of the two models is similar.} 
In the first stage, we pre-compute one (in \emph{Model A}) or a pair of temperatures (in \emph{Model B}) for the entire space of 
the model parameters. It is usually assumed that the surface albedo and its thermal emissivity are fixed. Then in \emph{Model A}, the independent variable  is the absorbed solar power density. In \emph{Model B}, the characteristics of the surface layer are added: its porosity, thickness, and average void size. Although the {\color{black} number}
of model parameters noticeably increases, the calculation time remains very short. In both models, the result of the first stage is a matrix containing the 
temperature values, 
calculated only once. For the cases used in this work, the implementation of the first stage on a regular desktop in a four-dimensional space of model parameters (illumination, porosity and layer thickness, particle size) takes 
{\color{black} a} fraction of a second.
In the second modelling stage, first, the intensity array is calculated for the selected shape model and time, and then the obtained temperature matrix is used and, for example, the 
gas production is calculated. This stage is also identical for both models. We emphasise that the calculation of illumination at the second stage is carried out independently of the first stage, i.e. one data set for temperature is used later to analyse different sets of model parameters. Concerning the computing cost of the second stage, in our current implementation on 
{\color{black} a} regular desktop there is no difference in computing time between \emph{Models A} and \emph{B}. For both models, we read-in the temperature from a pre-computed look-up table. The computing time is about 3 sec per computing core per time step for a complex shape with about 125,000 facets.
A general flowchart for performing gas production estimation using \emph{Model B} and a realistic shape model is shown in Fig.\ref{fig:Fig_flowchart}.

Note that this 
{\color{black} separation} of the computation stages distinguishes both models from 
{\color{black} another} simulation approach when a differential non-stationary heat equation is solved (see, for example, \citealt{Hu:2017}, \citealt{Hu2021}). In the latter case, it is necessary to 
{\color{black} describe}
the insolation as a function of time, that is, the position of a point on the surface must be chosen in order to find a solution. In addition, by solving the differential heat equation, we find a solution in the entire modelling domain that is usually not necessary for the 
{\color{black} analysis of observations}, whereas in \emph{Models A} and \emph{B} we find only the required values at the boundaries. These differences make the approach based on the numerical solution of the partial differential equation many times more complex and time-consuming. 
The difference in execution time 
{\color{black} is} 
several orders of magnitude. This does not mean that \emph{Model B} is always better. Along with the obvious advantages the using of a stationary approach has some constraints. These include its inability to determine the activity lag at dawn, or the continuing of the emission for a short period after sunset.  However, there is a wide class of problems when its use is preferable. Estimates of the sensitivity of the solution to the choice of model input parameters are 
{\color{black} examples} of this class. 

Summing up, we 
conclude that {\color{black} the computational complexity of} \emph{Models A} and \emph{B} {\color{black} is} 
the same, 
but the latter is 
more physical. \emph{Model B} allows us to analyse observations to obtain constraints on surface properties and does not contain free model parameters that {\color{black} have no physical meaning.}

{\color{black} Now we compare the results of Models \emph{A} and \emph{B} for some examples.}
First, we compare the temperatures and the corresponding calculated gas productions. For \emph{Model B}, calculations were performed for three particle sizes (20, 200, and 1000 microns) and layers having a relative thickness from 2 to 30 particle sizes. Radiative thermal conductivity was taken into account. The illumination range roughly corresponds to heliocentric distances from 3 to 1 au. The results of {\color{black} the comparison} 
are shown in Fig. \ref{fig:models_A2B}. In the left column, we present the ratio of gas production, and in the right column, we display the ratio of ice boundary temperatures. As noted above, \emph{Model A} gives higher gas production for a given level of insolation. Note that the surface temperature in \emph{Model B} is much higher than in \emph{Model A}, which means that the energy losses for thermal radiation are also higher. 
This leads to a decrease in the part of energy available for sublimation. The additional lowering is
associated with the weakening effect of the porous dust layer. Even for the thinnest layer, the permeability is noticeably lower than unity, and for the layer of the maximum thickness, it is only a few per cent. These effects are expected.
It is interesting to note that the 
scaling factor introduced artificially into \emph{Model A} appears in \emph{Model B} naturally and is 
physical{\color{black}ly meaningful.}
The general behaviour of the temperature ratio shown in the right column is also understandable: the covering dust layer causes an increase in the temperature of the 
{\color{black} sublimating} ice. This increase does not seem too large, but due to the exponential dependence of saturation vapour pressure on the temperature in the Clausius-Clapeyron equation, this increase is sufficient to maintain the energy balance.


When comparing the models, one has to remember that in \emph{Model B} the effective conductivity is the sum of solid and radiative conductivities between aggregates. 
{\color{black} The}
solid one is a decreasing function of the size of the aggregate (\citealp{Gundlach:2012}).
For the 
cases {\color{black} presented here}, its value is reduced by about 5 times \YUS{(check by the formula!)}. 
Radiative conductivity is proportional to the cube of the temperature and the size of the void, which is, in turn, proportional to the particle size.
Therefore, in the presented analysis radiative conductivity manifests itself differently from solid conductivity: 
{\color{black} it} increases with increasing illumination, inducing a rise in temperature. Note that, layers having the same dimensionless thickness differ very little in performance.
This non-trivial result is clearly seen from a comparison of the middle and lower rows in the figure, where the sizes of the aggregates and the dimensional thicknesses of the layers differ by a factor of five. It is the proportionality between the radiative thermal conductivity and the size of voids that explains the observed results.

The results prove that the difference between the models is significant and non-trivial. Gas production may differ several times 
{\color{black} between} different models. A relatively small increase in the temperature of the ice boundary (right panel of Fig.\ref{fig:models_A2B}) partially compensates for the weakening caused by the resistance of the porous layer. These effects are non-linear and cannot be easily parameterized.

In order to make the comparison of models even more visual and closer to observations, we performed additional calculations for the realistic shape of the nucleus of comet 67P. In particular, we used the SHAP7 model from (\citealp{Preusker:2017}) decimated to about 12500 triangular facets.  When determining the effective incoming energy flux on the surface, the effects of shading and re-radiation were taken into account (as was done in \citealt{Keller:2015b}).

The calculations were performed for \emph{Models A} and \emph{B} at 
perihelion (Fig. \ref{fig:models_A2B_peri}) and at a {\color{black} heliocentric} distance of 3.45 au (Fig. \ref{fig:models_A2B_345au}). For \emph{Model B}, the radius of porous aggregates was taken to be 100 µm. The thickness of the dust layer was 10 and 50 particle radii (i.e., 1 and 5 mm,  respectively). Radiative and contact thermal conductivities were taken into account as before. The attenuation of the gas flow in the layer (i.e. its permeability, which is a function of the thickness and size of the aggregate) was also taken into account.

We show the surface distribution of gas production (left column) and surface temperature (right column). We emphasise that it is the surface temperatures that are compared as 
{\color{black} an} observable characteristics. 
{\color{black} We use common scales for easy comparison.}
The ratio of the maximum gas production 
{\color{black} between} \emph{Models A} and \emph{B} is approximately 2.33 for a 1 mm thick layer, and 7.98 for a 5 mm thick layer. This {\color{black} increase of the difference for thicker layers is because of the lower permeability and the lower heat flux delivered to the icy layer.}
The latter effect is clearly visible from the comparison of surface temperatures (middle and bottom rows): a higher surface temperature means a decrease in the fraction of energy available for sublimation. It is noteworthy that these visible differences are observed precisely for all areas with a high degree of insolation, where 
{\color{black} most} gas release actually occurs. Areas with low insolation (which corresponds to low temperatures in \emph{Model B}) contribute little to total gas production. With a decrease in the direct incident energy flux by a factor of 50, gas production decreases by six orders of magnitude due to the exponential dependence of the sublimation rate on temperature. Since one-dimensional models are analysed, the boundaries of inactive areas do not change and retain their geometry. Comparison of surface temperatures (right column) again clarifies the well-known conclusion that \emph{Model A} may be a satisfactory fit for estimating gas production (albeit with the help of additional free parameters), but completely fails in predicting surface temperature\footnote{We note that a more satisfactory agreement with observations can be obtained by adding inhomogeneity to the surface structure. It is possible to imagine a model in which only a small part of a surface unit is active and this part is actually responsible for gas production, while the main (up to 90\%) part of this unit area is covered with a hot crust and is inactive at all. In such a model (assuming that the heat exchange between these fractions is negligibly small), the inactive part will determine some "effective" temperature of the "gray body" which can be much higher than the temperature of the active part where the ice sublimates.}. The supposed sublimation of ice from the surface inevitably lowers the temperature dramatically. The maximum surface temperature in this model at perihelion is about 205K, while in \emph{Models B} it is about 347K for a thin layer and 382K for a thicker one. As expected, the minimum temperatures (at points where the insolation is 2\% of the maximum) differ little and are approximately equal to 147K. In this case, the temperature drop in the layer in \emph{Models B} is small and amounts to about 1K for a thin layer and 3K for a thick one. These small variations and gradients reflect the insignificant role of sublimation in the overall energy balance 
{\color{black} when the energy input is low.} 

{\color{black} We} proceed to the analysis of models at 
large {\color{black} heliocentric} distance (Fig. \ref{fig:models_A2B_345au}). The differences described above and the trends remain.
At the same time, the 
{\color{black} ratio} between thermal emission and sublimation changes significantly. At such a large distance, the maximum surface temperature in \emph{Model  B}  is 
{\color{black} about} half 
of what it was at perihelion. The radiative thermal conductivity does not {\color{black} contribute} 
{\color{black} significantly}, and the effective thermal conductivity is very small everywhere on the surface. Consequently, the maximum temperature in \emph{Model  B}  is only 4K below the corresponding black body temperature. In this model, only about 5\% of the total energy remains for sublimation for the thick layer. Although the maximum surface temperature in \emph{Model A} is still lower, now this difference does not exceed 25K (right column). Energy losses due to thermal radiation differ by about one and a half times in \emph{Models A} and \emph{B}, and the determining factor that reduces gas production in model B is the resistance (permeability) of the porous dust layer. As a result, the maximum values of gas production in \emph{Model B}  are less than the maximum in \emph{Model A}  by 2.89 and 10.67 times for a thin ($10 R_A$) and thick ($ 50 R_A$ ) layer, respectively.

Concluding our comparison, we note that: 1) \emph{Models A} and \emph{B}  have the same computational complexity, 2) \emph{Model B}  is much more realistic, 3) differences between the corresponding results are observed for both gas production and surface temperature, 4) these differences remain significant for both high and low incoming solar energy (although the physical reasons for these differences vary).


\vspace{-10pt}
\section{Conclusion}

In the presented work, we considered for the first time the free molecular diffusion of gas in a porous random hierarchical layer composed of very large aggregates (their typical size is hundreds and thousands of microns). These aggregates, in turn, are composed of solid spherical monomers whose size hardly exceeds one micron. That is, each aggregate contains many billions of monomers. To simulate diffusion in such layers, a method based on the probabilistic description of the act of scattering of a test particle (molecule) by such an aggregate is proposed. As a result, we obtained the dependencies of the mean free path $MFP$ and the permeability $\Pi$ of such model layers.

These results made it possible to proceed to an assessment of energy transfer due to  radiation transfer (radiative thermal conductivity) in the porous layer. We tested various approaches used to quantify this characteristic.  A concise overview of commonly used approximate analytical approaches was presented. These approaches were compared with a much more complex and physically based model using the dense-medium radiative transfer (DMRT) method. Both thermal conductivities are non-linear functions of the medium temperature and its structural characteristics.

The estimates obtained for the  radiative thermal conductivity were included in the modified thermal models. In these models, in the stationary approximation, a system of nonlinear equations was solved that defines the energy balance at the outer (surface of the comet) and inner (water ice sublimation front) boundaries of the dust layer.

The calculations were performed for different layer thicknesses, different porosities, and different heliocentric distances. It has been shown that:

- for millimetre particles, the contribution of radiative thermal conductivity also dominates in comparison with the solid conductivity for all tested sets of model parameters;

- comparison of the model using the DMRT approach with idealised approximate models showed that the difference in gas production estimation when using simplified models can be several times.

All of the above convincingly proves that for porous random layers composed of large (submillimeter and millimetre) aggregates, it is necessary to take into account all the mechanisms of heat conduction when estimating gas production. The obtained results also indicate that earlier conclusions, for example, about gas heating by a hot non-isothermal layer of dust or about the pressure drop in the layer, require serious revision, because these conclusions strongly depend on the sublimating ice temperature which is, in turn, depends on the effective medium conductivity. These questions will be considered by us in future works.

\vspace{-6pt}
\section*{Data availability}
The data underlying this article will be shared on reasonable request to the corresponding author.

\vspace{-6pt}
\section*{Acknowledgements}
We acknowledge support from ESA through the Faculty of the European Space Astronomy Centre (ESAC) - Funding reference 050. J.~M. acknowledges funding from the European Union’s Horizon 2020 research and innovation program under grant agreement No 75390 CAstRA. M.R.~EL-M acknowledges support from the internal KU grant (8474000336-KU-SPSC). Yu.~S. also thanks the International Space Science Institute (Bern, Switzerland) for its support. 

\vspace{-6pt}

\bibliographystyle{mnras}
\bibliography{test.bib} 

\begin{thebibliography}{}
\makeatletter
\relax
\def\mn@urlcharsother{\let\do\@makeother \do\$\do\&\do\#\do\^\do\_\do\%\do\~}
\def\mn@doi{\begingroup\mn@urlcharsother \@ifnextchar [ {\mn@doi@}
  {\mn@doi@[]}}
\def\mn@doi@[#1]#2{\def\@tempa{#1}\ifx\@tempa\@empty \href
  {http://dx.doi.org/#2} {doi:#2}\else \href {http://dx.doi.org/#2} {#1}\fi
  \endgroup}
\def\mn@eprint#1#2{\mn@eprint@#1:#2::\@nil}
\def\mn@eprint@arXiv#1{\href {http://arxiv.org/abs/#1} {{\tt arXiv:#1}}}
\def\mn@eprint@dblp#1{\href {http://dblp.uni-trier.de/rec/bibtex/#1.xml}
  {dblp:#1}}
\def\mn@eprint@#1:#2:#3:#4\@nil{\def\@tempa {#1}\def\@tempb {#2}\def\@tempc
  {#3}\ifx \@tempc \@empty \let \@tempc \@tempb \let \@tempb \@tempa \fi \ifx
  \@tempb \@empty \def\@tempb {arXiv}\fi \@ifundefined
  {mn@eprint@\@tempb}{\@tempb:\@tempc}{\expandafter \expandafter \csname
  mn@eprint@\@tempb\endcsname \expandafter{\@tempc}}}

\bibitem[\protect\citeauthoryear{{A'Hearn} et~al.,}{{A'Hearn}
  et~al.}{2011}]{A'Hearn:2011}
{A'Hearn} M.~F.,  et~al., 2011, \mn@doi [Science] {10.1126/science.1204054},
  \href {https://ui.adsabs.harvard.edu/abs/2011Sci...332.1396A} {332, 1396}

\bibitem[\protect\citeauthoryear{Arakawa, Tanaka, Kataoka  \& Nakamoto}{Arakawa
  et~al.}{2017}]{Arakawa2017thermal}
Arakawa S.,  Tanaka H.,  Kataoka A.,   Nakamoto T.,  2017, Astronomy \&
  Astrophysics, 608, L7

\bibitem[\protect\citeauthoryear{Arakawa, Takemoto  \& Nakamoto}{Arakawa
  et~al.}{2019}]{Arakawa2019geometrical}
Arakawa S.,  Takemoto M.,   Nakamoto T.,  2019, Progress of Theoretical and
  Experimental Physics, 2019, 093E02

\bibitem[\protect\citeauthoryear{Balaji}{Balaji}{2014}]{balaji2014essentials}
Balaji C.,  2014, Essentials of radiation heat transfer.
 Vol. 1, Springer

\bibitem[\protect\citeauthoryear{{Blum} et~al.,}{{Blum}
  et~al.}{2017}]{Blum:2017}
{Blum} J.,  et~al., 2017, \mn@doi [\mnras] {10.1093/mnras/stx2741}, \href
  {https://ui.adsabs.harvard.edu/abs/2017MNRAS.469S.755B} {469, S755}

\bibitem[\protect\citeauthoryear{Bosworth}{Bosworth}{1952}]{Bosworth1952}
Bosworth R. C.~L.,  1952, Heat transfer phenomena: the flow of heat in physical
  systems.
Associated General Publications

\bibitem[\protect\citeauthoryear{Brown \& Ziegler}{Brown \&
  Ziegler}{1979}]{Brown:1979}
Brown G.~N.,  Ziegler W.~T.,  1979, Adv. Cryo. Eng., 25, 662

\bibitem[\protect\citeauthoryear{Cartigny, Yamada  \& Tien}{Cartigny
  et~al.}{1986}]{cartigny1986}
Cartigny J.~D.,  Yamada Y.,   Tien C.~L.,  1986, \mn@doi [Journal of Heat
  Transfer] {10.1115/1.3246979}, 108, 608

\bibitem[\protect\citeauthoryear{Chan \& Tien}{Chan \& Tien}{1974}]{Chan1974}
Chan C.~K.,  Tien C.~L.,  1974, \mn@doi [Journal of Heat Transfer]
  {10.1115/1.3450140}, 96, 52

\bibitem[\protect\citeauthoryear{Chen \& Churchill}{Chen \&
  Churchill}{1963}]{ChenChurchill1963}
Chen J.~C.,  Churchill S.~W.,  1963, AIChE Journal, 9, 35

\bibitem[\protect\citeauthoryear{Delgado}{Delgado}{2011}]{delgado2011heat}
Delgado J.~M.,  2011, Heat and mass transfer in porous media.
 Vol. 13, Springer

\bibitem[\protect\citeauthoryear{{Fanale} \& {Salvail}}{{Fanale} \&
  {Salvail}}{1984}]{Fanale:1984}
{Fanale} F.~P.,  {Salvail} J.~R.,  1984, \mn@doi [Icarus]
  {10.1016/0019-1035(84)90157-X}, 60, 476

\bibitem[\protect\citeauthoryear{{Fanale}, {Salvail}, {Matson}  \&
  {Brown}}{{Fanale} et~al.}{1990}]{Fanale_etal:1990}
{Fanale} F.~P.,  {Salvail} J.~R.,  {Matson} D.~L.,   {Brown} R.~H.,  1990,
  \mn@doi [\icarus] {10.1016/0019-1035(90)90185-C}, \href
  {https://ui.adsabs.harvard.edu/abs/1990Icar...88..193F} {88, 193}

\bibitem[\protect\citeauthoryear{Fletcher}{Fletcher}{1971}]{Fletcher1971AMM}
Fletcher R.,  1971. Theoretical Physics Division, Atomic Energy Research
  Establishment Harwell, UK

\bibitem[\protect\citeauthoryear{{Gundlach} \& {Blum}}{{Gundlach} \&
  {Blum}}{2012}]{Gundlach:2012}
{Gundlach} B.,  {Blum} J.,  2012, \mn@doi [\icarus]
  {10.1016/j.icarus.2012.03.013}, \href
  {https://ui.adsabs.harvard.edu/abs/2012Icar..219..618G} {219, 618}

\bibitem[\protect\citeauthoryear{{G{\"u}ttler} et~al.,}{{G{\"u}ttler}
  et~al.}{2019}]{Guettler2019}
{G{\"u}ttler} C.,  et~al., 2019, \mn@doi [\aap] {10.1051/0004-6361/201834751},
  \href {https://ui.adsabs.harvard.edu/abs/2019A&A...630A..24G} {630, A24}

\bibitem[\protect\citeauthoryear{Hamaker}{Hamaker}{1947}]{Hamaker1947}
Hamaker H.~C.,  1947, Philips Research Reports, 2, 420

\bibitem[\protect\citeauthoryear{{Hu} \& {Shi}}{{Hu} \& {Shi}}{2021}]{Hu2021}
{Hu} X.,  {Shi} X.,  2021, \mn@doi [\apj] {10.3847/1538-4357/abddbf}, \href
  {https://ui.adsabs.harvard.edu/abs/2021ApJ...910...10H} {910, 10}

\bibitem[\protect\citeauthoryear{{Hu} et~al.,}{{Hu} et~al.}{2017}]{Hu:2017}
{Hu} X.,  et~al., 2017, \mn@doi [A\&A] {10.1051/0004-6361/201629910}, 604, A114

\bibitem[\protect\citeauthoryear{Kaviany}{Kaviany}{2014}]{Kaviany:2014}
Kaviany M.,  2014, Heat transfer physics.
Cambridge University Press

\bibitem[\protect\citeauthoryear{{Keller}, {Delamere}, {Reitsema}, {Huebner}
  \& {Schmidt}}{{Keller} et~al.}{1987}]{Keller87}
{Keller} H.~U.,  {Delamere} W.~A.,  {Reitsema} H.~J.,  {Huebner} W.~F.,
  {Schmidt} H.~U.,  1987, \aap, \href
  {http://adsabs.harvard.edu/abs/1987A%26A...187..807K} {187, 807}

\bibitem[\protect\citeauthoryear{{Keller}, {Mottola, S.}, {Skorov, Y.}  \&
  {Jorda, L.}}{{Keller} et~al.}{2015a}]{Keller:2015a}
{Keller} H.~U.,  {Mottola, S.} {Skorov, Y.}  {Jorda, L.} 2015a, \mn@doi [A\&A]
  {10.1051/0004-6361/201526421}, 579, L5

\bibitem[\protect\citeauthoryear{{Keller} et~al.,}{{Keller}
  et~al.}{2015b}]{Keller:2015b}
{Keller} H.~U.,  et~al., 2015b, \mn@doi [A\&A] {10.1051/0004-6361/201525964},
  583, A34

\bibitem[\protect\citeauthoryear{{Keller} et~al.,}{{Keller}
  et~al.}{2017}]{Keller:2017}
{Keller} H.~U.,  et~al., 2017, \mn@doi [Mon. Not. R. Astron. Soc.]
  {10.1093/mnras/stx1726}, 469, S357

\bibitem[\protect\citeauthoryear{{Krause}, {Blum}, {Skorov}  \&
  {Trieloff}}{{Krause} et~al.}{2011}]{Krause2011}
{Krause} M.,  {Blum} J.,  {Skorov} Y.~V.,   {Trieloff} M.,  2011, \mn@doi
  [\icarus] {10.1016/j.icarus.2011.04.024}, \href
  {https://ui.adsabs.harvard.edu/abs/2011Icar..214..286K} {214, 286}

\bibitem[\protect\citeauthoryear{Laubitz}{Laubitz}{1959}]{laubitz1959thermal}
Laubitz M.,  1959, Canadian Journal of Physics, 37, 798

\bibitem[\protect\citeauthoryear{{Levasseur-Regourd}
  et~al.,}{{Levasseur-Regourd} et~al.}{2018}]{Levasseur-Regourd2018}
{Levasseur-Regourd} A.-C.,  et~al., 2018, \mn@doi [\ssr]
  {10.1007/s11214-018-0496-3}, \href
  {https://ui.adsabs.harvard.edu/abs/2018SSRv..214...64L} {214, 64}

\bibitem[\protect\citeauthoryear{Liang, Xu, Tsang, Andreadis  \&
  Josberger}{Liang et~al.}{2008}]{liang2008}
Liang D.,  Xu X.,  Tsang L.,  Andreadis K.~M.,   Josberger E.~G.,  2008, IEEE
  Transactions on Geoscience and Remote Sensing, 46, 3663

\bibitem[\protect\citeauthoryear{{Mannel} et~al.,}{{Mannel}
  et~al.}{2019}]{Mannel2019}
{Mannel} T.,  et~al., 2019, \mn@doi [\aap] {10.1051/0004-6361/201834851}, \href
  {https://ui.adsabs.harvard.edu/abs/2019A&A...630A..26M} {630, A26}

\bibitem[\protect\citeauthoryear{Markkanen \& Agarwal}{Markkanen \&
  Agarwal}{2019}]{markkanen2019}
Markkanen J.,  Agarwal J.,  2019, Astronomy \& Astrophysics, 631, A164

\bibitem[\protect\citeauthoryear{Markkanen \& Agarwal}{Markkanen \&
  Agarwal}{2020}]{markkanen2020}
Markkanen J.,  Agarwal J.,  2020, Astronomy \& Astrophysics, 643, A16

\bibitem[\protect\citeauthoryear{Mavko, Mukerji  \& Dvorkin}{Mavko
  et~al.}{2009}]{Mavko:2009}
Mavko G.,  Mukerji T.,   Dvorkin J.,  2009, The Rock Physics Handbook: Tools
  for Seismic Analysis of Porous Media, 2nd edn.
Cambridge University Press, Cambridge, \url
  {http://www.worldcat.org/isbn/0521861365}

\bibitem[\protect\citeauthoryear{{Mendis} \& {Brin}}{{Mendis} \&
  {Brin}}{1977}]{MendisBrin1977}
{Mendis} D.~A.,  {Brin} G.~D.,  1977, \mn@doi [Moon] {10.1007/BF00562645},
  \href {https://ui.adsabs.harvard.edu/abs/1977Moon...17..359M} {17, 359}

\bibitem[\protect\citeauthoryear{Merrill}{Merrill}{1969}]{Merrill:1969}
Merrill R.~B.,  1969, Thermal conduction through an evacuated idealized powder
  over the temperature range 100 to 500 K.
National Aeronautics and Space Administration

\bibitem[\protect\citeauthoryear{Modest}{Modest}{2013}]{modest2013}
Modest M.,  2013, Radiative Heat Transfer, Third Edit. ed.
Elsevier Inc., Oxford, UK

\bibitem[\protect\citeauthoryear{{Preusker} et~al.,}{{Preusker}
  et~al.}{2017}]{Preusker:2017}
{Preusker} F.,  et~al., 2017, \mn@doi [A\&A] {10.1051/0004-6361/201731798},
  607, L1

\bibitem[\protect\citeauthoryear{Ratcliffe}{Ratcliffe}{1963}]{Ratcliffe:1963}
Ratcliffe E.,  1963, Glass Technol, 4, 113

\bibitem[\protect\citeauthoryear{{Reshetnik}, {Skorov}, {Bentley}, {Rezac},
  {Hartogh}  \& {Blum}}{{Reshetnik} et~al.}{2022}]{Reshetnyk:2022SoSyR}
{Reshetnik} V.,  {Skorov} Y.,  {Bentley} M.,  {Rezac} L.,  {Hartogh} P.,
  {Blum} J.,  2022, \mn@doi [Solar System Research]
  {10.1134/S0038094622020071}, \href
  {https://ui.adsabs.harvard.edu/abs/2022SoSyR..56..100R} {56, 100}

\bibitem[\protect\citeauthoryear{{Reshetnyk}, {Skorov}, {Vasyuta}, {Bentley},
  {Rezac}, {Agarwal}  \& {Blum}}{{Reshetnyk}
  et~al.}{2021}]{Reshetnyk:2021SoSyR}
{Reshetnyk} V.,  {Skorov} Y.,  {Vasyuta} M.,  {Bentley} M.,  {Rezac} L.,
  {Agarwal} J.,   {Blum} J.,  2021, \mn@doi [Solar System Research]
  {10.1134/S0038094621020040}, \href
  {https://ui.adsabs.harvard.edu/abs/2021SoSyR..55..106R} {55, 106}

\bibitem[\protect\citeauthoryear{{Rosseland}}{{Rosseland}}{1936}]{Rosseland:1936}
{Rosseland} S.,  1936, {Theoretical astrophysics. Oxford: The Clarendon Press.}

\bibitem[\protect\citeauthoryear{Russell}{Russell}{1935}]{Russell:1935}
Russell H.,  1935, Journal of the American Ceramic Society, 18, 1

\bibitem[\protect\citeauthoryear{Sakatani, Ogawa, Iijima, Arakawa, Honda  \&
  Tanaka}{Sakatani et~al.}{2017}]{Sakatani2017thermal}
Sakatani N.,  Ogawa K.,  Iijima Y.,  Arakawa M.,  Honda R.,   Tanaka S.,  2017,
  in 48th Annual Lunar and Planetary Science Conference. No.~1964.
p.~1552

\bibitem[\protect\citeauthoryear{Schotte}{Schotte}{1960}]{schotte1960thermal}
Schotte W.,  1960, AIChE Journal, 6, 63

\bibitem[\protect\citeauthoryear{{Shen}, {Draine}  \& {Johnson}}{{Shen}
  et~al.}{2008}]{Shen2008ApJ}
{Shen} Y.,  {Draine} B.~T.,   {Johnson} E.~T.,  2008, \mn@doi [\apj]
  {10.1086/592765}, \href
  {https://ui.adsabs.harvard.edu/abs/2008ApJ...689..260S} {689, 260}

\bibitem[\protect\citeauthoryear{{Skorov}, {Rezac}, {Hartogh}  \&
  {Keller}}{{Skorov} et~al.}{2017}]{Skorov:2017}
{Skorov} Y.~V.,  {Rezac} L.,  {Hartogh} P.,   {Keller} H.~U.,  2017, \mn@doi
  [A\&A] {10.1051/0004-6361/201630000}, 600, A142

\bibitem[\protect\citeauthoryear{{Skorov}, {Reshetnyk}, {Rezac}, {Zhao},
  {Marschall}, {Blum}  \& {Hartogh}}{{Skorov} et~al.}{2018}]{Skorov:2018MNRAS}
{Skorov} Y.,  {Reshetnyk} V.,  {Rezac} L.,  {Zhao} Y.,  {Marschall} R.,  {Blum}
  J.,   {Hartogh} P.,  2018, \mn@doi [\mnras] {10.1093/mnras/sty1014}, \href
  {https://ui.adsabs.harvard.edu/abs/2018MNRAS.477.4896S} {477, 4896}

\bibitem[\protect\citeauthoryear{{Skorov}, {Reshetnyk}, {Bentley}, {Rezac},
  {Agarwal}  \& {Blum}}{{Skorov} et~al.}{2021}]{Skorov:2021}
{Skorov} Y.,  {Reshetnyk} V.,  {Bentley} M.,  {Rezac} L.,  {Agarwal} J.,
  {Blum} J.,  2021, \mn@doi [\mnras] {10.1093/mnras/staa3735}, \href
  {https://ui.adsabs.harvard.edu/abs/2021MNRAS.501.2635S} {501, 2635}

\bibitem[\protect\citeauthoryear{{Skorov}, {Reshetnyk}, {Bentley}, {Rezac},
  {Hartogh}  \& {Blum}}{{Skorov} et~al.}{2022}]{Skorov:2022}
{Skorov} Y.,  {Reshetnyk} V.,  {Bentley} M.~S.,  {Rezac} L.,  {Hartogh} P.,
  {Blum} J.,  2022, \mn@doi [\mnras] {10.1093/mnras/stab3760}, \href
  {https://ui.adsabs.harvard.edu/abs/2022MNRAS.510.5520S} {510, 5520}

\bibitem[\protect\citeauthoryear{{Skorov}, {Reshetnyk}, {K{\"u}ppers},
  {Bentley}, {Besse}  \& {Hartogh}}{{Skorov} et~al.}{2023}]{Skorov2023a}
{Skorov} Y.,  {Reshetnyk} V.,  {K{\"u}ppers} M.,  {Bentley} M.~S.,  {Besse} S.,
    {Hartogh} P.,  2023, \mn@doi [\mnras] {10.1093/mnras/stac3242}, \href
  {https://ui.adsabs.harvard.edu/abs/2023MNRAS.519...59S} {519, 59}

\bibitem[\protect\citeauthoryear{Sparrow \& Cess}{Sparrow \&
  Cess}{2018}]{sparrow2018radiation}
Sparrow E.~M.,  Cess R.~D.,  2018, Radiation Heat Transfer: Augmented Edition.
Routledge

\bibitem[\protect\citeauthoryear{Spohn et~al.,}{Spohn
  et~al.}{2015}]{Spohn:2015}
Spohn T.,  et~al., 2015, \mn@doi [Science] {10.1126/science.aab0464}, 349

\bibitem[\protect\citeauthoryear{{Sunshine} et~al.,}{{Sunshine}
  et~al.}{2006}]{Sunshine:2006}
{Sunshine} J.~M.,  et~al., 2006, \mn@doi [Science] {10.1126/science.1123632},
  \href {https://ui.adsabs.harvard.edu/abs/2006Sci...311.1453S} {311, 1453}

\bibitem[\protect\citeauthoryear{{Tazaki} \& {Dominik}}{{Tazaki} \&
  {Dominik}}{2022}]{Tazaki2022}
{Tazaki} R.,  {Dominik} C.,  2022, \mn@doi [\aap]
  {10.1051/0004-6361/202243485}, \href
  {https://ui.adsabs.harvard.edu/abs/2022A&A...663A..57T} {663, A57}

\bibitem[\protect\citeauthoryear{Thomas, Davidsson, Jorda, K{\"u}hrt,
  Marschall, Snodgrass  \& Rodrigo}{Thomas et~al.}{2021}]{Thomas2021cometary}
Thomas N.,  Davidsson B.,  Jorda L.,  K{\"u}hrt E.,  Marschall R.,  Snodgrass
  C.,   Rodrigo R.,  2021, Cometary Science: Insights from
  67P/Churyumov-Gerasimenko.
Space Sciences Series of ISSI, Springer Netherlands, \url
  {https://books.google.de/books?id=fnMjzgEACAAJ}

\bibitem[\protect\citeauthoryear{Tien}{Tien}{1988}]{Tien1988}
Tien C.~L.,  1988, \mn@doi [Journal of Heat Transfer] {10.1115/1.3250623}, 110,
  1230

\bibitem[\protect\citeauthoryear{Tien \& Vafai}{Tien \&
  Vafai}{1979}]{Tien:1979}
Tien C.,  Vafai K.,  1979, in 2nd Thermophysics and Heat Transfer Conference.
  p.~874

\bibitem[\protect\citeauthoryear{Tsang}{Tsang}{1985}]{Tsang1985}
Tsang L.,  1985, Theory of microwave remote sensing.
Wiley series in remote sensing, Wiley, New York

\bibitem[\protect\citeauthoryear{Tsang, Kong, Ding  \& Ao}{Tsang
  et~al.}{2001}]{Tsang2001}
Tsang L.,  Kong J.~A.,  Ding K.-H.,   Ao C.~O.,  2001, Scattering of
  Electromagnetic Waves: Numerical Simulations.
John Wiley {\&} Sons, Inc.

\bibitem[\protect\citeauthoryear{Tsang, Pan, Liang, Li, Cline  \& Tan}{Tsang
  et~al.}{2007}]{tsang2007}
Tsang L.,  Pan J.,  Liang D.,  Li Z.,  Cline D.~W.,   Tan Y.,  2007, IEEE
  Transactions on Geoscience and Remote Sensing, 45, 990

\bibitem[\protect\citeauthoryear{Van~der Held}{Van~der
  Held}{1952}]{vanderHeld1952}
Van~der Held E.,  1952, Applied Scientific Research, Section A, 3, 237

\bibitem[\protect\citeauthoryear{Vortmeyer}{Vortmeyer}{1978}]{vortmeyer1978radiation}
Vortmeyer D.,  1978, in International Heat Transfer Conference Digital Library.

\bibitem[\protect\citeauthoryear{Wesselink}{Wesselink}{1948}]{wesselink1948heat}
Wesselink A.,  1948, Bulletin of the Astronomical Institutes of the
  Netherlands, 10, 351

\bibitem[\protect\citeauthoryear{{Whipple}}{{Whipple}}{1950}]{Whipple:1950}
{Whipple} F.~L.,  1950, \mn@doi [Astrophys. J.] {10.1086/145272}, 111, 375

\bibitem[\protect\citeauthoryear{Yang, Howell  \& Klein}{Yang
  et~al.}{1983}]{Yang1983}
Yang Y.~S.,  Howell J.~R.,   Klein D.~E.,  1983, Journal of Heat Transfer, 105,
  325

\makeatother
\end{thebibliography}

\vspace{-10pt}

\setcounter{secnumdepth}{0}
\section{APPENDIX A: simulation of test particle diffusion}
\renewcommand\thefigure{\arabic{figure}A}    
\renewcommand\thetable{\arabic{table}A}    
\setcounter{figure}{0}   
\setcounter{table}{0}

\begin{figure}
\includegraphics[width=\columnwidth]{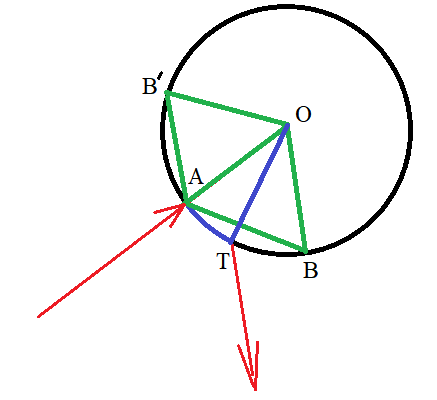}
\vspace{-10pt}
\caption{A sketch explaining how to simulate the interaction of a test particle with an mm-sized aggregate. The circle shows the boundary of the pseudo-monomer. Point A is the entry point, point T is the exit point. The arc BB' is the area defining the scattering nonlocality in the model or the area limiting the position of the possible exit.}
\vspace{-6pt}
\label{fig:Fig_1A}
\end{figure}

\begin{figure}
\centering
\begin{subfigure} {\columnwidth}
\includegraphics[width=\columnwidth]{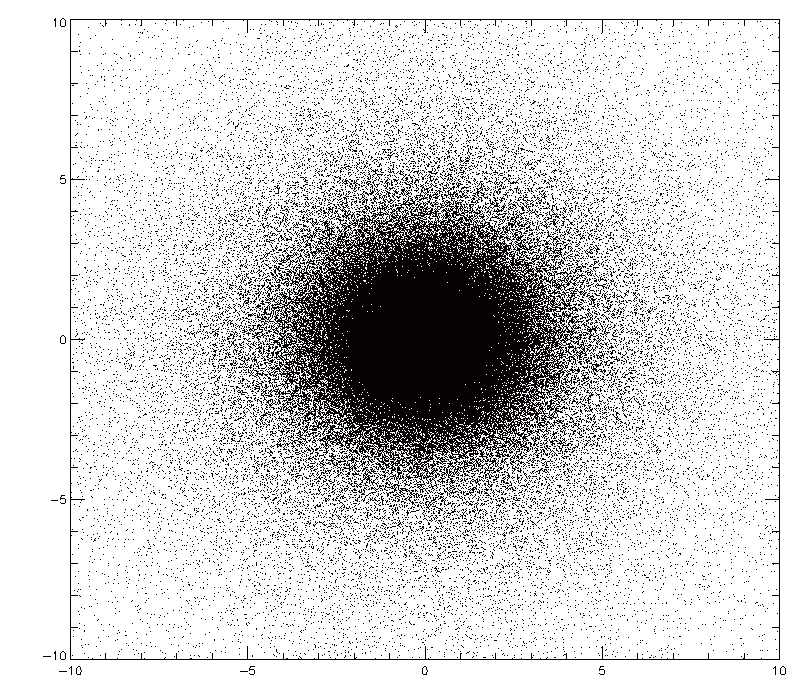}
\end{subfigure}
\hfill
\begin{subfigure} {\columnwidth}
\includegraphics[width=\columnwidth]{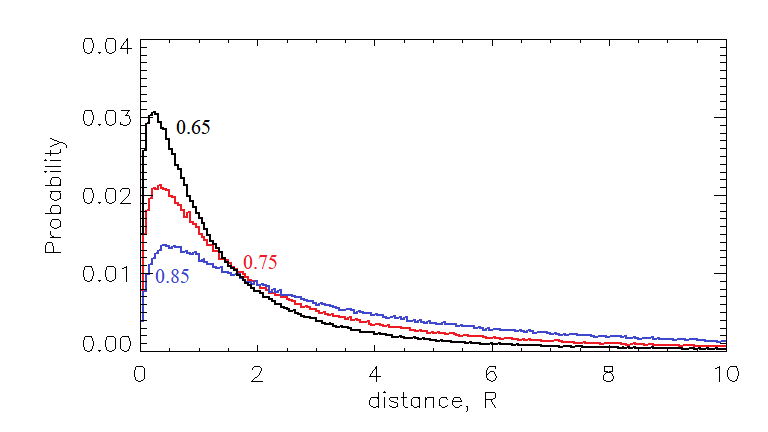}
\end{subfigure}
\caption{The top panel shows statistics on the distribution of exit points for a given entry point at the origin. The bottom panel shows the probability distribution of the position of the exit point (or what is the same distance from the entry point) as a function of the distance in radii of the monomers that make up the aggregate.}
\vspace{-6pt}
\label{fig:Fig_2A}
\end{figure}

As explained in Section 2, the completely deterministic approach to describe the elementary act of scattering of a test particle used in all previous studies is not applicable to layers containing very large  aggregates. The model layers of millimetre-sized aggregates must contain billions of spheres, which makes it extremely difficult to simulate the passage of millions of test particles. Possible ways to solve the  computational problem include the use of complex computational approaches using the ideas of parallel programming, dynamic modelling, and ray tracing in hierarchical structures. However, for our purposes, a much simpler approximate approach can be used. This allows us to maintain approximately the same computational complexity of the model that allowed us to conduct the research efficiently on a regular desktop. In addition, we focus on the further use of the obtained averaged statistical estimates in thermophysical models.

The basic idea goes back to the approach of describing complex porous aggregates as pseudo-monomers (\citealp{Skorov:2018MNRAS}). Keeping in mind the two significantly different porosity scales (inside and between aggregates), we describe the porous medium as a layer of spherical pseudo-monomers. Consider an aggregate with a radius of about a thousand monomer sizes and a test particle that has flown into the aggregate. From our previous permeability simulation for plane parallel cases, it is clear that the probability for a test particle to end up in the “inner” regions of the aggregate is negligible. The overwhelming number of elementary collisions-scatterings will occur in the outer “boundary” layer, say, with a thickness of about a hundred monomer sizes. Thus, it is natural to restrict our simulation to this region. Earlier, we considered both diffuse and specular scattering locally: the point of collision of the test particle with the sphere (the entry point) was also the exit point. 
In the new approach, the description of the test particle scattering act is no anymore approximating as elementary and local (point-scattering).

Let the particle enter the aggregate at point A (Fig. \ref{fig:Fig_1A}). Then, after the scattering chain, it will leave the aggregate at point T, which, due to the obvious symmetry of the problem, is characterized by the distance from the entry point. This is the first characteristic of the exit point. The second characteristic of the exit point is the angular orientation of the plane tangent to the point of the exit on the sphere (boundary of the pseudo-monomer). If we know the distance from the entry point and the radius of the pseudo-monomer this angle is completely determined. In the end, we define the   diffuse scattering into the external half-space with respect to this found plane.

Since the estimated thickness of this “penetration” layer is an order of magnitude smaller than the radius of the aggregate, it is natural to use the plane-parallel approximation to describe the motion of the test particle under consideration and to find the exit distance. In this simplification, we replace the spherical segment with a plane-parallel layer. Thus, we use the previous models of the porous layers to estimate the exit distance. In the first stage, we estimate the probability distribution function of the “bias”, i.e. the probability that the test particle will fly out of a circular segment of radius R (region BB' in Fig.\ref{fig:Fig_1A}). 
To do this, we simulate the movement of a particle in a homogeneous layer consisting of monomers. Such layers of different porosity were studied in (\citealp{Skorov:2021}, \citealp{Reshetnyk:2021SoSyR}). An example of a scatterplot for a 65\% porosity layer is shown in Fig. \ref{fig:Fig_2A} (top panel). 
It is clearly seen how rapidly the probability  decreases even at distances of about five monomer sizes. Examples of probability distribution functions for different porosity cases are shown in Fig. \ref{fig:Fig_2A} (bottom panel). It should be noted that the exact form of the function for emitted particles depends on the angle at which the particle enters the layer. As expected, a particle entering the layer at a large angle to the outer normal boundary of the layer is less likely to fly away near the starting point. Nevertheless, the calculations showed that these variations do not have a noticeable statistical bias, and in the future, we did not take them into account using a single function for the distance distribution.

Using the obtained distribution function, one can randomly generate the coordinate of the exit point. For this, it is convenient to use polar coordinates. The azimuthal angle is distributed uniformly from 0 to 360 due to the symmetry of the problem. To generate it, a standard C++ 32-bit pseudo-random number generator was used by the Mersenne vortex algorithm (The Mersenne Twister) with a size of 19937 bits. To generate a random radius, the classical algorithm of the elimination method (Bird) was used according to a probability function specified in a tabular way using pseudo-random numbers of a uniform distribution based on the Mersenne vortex algorithm.

After generating new coordinates on the pseudo-sphere, simulation of the motion of the test particle outside a pseudo-monomer was continued and a new collision point was calculated. The sequence of steps was repeated until the specified size of the statistical sample was reached. As in previous models, calculations usually continued until a hundred thousand particles left the layer. Based on the statistical processing of the obtained results, we obtained estimates for the mean free path of the test particle and the permeability of the layer as functions of its effective porosity.

\vspace{-10pt}
\section{APPENDIX B: Radiative transfer solution for heat transfer in particulate media. }
\renewcommand\thefigure{\arabic{figure}B}    
\renewcommand\thetable{\arabic{table}B} 
\renewcommand\theequation{\arabic{equation}B} 
\setcounter{figure}{0}   
\setcounter{table}{0}    
\setcounter{equation}{0} 

Let us consider radiative transport in a semi-infinite layer of randomly packed hard spheres. When the volume fraction of spheres is low, one can solve a radiative heat transfer problem using the radiative transfer equation (RTE) \citep{modest2013}. However, when the volume fraction exceeds a few percents, the RTE is no longer valid \citep{Tsang2001}. A common approach to extend the applicability range of the RTE is to account for the far-field interferences due to the correlated positions of spheres using the analytical Percus-Yevick pair distribution function for hard spheres \citep{cartigny1986}. Such a correction gives rise to the so-called dense medium radiative equation (DMRT) \citep{Tsang1985}. The DMRT resembles the standard RTE except for the phase function and scattering coefficients are modified. It is also possible to numerically compute the required phase function and the scattering coefficients including the near field effects which are excluded in the analytical correction \citep{markkanen2019}.  

Here, we will present  a Monte Carlo solution for heat transfer in a semi-infinite random medium based on the DMRT. We modify the 3D solver introduced in \citep{markkanen2019,markkanen2020} to one dimension in order to deal with large structures instead of small particles. Assuming local thermodynamic equilibrium the energy balance equation in three dimensions for simultaneous conduction and radiation without additional sources can be written as 

\begin{equation}
    \rho c_p \frac{\partial T}{\partial t} = \nabla \cdot (K_{\mathit{}} \nabla T ) - \nabla \cdot \mathbf q_r.
\end{equation}

Here, $K_{\mathit{}}$ is the heat conduction coefficient that does not include radiation (radiative heat conduction is included in the last term of the right side of the equation), and $\mathbf q_r$ is the radiative flux, $\rho$ is the density and $c_p$ is the   specific heat capacity of the material. At the interface of the medium and free space, the normal component ($\mathbf n \cdot$) of heat conduction vanishes 

\begin{equation}
\mathbf n \cdot K_{\mathit{ }} \nabla T = 0.
\end{equation}

For a one-dimensional planar medium, the energy balance equation is reduced to

\begin{equation}
    \rho c_p \frac{\partial T}{\partial t} = \frac{d}{dz} K_{\mathit{ }} \frac{dT}{dz} - \frac{dq_r}{dz}
    \label{eq_energybalance1D}
\end{equation}

where the divergence of the radiative heat flux is given by

\begin{equation}
    \frac{dq_r}{dz} = \int_0^\infty \kappa_\lambda (4\pi I_b - G) d\lambda.
    \label{eq_dqr}
\end{equation}

The first term on the right-hand side corresponds to emitted energy where $I_b$ is the black body intensity and $\kappa_\lambda$ is the absorption coefficient depending on the wavelength. The second term corresponds to the absorbed radiation with $G$ being the total spectral intensity, and it has the opposite sign as absorption increases the system's energy while emission decreases. 

We utilised the finite-element method to solve the equation (\ref{eq_energybalance1D}). The resulting weak formulation reads as: 

Find $T \in H^1(\Omega)$ such that 

\begin{equation}
\int_{\Omega} w\rho c_p \frac{\partial T}{\partial t}dl + \int_{\Omega} \frac{dw}{dz} K_{\mathit{ }} \frac{dT}{dz} dl = -\int_{\Omega} w\frac{dq_r}{dz} dl
\label{eq_weak}
\end{equation}

is valid for all $w \in H^1({\Omega})$. The boundary term in the weak formulation vanishes due to the imposed boundary condition. $H^1(\Omega)$ is a space of square integrable functions whose derivatives are also square integrable, and the domain $\Omega$ is one dimensional. 

Next, we apply the finite-difference formula to the time derivative 

\begin{equation}
    \frac{\partial T}{\partial t} \approx \frac{1}{\tau}(T_{t+1} - T_{t})
\end{equation}

where $\tau$ is the size of the time step and the subscript $t$ is its index. We use the implicit backward Euler time integration scheme, and expand the unknown function $T_t$ into the linear combination of the lowest order nodal functions $u^m$ as $T_t \approx \sum_m x_t^m u^m$ where $x_t^m$ are the unknown coefficients. Using Galerkin's method with the identical testing and basis function $w^n = u^m$, the equation for the temperature coefficients at the future time step ($t+1$) becomes

\begin{equation}
x_{t+1} = (M + \tau S)^{-1}(Mx_t + \tau F x_{t+1})  
\label{eq_iter}
\end{equation}

in which the mass and stiffness matrices are defined as

\begin{equation}
    M = \rho c_p\int_\Omega w^nu^m dl,
\end{equation}

\begin{equation}
    S = K \int_\Omega \frac{dw^n}{dz} \frac{du^m}{dz} dl
\end{equation}

and the force vector reads as 

\begin{equation}
    F = -\int_\Omega w^n \frac{dq_r}{dz}.
\end{equation}

The equation (\ref{eq_iter}) is non-linear as the force vector depends on the temperature distribution. Hence, we use an iterative approach to solve it for each time step.

To evaluate the force vector, we use the Monte Carlo technique to solve the DMRT. We assume that the medium consists of randomly packed spherical particles of equal sizes. With these assumptions, the input parameters are the sphere radius, their refractive indices, and the volume fraction $\phi$. We use the Lorenz-Mie theory to calculate the scattering matrix $M$, scattering and absorption cross sections $C_{\rm sca}$ and $C_{\rm abs}$. Then, the DMRT theory based on the quasicrystalline approximation and the Percus-Yevick pair distribution function for hard spheres \citep{tsang2007,liang2008} are used to compute the modified phase function $M^{\rm DMRT}$, the mean free path $\ell$, the single scattering albedo $SSA$ and the effective refractive index $m_{\rm eff}$.    


The Monte Carlo solver traces rays associated for a given wavelength range $\Delta_\lambda$. For an external source, the rays are generated having a specific direction and a flux density. For thermally emitted radiation, the rays are generated inside the medium having random positions $z$ and directions $\mathbf{\hat{k}}$. The emitted power for each ray is given by 

\begin{equation}
E^e = \int_{\Delta_\lambda} \frac{4N_d}{N_{\rm ray}} \pi B(\lambda, T(z), m_{\rm eff}) C_{\rm abs} d\lambda     
\end{equation}

where $N_{\rm d}$ is the number density of spheres and $N_{\rm ray}$ is the total number of rays. The Planck function in a medium is given by

\begin{equation}
    B_\lambda(\lambda,T,m) = \frac{2hc^2 Re\{m\}^2}{\lambda^5} 
    \Bigl[ \exp (\frac{hc}{\lambda k_B T})-1 \Bigr]^{-1}
\end{equation}

where $h$ is the Planck constant, $c$ is the speed of light and $k_B$ is the Boltzmann constant. The solver uses $E^e$ to update $\kappa_\lambda 4\pi I_b$ in equation (\ref{eq_dqr}).  

A propagation distance for a ray is drawn from the exponential distribution as 

\begin{equation}
    d = -\ell \log \Phi
    \label{eq_dist}
\end{equation}

where $\Phi$ is a uniform random number within [0,1]. Then, the ray is scattered or absorbed depending of the probability defined by the SSA. The absorbed rays are used to update $\kappa_\lambda G$ in equation (\ref{eq_dqr}). If the ray is scattered, a new propagation direction is drawn using the cumulative probability density function defined by the modified phase function $M^{\rm DMRT}$, and a new propagation distance is generated from equation (\ref{eq_dist}). If the ray hits the boundary it is reflected and refracted according to Snel's law and the Fresnel coefficients defined by $m_{\rm eff}$. 

Assuming that only radiation carries energy we can estimate $K_{\mathit{rad}}$ by fitting the temperature distributions inside the medium computed using the 1D heat diffusion model with $K_{\mathit{rad}}$ as a free parameter to the temperature distribution from the Monte Carlo DMRT solution.



\vspace{-10pt}

\section{APPENDIX C: Modernisation of the thermal model}
\renewcommand\thefigure{\arabic{figure}C}    
\renewcommand\thetable{\arabic{table}C} 
\renewcommand\theequation{\arabic{equation}C} 
\setcounter{figure}{0}   
\setcounter{table}{0}    
\setcounter{equation}{0}   

\begin{table*}
\caption{Model values for ices and thermal conductivity. }
\label{Table_1C}
\begin{tabular}{llll}
\hline
\multirow{2}{*}{H$_2$$\mathrm{O}$} & $P(T)$   & $3.56\times10^{12}\times \exp(-6141.7/T)$ [Pa] & \citealp{Fanale:1984}\\
                       & $H$ & $2.75\times10^{6}$ [J/kg]                        & \citealp{Fanale:1984}\\
\multirow{2}{*}{CO$_2$}   & $P(T)$    & $1.23\times10^{12} \times \exp(-3167.8/T)$ [Pa]    & \citealp{Fanale_etal:1990}\\
                       & $H$ & $5.7\times10^{5}$ [J/kg]                                & \citealp{Mavko:2009}\\
\multirow{2}{*}{CO}    & $P(T)$    & $1.26\times10^{9} \times \exp(-764.2/T)$ [Pa]      & Fanale    et  al.,  1990\\
                       & $H$ & $3\times10^{5}$ [J/kg]                                   & \citealp{Brown:1979} \\
Dust bulk              & $K_{bulk}$& {$1.26 \times 10^{-3} \times T_{dust}+9.94\times10^{-1}$}            & \citealp{Ratcliffe:1963} \\
conductivity    & & &   \\ 
\hline
\end{tabular}
\end{table*}

Analysing the influence of uncertainties in the values of microstructural and thermophysical parameters of the near-surface region of the cometary nucleus on gas production, it is necessary to use a heat transfer model that relates the characteristics of the layer and gas production.
$Model B$, which we first used in (\citealt{Keller:2015b}), explicitly takes into account both the porous dust layer and its effective thermal conductivity based on the energy balance at the nucleus surface and at the ice sublimation front. As noted, it is based on the assumption that a quasi-stationary temperature distribution takes place in the near-surface layer, i.e. the general heat conduction equation (see Equations 1B and 3B) is reduced to a stationary form

\begin{equation}
\nabla \cdot \biggr( K_{\mathit{eff}} \nabla T(x) \biggr) = 0.
\label{eq:1C}
\end{equation}

or equivalently 

\begin{equation}
K_{\mathit{eff}} \nabla T(x) = C = Const.
\label{eq:1Ce}
\end{equation}

Obviously, if our goal is not to find the complete temperature profile in the layer, but to determine its values at the layer boundaries, then the problem is reduced to solving a system of two nonlinear algebraic equations for the variables $T_s$ and $T_i$ - temperatures on the surface and on the sublimation front, respectively

\begin{equation}
(1-A_v) I_{\mathit{eff}}= \epsilon \sigma T_{\mathit{surface}}^4 + K_{\mathit{eff}}\frac{dT}{dx}\bigg\rvert_{x=0}
\label{eq:2C}
\end{equation}

\begin{equation}
K_{\mathit{eff}}\frac{dT}{dx}\bigg\rvert_{x=L^{+}} - K_{\mathit{eff}}\frac{dT}{dx}\bigg\rvert_{x=L^{-}}= \Pi Z(T_i)H  
\label{eq:3C}
\end{equation}

 \noindent where $I_{\mathit{eff}}$ is the solar irradiation, $A_v$ is the bolometric Bond albedo of the dirty ice, $\epsilon$ is the emissivity, $\sigma$ is the Stefan-Boltzmann constant, $Z$ is the sublimation rate, $H$ is the latent sublimation heat, and $\Pi$ is the layer gas permeability that is a function of the layer thickness, porosity and particle size. The sublimation rate is given by the Hertz-Knudsen formula, $Z(T) =P(T)/(0.5\pi v_{th})$, where $P(T)$ is the saturation vapour pressure, $v_{th}(T)=\sqrt{8RT/\pi \mu}$ is the thermal velocity of the vapour molecules and $\mu$ is the molar mass. 

Based on $Model B$ we study the gas production as a function of the layer porosity, thickness, micro-structure and thermo-physical characteristics. The tested parameters are aggregate size, layer thickness, and porosity. The solid and aggregate thermal conductivities are calculated by following \citep{Gundlach:2012}. The main thermophysical parameters are shown in Table 1C.

Let us now consider how the above equations are modified for the cases when $K_{\mathit{eff}}$ includes radiative  thermal conductivity.

In this case, we can write a general expression for effective conductivity as 

\begin{equation} 
K_{\mathit{eff}}(T) = a_1 + a_2  T^3
\end{equation}

where $a_1$ is the constant thermal conductivity of the solid phase and $a_2$ is the constant factor before the cube of temperature in the formulas of radiative conductivity in Section 4. 

Integrating Eq.\ref{eq:1C} we get

\begin{equation}
(a_1 + a_2 T^3 ) \frac{dT}{dx} = C = Const   
\end{equation}

where $C$ is the heat flux, and its constancy reflects the absence of energy sources or sinks inside the layer and the stationary temperature distribution. It is this flow that enters the equations (\ref{eq:2C} - \ref{eq:3C}).

Performing separation of variables and integration, get

\begin{equation}
 C =    \Biggr( a_1 T  + \frac{1}{4} a_2 T^4 \Biggr) \Biggr|_{T_i}^{T_s} L^{-1} 
\end{equation}

where $L$ is the layer thickness. The relation obtained allows using $Model B$ for the case of very large aggregates, taking into account the radiative heat conduction. To solve a nonlinear algebraic system, the same method is used as in (\citealp{Skorov:2022}).

\bsp	
\label{lastpage}
\end{document}